\documentclass[11pt]{article}
\usepackage{theorem,amsfonts}
\usepackage{bezier}

{\theoremstyle{break}
\theorembodyfont{\normalfont}

}
\makeatletter
\@addtoreset{equation}{section}
\makeatother

\renewcommand{\theequation}{\thesection.\arabic{equation}}

\font\tengoth=eufm10 \font\sevengoth=eufm7 \font\fivegoth=eufm5
\newfam\gothfam

\textfont\gothfam=\tengoth \scriptfont\gothfam=\sevengoth
  \scriptscriptfont\gothfam=\fivegoth
  \def\goth{\fam\gothfam}    
\font\frak=eufm10 scaled\magstep1

\def\goth#1{\hbox{{\frak#1}}}

\def\be{\begin{equation}}
\def\ee{\end{equation}}
\def\bea{\begin{eqnarray}}
\def\eea{\end{eqnarray}}

\def\Re{{\mathbb R}}
\def\Ce{{\mathbb C}}
\def\He{{\mathbb H}}
\def\Oe{{\mathbb O}}
\def\Ze{{\mathbb Z}}

\def\k{\kappa}
\def\Invol{\Pi}
\def\dual{{\cal D}}  

\def\stH{{\cal H}}
\def\stP{{\cal P}}
\def\stK{{\cal K}}

\def\c{C_{\k_1}}
\def\cc{C_{\k_2}}
\def\ccc{C_{\k_1\k_2}}
\def\s{S_{\k_1}}
\def\ss{S_{\k_2}}
\def\sss{S_{\k_1\k_2}}
\def\v{V_{\k_1}}
\def\vv{V_{\k_2}}

\def\T{T_{\k_1}}
\def\TT{T_{\k_2}}
\def\TTT{T_{\k_1\k_2}}

\def\carea{C_{\k_1^2 \k_2}}
\def\sarea{S_{\k_1^2 \k_2}}
\def\tarea{T_{\k_1^2 \k_2}}
\def\ccoarea{C_{\k_1 \k_2^2}}
\def\scoarea{S_{\k_1 \k_2^2}}
\def\tcoarea{T_{\k_1 \k_2^2}}

\def\apa{a P_a}
\def\bpb{b P_b}
\def\cpc{c P_c}
\def\AJA{A J_A}
\def\BJB{B J_B}
\def\CJC{C J_C}

\def\half#1{\frac{#1}{2}}

\def\exc{e}
\def\EXC{E}
\def\area{{\cal{S}}}
\def\coarea{s}

\def\myexp#1{e^{#1}}
\def\mifrac#1#2{\frac{#1}{#2}}
\def\overto#1{\stackrel{#1}{\to}}

\def\orto{{\,\line(0,1){8}\line(1,0){8}\!\!\! \cdot}\,\,}
\def\alTusi{al-\d{T}\={u}s\={\i}\ }

\def\ta{\tau_a}\def\tb{\tau_b}\def\tc{\tau_c}\def\tp{\tau_p}
\def\xA{\chi_A}\def\xB{\chi_B}\def\xC{\chi_C}\def\xP{\chi_P}
\def\tiempo{\tau}
\def\closeminus{\!\!-\!\!}

\def\CKPointSpace{ S^2_{[\k_1],\k_2}}
\def\CKLineSpace{{ S^2_{\k_1,[\k_2]}}}

\begin{document}

\noindent 
\ 
\bigskip

\begin{center}
{\LARGE{\bf{Trigonometry of spacetimes: \\  a new self-dual
approach to a\\[0.1cm]  curvature/signature (in)dependent
trigonometry}}}
\end{center}

\bigskip

\begin{center}
Francisco J. Herranz$^\dagger$,
Ram\'on Ortega$^\star$,
and Mariano Santander$^\star$
\end{center}

\begin{center}
{\it $^\dagger$ Departamento de F\'{\i}sica, Escuela
Polit\'ecnica Superior\\
Universidad de Burgos, E--09006 Burgos, Spain}
\end{center}

\begin{center}
{\it $^{\star}$ Departamento de F\'{\i}sica Te\'orica,
Facultad de Ciencias\\ Universidad de Valladolid,
E--47011 Valladolid, Spain}
\end{center}

\bigskip

\begin{abstract}
\noindent
A new method to obtain trigonometry for the real
spaces of constant curvature and metric  
of any (even degenerate) signature is presented. The method could be
described as `curvature/signature (in)dependent trigonometry' and 
encapsulates trigonometry for all these spaces into a
single {\em basic trigonometric group equation}. This brings
to its logical end the idea of an `absolute trigonometry',
and provides equations which hold true for  the
nine two-dimensional spaces of constant curvature and any
signature.  This family of spaces includes both
relativistic and non-relativistic homogeneous spacetimes; 
therefore a complete discussion of
trigonometry in the six de Sitter, minkowskian, Newton--Hooke and galilean
spacetimes follow as particular instances of the general
approach.

Distinctive traits of the method are `universality' and
`(self-)duality': every equation is meaningful for the
nine spaces at once, and displays explicitly invariance
under a duality transformation relating the nine spaces
amongst themselves. These basic structural properties
allow a complete study of trigonometry  and in fact {\em
any} equation previously known for the
three classical (riemannian) spaces also has a version
for the remaining six `spacetimes'; in most cases these
equations are new. The derivation of
the single basic trigonometric equation at group level,
its translation to a set of equations (cosine, sine and
dual cosine laws) and the natural apparition of angular and lateral excesses,
area and coarea are explicitly discussed in detail. 

The exposition also aims to introduce the main ideas of
this direct group theoretical way to trigonometry; this
can be successfully applied for other rank-one spaces as
well (e.g. the complex type, as the quantum space of
states), and may well provide a path to systematically
study trigonometry for any homogeneous symmetric space.
\end{abstract}

\bigskip

\newpage


\section*{Contents}

\noindent
1. Introduction\hfill \pageref{Sec:i}

\noindent
2. Cayley--Klein geometries and spaces of real type    \hfill
\pageref{Sec:ii}

2.1. Cayley--Klein geometries in dimension $N$\hfill \pageref{Sec:ii:i}

2.2. The nine two-dimensional real Cayley--Klein geometries\hfill
\pageref{Sec:ii:ii}

2.3. Spacetimes as  Cayley--Klein spaces\hfill \pageref{Sec:ii:iii}

2.4. Realization of the spaces of points\hfill \pageref{Sec:ii:iv}

\noindent
3. The compatibility conditions for a triangular
loop\hfill \pageref{Sec:iii}

3.1. Loop excesses and  loop equations\hfill \pageref{Sec:iii:i}

3.2. The basic trigonometric identity\hfill \pageref{Sec:iii:ii}

\noindent
4. The basic equations of trigonometry in the nine Cayley--Klein
spaces\hfill \pageref{Sec:iv}

4.1. Alternative forms for the cosine theorems\hfill \pageref{Sec:iv:i}

4.2. Dependence and sets of basic equations\hfill \pageref{Sec:iv:ii}

4.3. Relation with the usual approach and with absolute
trigonometry\hfill \pageref{Sec:iv:iii}

4.4. A compact notation\hfill \pageref{Sec:iv:iv}

4.5. Area and coarea and the dualities length/area
and angle/coarea\hfill \pageref{Sec:iv:v}

4.6. The trigonometric  equations in the {\em minimal}
form\hfill \pageref{Sec:iv:vi}

\noindent
5. A trigonometric bestiarium\hfill \pageref{Sec:v}

5.1. Equations of  Euler, Gauss--Delambre--Mollweide and
  Napier\hfill \pageref{Sec:v:i}

5.2. Equations  for area and coarea\hfill \pageref{Sec:v:ii}

5.3. Some historical comments\hfill \pageref{Sec:v:iii}

5.4. Existence conditions\hfill \pageref{Sec:v:iv}

\noindent
6. Other types of triangles\hfill \pageref{Sec:vi}

6.1. Second-kind triangles\hfill \pageref{Sec:vi:i}

6.2. Orthogonal triangles\hfill \pageref{Sec:vi:ii}

\noindent
7. On the trigonometry of homogeneous spacetimes\hfill
\pageref{Sec:vii}

\noindent
8.  Concluding remarks\hfill \pageref{Sec:viii}

\noindent
Acknowledgments\hfill \pageref{Sec:Ack}

\noindent
Appendix: Some relations for the trigonometric functions\hfill
\pageref{Sec:Ap}

\noindent
References\hfill \pageref{Sec:Bib}

\bigskip

\newpage


\section{Introduction}
\label{Sec:i}

Trigonometry of relativistic homogeneous, constant curvature 
models of spacetimes (anti-de Sitter, Minkows\-ki and de Sitter) is
the most elementary part of the geometry in these spacetimes.
However, it has yet to become part of common mathematical or
theoretical physicist knowledge. Trigonometry in minkowskian
spacetime was first explicitly studied by Birman and Nomizu
\cite{BN} and except for some  results in Yaglom's book
\cite{Yaglom} ---under the names of cohyperbolic and doubly
hyperbolic geometries--- we have not found any explicit formulation
for the trigonometry in either anti-de Sitter or de Sitter
spacetimes, in spite of the very basic nature and strong current
interest in these spaces. Thus a first and short term aim of this
paper is to fill in this gap.

There is also a second, more long term aim. Trigonometry,  the study
of the simplest geometrical configuration in a given space, should
be a basic building block within the specific study of the geometry
of homogeneous symmetric spaces.  Hence this paper should also (and
in the long run, mainly) be read as a step within the general
programme of studying trigonometry of symmetric spaces (see
\cite{Hsiang}--\cite{Aslaksen2}).

Within this perspective, the final and primary aim of this work is
to introduce the ideas and methods of a group theoretical
derivation to trigonometry which we believe to be new. This
approach does not consider trigonometry for a {\em single} space
(for, say, the anti-de Sitter spacetime), but instead is addressed
to providing simultaneously the trigonometry of a {\em whole
family} of spaces. This approach carries to its logical end the
`absolute trigonometry' ---first discussed by Bolyai and then
continued by de Tilly amongst others
\cite{Bonola, Martin}--- which covered simultaneously the three
classical spaces of constant curvature  (sphere, euclidean space and
Lobachewski hyperbolic space). This `absolute trigonometry' can be
considered as a first, albeit partial step in the  direction we are
pursuing here.

There are several distinctive traits in this approach. First,
economy of thought:  a single (parameter dependent) computation
covers at once the trigonometry of several spaces. Second, a clear
view is obtained for relationships between different spaces in the
same family, such as several {\em dualities}; otherwise some of
these may easily pass unnoticed, yet they may provide new
hindsights. Third, limiting (contracted) cases, corresponding to
vanishing curvature and/or degenerate metric, are included and
described at the same level as the generic ones, thus making
completely redundant a separate study of contractions. These traits
apply not only for trigonometry, but also for the study of most
properties of geometries, groups and algebras within each family
\cite{CKdos}--\cite{expansion}; for instance superintegrable
potentials with constants of motion quadratic in the momenta can be
studied simultaneously for the sphere, the euclidean plane and the
hyperbolic plane
\cite{MfrMsSupS2H2}.

All symmetric homogeneous spaces can be classed into several natural
families \cite{FreudGoslar96,MSBurgos,MSClasLieGroups}, each with
their Lie groups of motion, Lie algebras, etc., which depend on some
parameters distinguishing family members. In the simplest cases (the
spaces of real type, associated to the quasi-orthogonal Lie groups
obtained by contraction of $SO(N)$ and $SO(p,q)$) these parameters
determine the {\em curvatures} and/or the {\em signatures} of the
fundamental metric for each space in the family. Additional
parameters in other families label a division algebra ($\Ce, \He,
\Oe$) or a pseudo-division variant coordinatising the space.

The method we are proposing should furnish trigonometry for all
these families of spaces. So far this goal has been completely
accomplished for {\em all} rank-one homogeneous symmetric
spaces, of  real type as well as of complex, quaternionic or
exceptional octonionic (Cayley) type. As it seems impossible to
squeeze all this material ---without seriously impairing the
possibility of a clear exposition--- for both the real and the
complex spaces into a single paper, we feel it justified to
devote a first paper to the introduction of the method and  the
discussion of the  trigonometry of the nine two-dimensional
spaces of real type. This will also serve as a background to
underlie a follow up forthcoming paper
\cite{HermTrig} devoted to the trigonometry of complex spaces.
Therefore in this paper we restrict to a complete and
detailed discussion of trigonometry of the rank-one symmetric
homogeneous spaces of real type, also called quadratic or orthogonal
Cayley--Klein spaces (see e.g.\ \cite{SanHerrDubna96,
AzcHerPerSanCohomSO}). Any three points in any rank-one real type
homogeneous symmetric space are always contained in a
two-dimensional (2D) totally geodesic submanifold, so considering
only 2D spaces (planes) is no restriction at all.

There are $3^N$ real Cayley--Klein spaces in dimension
$N$, thus  nine  2D real quadratic Cayley--Klein  spaces
\cite{Yaglom}:  the sphere, euclidean and hyperbolic planes, the
co-euclidean, galilean and co-minkowskian planes and finally the
co-hyperbolic, minkowskian and doubly hyperbolic planes. Out of this
list only the three spaces mentioned in the first place belong to
the restricted family of the  so-called {\em two-point homogeneous
spaces} whose trigonometry is very well known. The remaining six
spaces are not two-point homogeneous according to the usual
definition \cite{Wang} (because the action of the isotropy
group on the tangent space `unit sphere' is not transitive),
but together with the three previous ones they provide a
natural frame for a joint study of  trigonometry. Within a
concrete physical interpretation these six spaces are the
$(1+1)$D symmetric homogeneous spacetimes: oscillating (or
anti) Newton--Hooke, galilean, expanding Newton--Hooke 
$(1+1)$D spacetimes, and anti-de Sitter, minkowskian and de
Sitter $(1+1)$D spacetimes. The trigonometry of the two
constant curvature counterparts of the special relativity
spacetime, mentioned as a short term first aim of this work,
follows as a sideline byproduct. The required information on
Cayley--Klein  spaces of real type, in particular on the nine
2D spaces, is given in Section  2.

The method we propose is presented in Section 3. It  embodies the
trigonometry for the whole biparametric family of these real 2D
Cayley--Klein  spaces into a {\em single} group equation for the
parameter-dependent group of motions of the corresponding space.
Dealing with a whole family at once, this equation, that we call the
{\em basic trigonometric identity}, gives a  perspective on some 
relationships between different spaces which goes beyond the
treatments devoted to the study of a single space; this is so
because these relationships involve  simultaneously several (and at
least two) different spaces. The simplest such relations are the
{\em contractions} (e.g.\ from anti-de Sitter or de Sitter
trigonometries to the minskowskian one), whose description is
built-in automatically in the Cayley--Klein  scheme. However, there
are others, such as a fundamental {\em duality} (or polarity) and
an interesting {\em triality}. Duality is the main structural
backbone in our approach, and the requirement to explicitly
maintaining duality in all expressions and at all stages acts as a
kind of method `fingerprint'. This duality should not be confused
with the ordinary Cartan duality for  pairs of symmetric spaces
\cite{Helgason} which also has a very natural description in our
approach.

The basic trigonometric equation is also directly related to other
product formulas, which embody the Gauss--Bonnet theorem for
triangles as well as its dual theorem; the proof for these
equations is also given as they appear as an integral part in the
derivation of the basic trigonometric identity. Each of these
formulas contains trigonometry in a nutshell, but to unveil
explicitly the trigonometric equations, it is better to start from
the basic trigonometric identity. This procedure allows a very
rapid browsing through the complete {\em zoo} of trigonometric
equations for the nine spaces with an effort definitely lower than
that required for studying any single case, because duality acts as
a kind of `superstructural' requirement. All the equations we
obtain for any of the nine 2D spaces in the
 real Cayley--Klein  family are very well known in the three
constant curvature {\em riemannian} cases (spherical, euclidean and
hyperbolic) but we have not found any reference to the anti-de
Sitter and de Sitter versions of most of these formulas, especially
those involving areas and coareas of triangular loops. All this
material is covered in Sections 4 and 5, and includes a complete
{\it bestiarium} of equations valid in all the nine spaces, as well
as some brief historical comments to relate the general equations
to their spherical or euclidean counterparts (which have been known
for centuries). A length/area duality  recently pointed out by
Arnol'd
\cite{ArnGeomSphCur} also follows naturally from our approach,
suggesting the extension of the spherical triality introduced there
to a triality between the hyperbolic plane, the anti-de Sitter
spacetime and the de Sitter spacetime which would be worth
studying.

In the generic 2D Cayley--Klein   space there may be two types of
lines (time- and space-like in the kinematical interpretation). The
main body of the paper deals with triangles whose sides are of a
single particular type, but a brief report of the results obtained
for the remaining types of triangles is given in Section 6; these
include triangles with two orthogonal sides. In Section 7 we
translate some of the results to the kinematical language, and
offer several trigonometric relations for the de Sitter and anti-de
Sitter spacetimes, as well as for their non-relativistic analogous.

There is a small price to pay for this wide scope: the use of a
non-standard notation in which the circular, hyperbolic and their
common intermediate {\em parabolic} functions appear altogether
under the guise of `labeled' trigonometric functions, which reduce
to the two familiar cases for the label values $\k=1, -1$ (see e.g.\
\cite{CKdos,Poisson}).  The main part of the paper is couched
using these functions, but to help the reader get a quick
appraisal, we have included several tables in which a sample of
the results are displayed in the conventional notation; these
examples may help to translate any of the equations we give
either to their `natural' form, where the non-zero labels are
reduced to their standard $1, -1$ values by using `natural'
units as in table
\ref{table:BasicTrigEqns},  or to a form involving explicitly
the constants determining the curvature and signature of each
space, as  the universe (time) radius $\tiempo$  and the
relativistic constant
$c$ in table
\ref{table:EqnsSixSpacetimes}. A Section with some final comments
and prospects for continuation of this work closes the paper.

Although it is not our aim to cover here any applications, we should
point to the relevance of many of the complicated trigonometric
equations whose `general' form we derive in several fields. For
instance they appear in the Zamolodchikov solution for tetrahedral
equations as factorization condition for the $S$-matrix  in
$(1+2)$D \cite{Baxter, Zamol}, both reproduced in \cite{Jimbo}.
The extension of the Moyal type formulation of Quantum
Mechanics to spaces with constant curvature also involves many of the
complicated equations for spherical or hyperbolic area in
terms of sides. Trigonometry in the relativistic de
Sitter spacetimes can be also used to get understanding
of their global structure, the presence of horizons,
etc., just as hyperbolic trigonometry is essential for a
detailed understanding of the hyperbolic plane. 


\section{Cayley--Klein geometries and spaces of real type}
\label{Sec:ii}


\subsection{Cayley--Klein geometries in dimension $N$}
\label{Sec:ii:i}

We only give a brief summary of the $N$-dimensional Cayley--Klein
(herafter CK) geometries, enough to allow a meaningful discussion of
trigonometry in $(1+3)$D spacetimes. For more specific
details, the reader can consult
\cite{SanHerrDubna96, AzcHerPerSanCohomSO, Jaca}.

The real Lie algebra $\goth{so}(N+1)$ has a
${\Ze}_2^{\otimes N}$ group of commuting involutive automorphisms
determining a grading. A particular subfamily of all graded
contractions of $\goth{so}(N+1)$ depend on $N$ real parameters $\k_1,
\k_2, \dots, \k_N$ and are called  orthogonal CK algebras, since they
are exactly the motion algebras of the geometries of a real space with
a projective metric in the CK sense
\cite{Sommer,Yas}. In this theory, the pencil of points in a line
can be either elliptic/parabolic/hyperbolic, the pencil of lines
through a point in a 2-plane can be also 
elliptic/parabolic/hyperbolic, the same happens for the pencil
of 2-planes through a line in a 3-plane, etc. Each alternative
is described by one of the real  constants,
$\k_1, \k_2, \k_3,\dots$, and the elliptic/parabolic/hyperbolic
character corresponds in each case to $\k_1 ,  \k_2, \k_3,
\dots$ being 
$>0/=0/<0$.

These algebras depending on $N$ real coefficients are denoted as
$\goth{so}_{\k_1,\k_2, \dots, \k_N}(N+1)$. Their $(N+1)N/2$
generators are
$P_i$, $J_{ij}$, $i,j=1, \dots N$, $i<j$, and have a vector
representation by
$(N+1) \times (N+1)$ real matrices:
\be
P_i=-\k_{0i}e_{0i}+e_{i0}  \qquad J_{ij}=-\k_{ij}e_{ij}+e_{ji} 
\label{zzz:ccd}
\ee
where the $\k$ with two indices are defined as
$\k_{ab}:=\k_{a+1}\k_{a+2}\dots\k_b$ and $e_{ab}$ is the matrix
with a single  non-zero entry, 1,  in the row $a$ and column
$b$, $a, b = 0, 1, \dots, N$. These expressions suggest a close association
$P_i \leftrightarrow \k_{0i}=\k_1\k_2\dots\k_i$ and
$J_{ij}\leftrightarrow \k_{ij}=\k_i\k_{i+1}\dots\k_j$;  the $\k$
associated to each generator will be called its label. 

By
matrix exponentiation this matrix realization generates a group
of matrices of order
$N+1$ denoted $SO_{\k_1,\k_2, \dots, \k_N}(N+1)$. When all constants
are different from zero the group  $SO_{\k_1,\k_2, \dots, \k_N}(N+1)$
is a simple group isomorphic to a (pseudo) orthogonal group $SO(p, q)$
with $(p, q)$ being the number of (positive, negative) terms in the
sequence $(1, \k_1, \k_1\k_2,\k_1\k_2\k_3, \dots,
\k_1\k_2\cdots\k_N)$. When some constants are equal to zero
$SO_{\k_1,\k_2, \dots, \k_N}(N+1)$ is a non-simple contraction of
$SO(p, q)$; this group is near enough to a simple group to warrant the name
quasi-simple. By suitable scale changes in the generators, all
constants $\k_i$ can be reduced to their `canonical' values $1, 0,
-1$. The notation has been chosen so that when all constants $\k_i$
are equal to $1$, the group $SO_{\k_1,\k_2, \dots, \k_N}(N+1)$ reduces
to $SO(N+1)$.

Each constant $\k_i$ (or $\k_{ab}$) is linked to an
involutive automorphism of the Lie algebra, so a symmetric
homogeneous space can be obtained as the coset space
$SO_{\k_1,\k_2, \dots, \k_N}(N+1)/H_{(i)}$ where the subgroup
$H_{(i)}$ is generated by the elements in the Lie algebra which
are invariant under the involution associated to the constant
$\k_i$. In particular, the involution associated to
$\k_1$ has as the subgroup $H_{(1)}$ the CK group $SO_{\k_2, \dots,
\k_N}(N)$ generated by the $J_{ij}, i,j=1, \dots N, i<j$.

The more relevant point to stress is the double role the constants
$\k_i$ play. From the projective viewpoint they determine the nature
of the pencil of points in a line, of a pencil of lines in a 2-plane,
etc. From the metric viewpoint these constants determine the
curvature and the signature of the quadratic metric naturally
induced in each of the spaces
$SO_{\k_1,\k_2, \dots, \k_N}(N+1)/H_{(i)}$ by the
`Killing--Cartan' metric in the group (see \cite{Jaca} for
details). Specifically,
$\k_i$ is the {\em  curvature} (`constant' in a suitable sense)
of this homogeneous space, and the remaining
$\k$ constants  determine the signature of the metric. For the
particular homogeneous space
$SO_{\k_1,\k_2, \dots, \k_N}(N+1)/SO_{\k_2, \dots, \k_N}(N)$
the curvature is constant in the usual sense and equals to $\k_1$,
while the signature of the metric is given by the diagonal matrix with
entries $(1, \k_2, \k_2\k_3, \dots, \k_2\cdots\k_N)$.

Let us comment on some of the spaces in the family $SO_{\k_1,\k_2,
\dots, \k_N}(N+1)/SO_{\k_2, \dots, \k_N}(N)$. When all constants are
equal to $1$, this space is the standard $N$D sphere $S^N$
of curvature $1$; the values $\{ \k_1, \k_2, \dots, \k_N\} = \{\k>0,
1, \dots ,1\}$ lead to the $N$D sphere of (positive) curvature $\k$.
Other choices are $\{\k<0, 1, \dots , 1\}$, leading to the $N$D
Lobachewski space $H^N$ of (negative) curvature $\k$,  or
$\{ 0, 1,\dots 1\}$ which leads to the flat $N$D euclidean space
$E^N$. Thus the three classical homogeneous spaces of constant
curvature and definite positive metric are all included in the
CK family. But there are many others. For instance, the $(1+3)$D
Minkowski spacetime of special relativity corresponds to $\{0,
-1, 1, 1\}$: it is a flat space with a metric of $(1, 3)$ type.
The projective interpretation also agrees with the choice of
$\k_i$ for Minkowski spacetime: the pencil of events in a
time-like line is parabolic (separation measured by proper time,
$\k_1=0$), the pencil of time-like lines through an event in a
time-like 2-plane is hyperbolic (separation measured by the
rapidity, $\k_2<0$), while the pencils of time-like 2-planes in
a time-like 3-plane and of time-like 3-planes in the
$(1+3)$D spacetime are clearly elliptic (separation measured by
plane and dihedral space angles, $\k_3 >0$, $\k_4 >0$). The
anti-de Sitter spacetime has
$\k_1>0$ (that is, the pencil of events on a time-like line is
elliptic), and the same remaining constants as  the minkowskian
one, etc.  

The CK algebras, groups and spaces have many subgroups, subalgebras and
subspaces of the same CK type. Generically, any translation in the
space $SO_{\k_1,\k_2, \dots, \k_N}(N+1)/SO_{\k_2, \dots, \k_N}(N)$ is
conjugated to   (at least) one of the $N$ basic translations $P_1,
P_2, \dots ,P_N$. Three points in such a space will determine three
translations, which will translate along lines contained in a 2-plane,
and the rotation generator in this plane will be also generically
conjugated to (at least) one of the
$N(N-1)/2$ rotation generators $J_{ij}$. Generically and up to a group
motion, these three points can always be assumed to lie on the 2-plane
generated by say $P_m, P_n, m<n$; these generators together with
the rotation generator $J_{mn}$ close a CK subalgebra
$\goth{so}_{\k_{0m},\k_{mn}}(3)$. Thus consideration  of the 2D case
will suffice for a complete study of trigonometry, and therefore in
the rest of the paper we  will only discuss the
2D spaces, described by two constants $\k_1, \k_2$.
When considering trigonometry in spaces of higher dimension, the two
constants should be understood in the light of these remarks.


\subsection{The nine two-dimensional real Cayley--Klein geometries}
\label{Sec:ii:ii}

The motion groups  of  the
nine 2D CK geometries of real type can be  described in a unified
setting by means  of two real coefficients
$\k_1$,  $\k_2$ and are collectively denoted $SO_{\k_1,\k_2}(3)$. The
generators  $\{P_1,P_2,J_{12}\}$ of the  corresponding Lie  algebras
$\goth{so}_{\k_1,\k_2}(3)$ have Lie commutators:
\be
[J_{12},P_1]=P_2 \qquad [J_{12},P_2]=-\k_2 P_1 \qquad
[P_1,P_2]=\k_1 J_{12} .
\label{eq:OrthoCKBiDConmRels}
\ee
There is a single Lie algebra  Casimir coming from 
the Killing--Cartan
form:
\be
{\cal C}=P_2^2+\k_2 P_1^2 +\k_1 J_{12}^2 .
\label{eq:OrthoCKBiDCasimir}\ee

The CK algebras in the quasi-orthogonal family
$\goth{so}_{\k_1,\k_2}(3)$
can be endowed with a ${\Ze}_2\otimes {\Ze}_2$ group of commuting
automorphisms generated by:
\bea
\Invol_{(1)} &:& (P_1,P_2,J_{12})\to (-P_1,-P_2,J_{12})\cr
\Invol_{(2)} &:& (P_1,P_2,J_{12})\to (P_1,-P_2,-J_{12}).
\label{zzz:bc}
\eea
The two remaining involutions are  the composition
$\Invol_{(02)}=\Invol_{(1)}\cdot \Invol_{(2)}$ and the identity.
Each involution $\Invol$ determines  a subalgebra of
$\goth{so}_{\k_1,\k_2}(3)$, denoted $\goth{h}$,
whose elements are invariant under $\Invol$; the subgroups generated
by these subalgebras will be denoted  by $H$  with the
same subindices as the involution itself.

The elements defining a  2D CK geometry are as follows
\cite{CKdos,Vulpi}:

\noindent
$\bullet$ The {\it plane} as the set of points corresponds to the 2D
symmetrical homogeneous space
\be
\CKPointSpace\equiv SO_{\k_1,\k_2}(3)/H_{(1)}\equiv
SO_{\k_1,\k_2}(3)/SO_{\k_2}(2)
\qquad  H_{(1)}=\langle J_{12}\rangle\approx SO_{\k_2}(2) .
\label{eq:Ortho:CKBiDConmRels}
\ee
The generator  $J_{12}$ leaves a point $O$ (the origin) invariant, thus
$J_{12}$ acts as the rotation around $O$. The involution
$\Invol_{(1)}$ is the reflection around the origin. In this space
$P_1$ and $P_2$ generate translations which move the origin point in
two basic directions.

\noindent
$\bullet$ The set of {\it lines} is identified as the 2D symmetrical
homogeneous space
\be
\CKLineSpace\equiv SO_{\k_1,\k_2}(3)/H_{(2)}\equiv
SO_{\k_1,\k_2}(3)/SO_{\k_1}(2)
\qquad  H_{(2)}=\langle P_1\rangle\approx SO_{\k_1}(2) .
\label{eq:Ortho:CKspaceFirstKindLines}
\ee
In this space, the generator $P_1$ leaves invariant the `origin'
line $l_1$, which is moved in two basic directions by   $J_{12}$ and
$P_2$. Therefore, within $\CKLineSpace$,
$P_1$ should be interpreted as the generator of `rotations' around
$l_1$, and
the involution  $\Invol_{(2)}$ is the reflexion in $l_1$.

\noindent
$\bullet$ There is a second set of lines corresponding to the 2D
symmetrical homogeneous space
\be
SO_{\k_1,\k_2}(3)/H_{(02)}\equiv
SO_{\k_1,\k_2}(3)/SO_{\k_1\k_2}(2)
\qquad  H_{(02)}=\langle P_2\rangle\approx SO_{\k_1\k_2}(2) .
\label{eq:Ortho:CKspaceSecKindLines}
\ee
In this case, it is the generator $P_2$ that leaves invariant an
`origin' line $l_2$ (considered as the elementary `point') in  
this space while $J_{12}$ and $P_1$ do move $l_2$. The
involution
$\Invol_{(02)}$ is the reflexion in the line $l_2$.

In order to distinguish the two sets of lines we  call the elements
of  $\CKLineSpace$ {\it  lines of first-kind} while the 
elements of the space 
$SO_{\k_1,\k_2}(3)/H_{(02)}$ will be called {\it lines of
second-kind}. By a   {\it two-dimensional CK
geometry} we will understand the set of three symmetrical homogeneous
spaces of points, lines of first-kind and lines of second-kind. The
group $SO_{\k_1,\k_2}(3)$ acts transitively on each of these
spaces. 

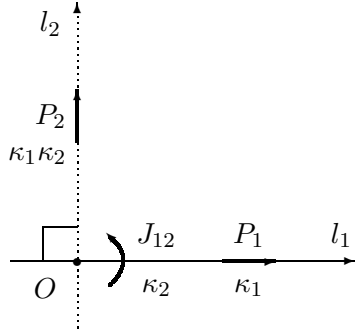
\begin{figure}[ht]

\begin{center}
\begin{picture}(130,120)
\put(25,25){\circle*{3}}
\put(13,15){\makebox(0,0){$O$}}
\put(12,25){\line(0,1){13}}
\put(12,38){\line(1,0){13}}
\put(0,25){\vector(1,0){130}}
\put(125,35){\makebox(0,0){$l_1$}}
\put(90,35){\makebox(0,0){$P_1$}}
\put(90,15){\makebox(0,0){$\k_1$}}
\qbezier[50](25,0)(25,60)(25,120)
\put(25,120){\vector(0,1){3}}
\put(15,115){\makebox(0,0){$l_2$}}
\put(15,80){\makebox(0,0){$P_2$}}
\put(10,65){\makebox(0,0){$\k_1\k_2$}}
\put(55,35){\makebox(0,0){$J_{12}$}}
\put(55,15){\makebox(0,0){$\k_2$}}
\linethickness{1pt}
\qbezier(37,16)(47,26)(37,34)
\put(37,35){\vector(-1,1){1}}
\put(25,70){\vector(0,1){20}}
\put(80,25){\vector(1,0){20}}
\end{picture}
\end{center}
\caption{Generators and their associated labels  in a
2D CK geometry.}  
\label{fig:CKspace}

\end{figure}

All properties of the two spaces of lines  
can be transcribed in terms of the space $\CKPointSpace$  itself,
and in this  interpretation the lines of first- or second-kind can
be seen as certain 1D  submanifolds of $\CKPointSpace$ rather than
`points' in the spaces $\CKLineSpace$ or
$SO_{\k_1,\k_2}(3)/H_{(02)}$. In the following we will interpret
everything in terms of the space
$\CKPointSpace$, where $l_1$ and $l_2$ should be considered as two
`orthogonal' lines meeting in $O$. The notation has been chosen to
implicitly suggest this, with $P_1$ translating along the line
$l_1$ (resp. $P_2$ along $l_2$) as shown in figure \ref{fig:CKspace}.

The coefficients $\k_1, \k_2$ play a twofold role. The
space  $\CKPointSpace$  has a {\em quadratic metric} coming from the
Casimir (\ref{eq:OrthoCKBiDCasimir}), whose signature  corresponds to
the diagonal matrix
$\mbox{diag}(1,\k_2)$. This metric is riemannian
(definite positive) for
$\k_2>0$, lorentzian (indefinite) for $\k_2<0$ and degenerate
for $\k_2=0$. Next, the same space has a canonical conexion (as any
symmetric space \cite{Nomizu}) which is compatible  with the metric, and
has {\em constant curvature} equal to $\k_1$; the notation
$\CKPointSpace$ for the space is intended to recall  the analogy with
the sphere $S^2$ to which $\CKPointSpace$ reduces when all the $\k_i$
constants are equal to $1$. The list of $\k_i$ values is appended to the
symbol for the space; the $\k_i$ in square brackets is the constant
curvature, and the remaining constant determines the signature.
Alternatively, the coefficients
$\k_1, \k_2$ determine the {\em kind of measures of separation } amongst
points and lines in the Klein sense
\cite{Yaglom,CKdos}:

\noindent
$\bullet$ The pencil of points on a first-kind line is
elliptical/parabolical/hyperbolical according to whether $\k_1$ is
greater than/equal to/lesser than zero.

\noindent
$\bullet$ Likewise for the pencil of points on a second-kind line
depending on the product $\k_1\k_2$.

\noindent
$\bullet$ Likewise for the pencil of lines through a point
according to $\k_2$.

For $\k_1$ positive/zero/negative the isotropy subgroup
 $H_{(2)}$ is  $SO(2)/\Re/SO(1,1)$, and the same
happens for 
$H_{(1)}$ (resp. $H_{(02)}$) according to the value of
$\k_2$ (resp.
$\k_1\k_2\equiv\k_{02}$). Any of these three subgroups would lead to
the 1D CK geometries with Lie algebra $\goth{so}_{\k}(2)$ and trivial
isotropy subalgebra, which can be identified as the circunference
($SO(2)$ for
$\k>0$), the real line ($\Re\equiv ISO(1)$ for $\k=0$) and  the
hyperbolic line  ($SO(1,1)$ for $\k<0$).

Whenever the coefficient $\k_1$ (resp.\ $\k_2$) is different from zero,
a suitable choice of length unit (resp.\ angle unit) allows us to 
reduce it to either $+ 1$ or $-1$. By taking into account  the
possible values for
$(\k_1,\k_2)$,  we obtain nine 2D real CK geometries, whose  groups of
motion, commutation rules (\ref{eq:OrthoCKBiDConmRels}) and Casimirs
(\ref{eq:OrthoCKBiDCasimir}) are
written explicitly in table
\ref{table:NineCKGeometries} together with the isotropy subgroups
$H_{(1)}$,
$H_{(2)}$ and
$H_{(02)}$ of their associated symmetric  spaces.

\begin{table}[h]
{\footnotesize
 \noindent
\caption{{The nine two-dimensional CK
geometries.}}\label{table:NineCKGeometries}
\smallskip
\noindent\hfill
\begin{tabular}{llll}
\hline
 &\multicolumn{3}{c}{Measure of distance}\\
\cline{2-4}
Measure&Elliptic&Parabolic&Hyperbolic\\
of angle&$\k_1=1$&$\k_1=0$&$\k_1=-1$\\
\hline
&Elliptic&Euclidean&Hyperbolic\\
&$SO(3)$&$ISO(2)$&$SO(2,1)$\\
Elliptic&$[J_{12},P_1]=P_2$&$[J_{12},P_1]=P_2$&$[J_{12},P_1]=P_2$\\
$\k_2=1$&$[J_{12},P_2]=-P_1$&$[J_{12},P_2]=-P_1$&$[J_{12},P_2]=-P_1$\\
&$[P_1,P_2]=J_{12}$&$[P_1,P_2]=0$&$[P_1,P_2]=-J_{12}$\\
&${\cal C}=P_2^2+P_1^2+J_{12}^2$&${\cal C}=P_2^2+P_1^2$&
${\cal C}=P_2^2+P_1^2-J_{12}^2$\\
&$H_{(1)}=SO(2)$&$H_{(1)}=SO(2)$&$H_{(1)}=SO(2)$\\
&$H_{(2)}=SO(2)$&$H_{(2)}=\Re$&$H_{(2)}=SO(1,1)$\\
&$H_{(02)}=SO(2)$&$H_{(02)}=\Re$&$H_{(02)}=SO(1,1)$\\
\hline
&Co-Euclidean&Galilean&Co-Minkowskian\\
&Oscillating NH & &Expanding NH\\
&$ISO(2)$&$IISO(1)$&$ISO(1,1)$\\
Parabolic&$[J_{12},P_1]=P_2$&$[J_{12},P_1]=P_2$&$[J_{12},P_1]=P_2$\\
$\k_2=0$&$[J_{12},P_2]=0$&$[J_{12},P_2]=0$&$[J_{12},P_2]=0$\\
&$[P_1,P_2]=J_{12}$&$[P_1,P_2]=0$&$[P_1,P_2]=-J_{12}$\\
&${\cal C}=P_2^2+ J_{12}^2$&${\cal C}=P_2^2 $&
${\cal C}=P_2^2 -J_{12}^2$\\
&$H_{(1)}=\Re$&$H_{(1)}=\Re$&$H_{(1)}=\Re$\\
&$H_{(2)}=SO(2)$&$H_{(2)}=\Re$&$H_{(2)}=SO(1,1)$\\
&$H_{(02)}=\Re$&$H_{(02)}=\Re$&$H_{(02)}=\Re$\\
\hline
&Co-Hyperbolic&Minkowskian&Doubly Hyperbolic\\
&Anti-de Sitter& &De Sitter\\
&$SO(2,1)$&$ISO(1,1)$&$SO(2,1)$\\
Hyperbolic&$[J_{12},P_1]=P_2$&$[J_{12},P_1]=P_2$&$[J_{12},P_1]=P_2$\\
$\k_2=-1$&$[J_{12},P_2]=P_1$&$[J_{12},P_2]=P_1$&$[J_{12},P_2]=P_1$\\
&$[P_1,P_2]=J_{12}$&$[P_1,P_2]=0$&$[P_1,P_2]=-J_{12}$\\
&${\cal C}=P_2^2-P_1^2+J_{12}^2$&${\cal C}=P_2^2-P_1^2$&
${\cal C}=P_2^2-P_1^2-J_{12}^2$\\
&$H_{(1)}=SO(1,1)$&$H_{(1)}=SO(1,1)$&$H_{(1)}=SO(1,1)$\\
&$H_{(2)}=SO(2)$&$H_{(2)}=\Re$&$H_{(2)}=SO(1,1)$\\
&$H_{(02)}=SO(1,1)$&$H_{(02)}=\Re$&$H_{(02)}=SO(2)$\\
\hline
\end{tabular}\hfill}
\end{table}

The same Lie group (up to isomorphism) can be the group of motion of
more than one  CK geometry. For instance, the euclidean group
$ISO(2)$ is the group of motion of two CK geometries, characterized by
two different sets for the three associated symmetric spaces. The
space of points is $ISO(2)/SO(2)$ for  euclidean geometry, but
$ISO(2)/\Re$ for the co-euclidean one; this last space can be
identified as the space of (first-kind) lines in the euclidean plane.
The same happens with $ISO(1,1)$. The simple group $SO(2,1)$ is
associated to {\em three} CK geometries, namely hyperbolic,
co-hyperbolic and doubly hyperbolic.

A fundamental property of the
scheme of CK geometries is the existence of an  `automorphism' of
the whole family, called {\em ordinary duality}
$\dual$. This is well defined for any dimension, and in the 2D case it
is given by:
\be
\dual : (P_1,P_2,J_{12}) \to (-J_{12},-P_2,-P_1)\qquad
\dual : (\k_1, \k_2 ) \to (\k_2, \k_1) .
\label{eq:OrthoCKBiDDuality}
\ee
The map $\dual$ leaves the general   commutation rules
(\ref{eq:OrthoCKBiDConmRels}) invariant while it interchanges the space
of points with the space of first-kind lines,
$\CKPointSpace\leftrightarrow \CKLineSpace$, and the corresponding
curvatures $\k_1\leftrightarrow\k_2$, preserving the space of
second-kind lines.  Note that  $\dual$ relates in general {\em two}
different CK geometries, which are placed in symmetrical positions
relative to the main diagonal in table
\ref{table:NineCKGeometries}. For instance, duality interchanges the
euclidean, hyperbolic and minkowskian geometries with the 
co-euclidean, co-hyperbolic and co-minkowskian ones, while elliptic,
galilean and doubly hyperbolic are self-dual geometries. This suggests
also a kind of duality between {\em curvature} and {\em signature}
which seems worth studying.

The non-generic situation where a coefficient $\k_i$ vanishes
corresponds to an In\"on\"u--Wigner  contraction \cite{IW}.  The limit
$\k_1\to 0$ is a local-contraction (around a point); it carries the
geometries of the first and third columns of table
\ref{table:NineCKGeometries} to the middle one, which have `flat'
spaces of points. The limit $\k_2\to 0$ is an axial-contraction
(around a line), which carries the geometries of the first and third
rows to the middle one. However, note that our approach to
contractions is indeed  built-in in the general expressions associated
to the CK geometries and groups. This means that we deal with 
expressions that contain explicitly  the constants $\k_i$ (which
determine the curvature and signature of the space) and which can
be also understood as contraction parameters so that a contraction
is simply  equivalent to putting $\k_i=0$ in the desired relation.


\subsection{Spacetimes as  Cayley--Klein spaces}
\label{Sec:ii:iii}

It is rather remarkable that the meaningful  kinematical spacetimes
and their invariance groups \cite{BLL} arise as particular CK
spaces and groups. Let $\stH$, $\stP$ and $\stK$ be the generators  of
time translations, space translations and boosts, respectively, in the
most simple $(1+1)$D homogeneous spacetime. Under the identification
\be
P_1\equiv \stH\qquad P_2\equiv \stP\qquad J_{12}\equiv \stK
\label{eq:OrthoCKBiDKimematInt}
\ee
the six CK groups with $\k_2\le 0$  (second and third rows of table
\ref{table:NineCKGeometries}; NH means Newton--Hooke) are the motion
groups of $(1+1)$D spacetimes.  The physical reading of the
three CK spaces within each of these CK geometries is:

\noindent
$\bullet$ $\CKPointSpace$ is a $(1+1)$D spacetime, and  points in
$\CKPointSpace$  are  {\em  spacetime events}; the spacetime
curvature equals $\k_1$ and is related to the usual
universe radius $\tiempo$ (measured in time units) by $\k_1=\pm
1/\tiempo^2$. Relativistic spacetimes occur for $\k_2<0$ (the signature
of the minkowskian type metric is
$\mbox{diag}(1, {-1}/{c^2})$) and their non-relativistic limits
correspond to $\k_2=0$.

\noindent
$\bullet$ The space of first-kind lines $\CKLineSpace$ corresponds to
the space of {\em time-like lines}. Here the coefficient  $\k_2$ can
be read as the curvature of the space of time-like lines, linked to the
fundamental relativistic constant $c$ as $\k_2={-1}/{c^2}$. From this
point of view, the passage from non-relativistic theories to relativistic
ones can be looked as the introduction of a non-zero, negative,
curvature in the space of time-like lines, which previously was flat.

\noindent
$\bullet$ The space of second-kind lines  
$SO_{\k_1,\k_2}(3)/H_{(02)}$ is the 2D space of {\em
space-like lines}.

According to the signs of the pair $(\k_1,\k_2)$ we have three
homogeneous `absolute-time' spacetimes for $\k_2=0$, namely,
oscillating (or anti) Newton--Hooke for $\k_1>0$,  galilean for $\k_1=0$
and expanding Newton--Hooke for $\k_1<0$; they are
degenerate riemannian spacetimes with constant curvature $\k_1$ and
a degenerate (`absolute-time') metric, which is the $c=\infty$ limit of
the time metric in relativity. For $\k_2=-1/c^2<0$ we find three
`relative-time' spacetimes: anti-de Sitter $(\k_1>0,-1/c^2)$,
minkowskian $(0,-1/c^2)$ and de Sitter $(\k_1<0,-1/c^2)$;
these are pseudoriemmanian spacetimes  with a metric of lorentzian
type and constant curvature $\k_1$. Note that our choice of metric
lets the curvature of anti-de Sitter be {\em positive}, which is
opposite to the standard choice (where the metric is 
taken with the opposite sign) but fits to the closeness
of time-like lines, which have been taken as the `basic'
ones; space-like lines are open in anti-de Sitter. The
limits $\k_1\to 0 \equiv
\tiempo\to \infty$ and $\k_2\to 0 \equiv c\to
\infty$ correspond to a spacetime contraction and  a speed-space
contraction,
respectively.

The three remaining geometries with $\k_2>0$ do not admit such a
kinematical interpretation.  They are the well known riemannian spaces
with constant curvature $\k_1$. In these cases, the sets of first- and
second-kind lines coincide, because in these cases (and only these)
the generators $P_1$ and $P_2$ are conjugated under the full motion
group:
\be
P_2=  \myexp{\frac{\pi}{2\sqrt{\k_2}} J_{12}}P_1\myexp{-
\frac{\pi}{2\sqrt{\k_2}} J_{12} }.
\ee
This is why only these three spaces fulfil the usual 
definition of two-point homogeneity; as we show in this paper there is no
compelling reason (and some drawbacks) to restrict any joint study
only to these three cases.


\subsection{Realization of the spaces of points}
\label{Sec:ii:iv}

The following 3D real matrix representation of the CK  algebra
$\goth{so}_{\k_1,\k_2}(3)$:
\be
P_1=\left(\begin{array}{ccc}
0&-\k_1&0 \cr 1&0&0\cr 0&0&0 \end{array}\right) \quad
P_2=\left(\begin{array}{ccc}
0&0&-\k_1\k_2 \cr 0&0&0\cr 1&0&0 \end{array}\right) \quad
J_{12}=\left(\begin{array}{ccc}
0&0&0 \cr 0&0&-\k_2\cr 0&1&0 \end{array}\right)
\label{zzz:bd}
\ee
gives rise to a natural realization of the CK group $SO_{\k_1,\k_2}(3)$
as a group of linear transformations in an ambient linear space
$\Re^3=(x^0,x^1,x^2)$ in which $SO_{\k_1,\k_2}(3)$ acts as  the
group of linear isometries of a bilinear form with matrix:
\be
\Lambda=\left(\begin{array}{ccc}
1&0&0 \cr 0&\k_1&0\cr 0&0&\k_1\k_2 \end{array}\right).
\label{eq:OrthoCKBiDMetricMatrix}
\ee

The exponential  of the matrices (\ref{zzz:bd})  leads to  a
representation of the one-parametric subgroups $H_{(2)}$, $H_{(02)}$ and
$H_{(1)}$ generated by
$P_1$, $P_2$ and $J_{12}$ as:
\bea
&&\exp(\alpha P_1)=\left(\begin{array}{ccc}
\c(\alpha)&-\k_1\s(\alpha)&0 \cr
\s(\alpha)&\c(\alpha)&0\cr
0&0&1
\end{array}\right) \cr
&&\exp(\beta P_2)=\left(\begin{array}{ccc}
\ccc(\beta)&0&-\k_1\k_2\sss(\beta)\cr
0&1&0\cr
\sss(\beta)&0&\ccc(\beta)
\end{array}\right) \cr
&&\exp(\gamma J_{12})=\left(\begin{array}{ccc}
1&0&0\cr
0&\cc(\gamma)&-\k_2\ss(\gamma)\cr
0&\ss(\gamma)&\cc(\gamma)
\end{array}\right)
\label{eq:OrthoCKBiDOneParamSubgroups}
\eea
where the generalized cosine $C_\k(x)$ and sine
$S_\k(x)$ functions are defined by \cite{CKdos}--\cite{Poisson}:
\be
C_{\k}(x) :=\sum_{l=0}^{\infty}(-\k)^l\frac {x^{2l}}{(2l)!}
=\left\{
\begin{array}{ll}
  \cos {\sqrt{\k}\, x} &\quad  \k >0 \cr
  1  &\quad \k  =0 \cr
\cosh {\sqrt{-\k}\, x} &\quad \k <0
\end{array}\right.
\label{eq:OrthoCKCosine}
\ee
\be
S_{\k}(x) :=\sum_{l=0}^{\infty}(-\k)^l\frac {x^{2l+1}}{(2l+1)!}
=\left\{
\begin{array}{ll}
    \frac{1}{\sqrt{\k}} \sin {\sqrt{\k}\, x} &\quad  \k >0 \cr
  x &\quad \k  =0 \cr
\frac{1}{\sqrt{-\k}} \sinh {\sqrt{-\k}\, x} &\quad \k <0
\end{array}\right. .
\label{zzz:bh}
\ee

Two other useful curvature-dependent  functions are the `versed sine'
$V_\k(x)$ (note that $V_{\k}(x)$ is well defined even if $\k=0$) and
the tangent $T_\k(x)$ given by:
\be
V_{\k}(x) :=\frac 1\k(1-C_\k(x))\qquad
T_{\k}(x) :=\frac{S_\k(x)}{C_\k(x)} .
\label{eq:OrthoCKTangentVersine}
\ee

These generalized trigonometric functions coincide with  the usual
elliptic and hyperbolic ones for   $\k=1$ and $\k=-1$ respectively; the
case $\k=0$  provides the parabolic or galilean functions:  $C_{0}(x)=1$,
$S_{0}(x)=x$ and $V_{0}(x)=x^2/2$. This slightly non-standard notation
is the price one should pay for the ability of describing all spaces at
once, but the effort is worth it.  Several identities for these
functions, which are necessary for further development, are included in
the Appendix.

A generic element $R\in SO_{\k_1,\k_2}(3)$ can be written as the product
of the matrices (\ref{eq:OrthoCKBiDOneParamSubgroups})
and satisfies
\be
R^T\, \Lambda\, R=\Lambda .
\label{zzz:bk}
\ee
The action of $SO_{\k_1,\k_2}(3)$ on $\Re^3$ is linear but not
transitive, since it conserves the quadratic form  $(x^0)^2+\k_1 (x^1)^2+
\k_1\k_2 (x^2)^2$. The subgroup $H_{(1)}$, whose matrix representation
is
$\exp(\gamma J_{12})$ (\ref{eq:OrthoCKBiDOneParamSubgroups}
),  is the isotropy subgroup of the point
$O\equiv (1,0,0)$, that is, the origin in the space $\CKPointSpace$.
The action becomes transitive on the orbit in $\Re^3$ of the
point $O$, which is contained in the `sphere' $\Sigma$:
\be
\Sigma \equiv   (x^0)^2+\k_1 (x^1)^2+  \k_1\k_2 (x^2)^2=1 .
\label{zzz:bl}
\ee
This orbit can be identified with the space of points 
$\CKPointSpace\equiv SO_{\k_1,\k_2}(3)/SO_{\k_2}(2)$ of the CK
geometry and the coordinates $(x^0,x^1,x^2)$  are
the {\em Weierstrass coordinates}, while
$(x^1/x^0,x^2/x^0)$ are the {\em Beltrami coordinates}.
We remark that this scheme includes under a common
description all the familiar embeddings (the so called
vector models) of the sphere, hyperbolic plane, anti-de
Sitter and de Sitter spaces in a linear 3D ambient
space, with a flat metric of either euclidean or
lorentzian type;  the induced metric on the CK  sphere
$\Sigma$ should be defined as the quotient by $\k_1$ of the
restriction of the flat ambient metric
$dl^2 = (dx^0)^2 + \k_1(dx^1)^2 + \k_1\k_2 (dx^2)^2 $
(\ref{eq:OrthoCKBiDMetricMatrix}); this is always well defined
because the restriction of the flat metric $dl^2$ to the CK
sphere contains
$\k_1$ as a factor \cite{Jaca}. 

The expressions for the metric in parallel $a, h$ and polar
coordinates $r, \chi$ relative to the origin point $O$ and
line $l_1$ are:
\be
  ds^2 = \ccc^2(h) \, d a^2 + \k_2 d h^2  {\qquad}
  ds^2 = d r^2 + \k_2\s^2 ( r) \, d \chi^2  \label{eq:CKBiDMetric}
\ee
and the canonical conexion is given by the non-zero conexion
coefficients (in parallel and polar coordinates)
 \begin{eqnarray}
 &&\Gamma^{h}_{ a a} =  {\k_1}\,\sss(h)\ccc(h)  \qquad
   \Gamma^{a}_{a h} = -\k_1\k_2\,\TTT(h)             \cr
 &&\Gamma^{r}_{\chi\chi} = -\k_2 \s(r)\c(r)       \qquad
   \Gamma^{\chi}_{r \chi} =  1/\T(r) .         
 \end{eqnarray}


\section{The compatibility conditions for a triangular loop}
\label{Sec:iii}

We now come to the main objective of this paper, which is the study of
trigonometry of the nine CK
spaces introduced in the previous Section.

In the euclidean plane three points always determine unambiguosly a
triangle, because any two points are connected by a single geodesic
segment, and all triangles are in an obvious sense of the same type. In
the sphere three points do not directly determine a triangle, because
two generic points can be joined by two geodesic segments (both lying
on the same geodesic), yet all triangles are of the same type. In the
minkowskian plane two points can always be joined by a single geodesic
segment (like in the euclidean plane), but this segment can be of two
generic and one non-generic types (time-like,  space-like and
isotropic) so here  three points do determine unambiguously a
triangle, but not all triangles are of the same type, and triangles
with three time-like sides can coexist with triangles with mixed
sides. Finally in the anti-de Sitter spacetime both complications
may appear together: there are three types of sides, and two points
with time-like separation can be joined, like in the sphere, by two
different time-like geodesic segments.

To avoid unnecessary complications and such boring attention to
details, it is better to introduce the concept  of {\em triangular
loop}, which affords a well defined replacement  of the unprecise idea
of `triangle as three points'. A triangular loop can be considered
either as a triangular {\em point loop} or as a triangular {\em line
loop}, and we will need the simultaneous consideration of both
aspects. Furthermore, and according to the type of the `sides', there
are several different types of triangular loops, which merge into a
single type in the riemannian case $\k_2>0$.

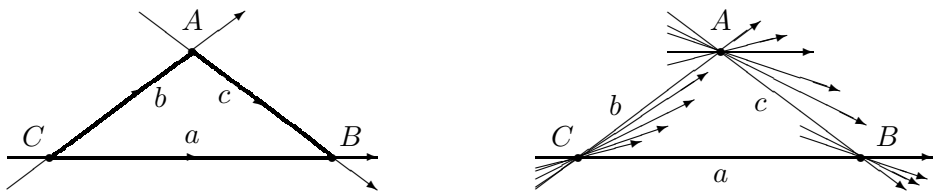
\begin{figure}[ht]

\begin{center}
\begin{picture}(350,85)
\put(0,25){\vector(1,0){140}}
\put(0,13){\vector(4,3){90}}
\put(50,80){\vector(4,-3){90}}
\put(16,25){\circle*{3}}
\put(123,25){\circle*{3}}
\put(70,65){\circle*{3}}
\put(10,33){\makebox(0,0){$C$}}
\put(130,33){\makebox(0,0){$B$}}
\put(70,77){\makebox(0,0){$A$}}
\put(70,33){\makebox(0,0){$a$}}
\put(82,48){\makebox(0,0){$c$}}
\put(58,48){\makebox(0,0){$b$}}
\put(200,25){\vector(1,0){150}}
\put(200,13){\vector(4,3){85}}
\put(250,80){\vector(4,-3){90}}
\put(200,21){\vector(4,1){40}}
\put(200,20){\vector(3,1){50}}
\put(200,17){\vector(2,1){60}}
\put(200,14){\vector(3,2){65}}
\put(250,60){\vector(4,1){45}}
\put(250,65){\vector(1,0){55}}
\put(250,72){\vector(3,-1){65}}
\put(250,75){\vector(2,-1){75}}
\put(300,33){\vector(3,-1){48}}
\put(300,37){\vector(2,-1){45}}
\put(216,25){\circle*{3}}
\put(323,25){\circle*{3}}
\put(270,65){\circle*{3}}
\put(210,33){\makebox(0,0){$C$}}
\put(333,33){\makebox(0,0){$B$}}
\put(270,77){\makebox(0,0){$A$}}
\put(270,18){\makebox(0,0){$a$}}
\put(285,45){\makebox(0,0){$c$}}
\put(230,45){\makebox(0,0){$b$}}
\put(69,25){\vector(1,0){3}}
\put(48,49){\vector(4,3){3}}
\put(94,47){\vector(4,-3){3}}
\linethickness{1pt}
\qbezier(16,25)(42,25)(68,25)
\qbezier(69,25)(96,25)(123,25)
\qbezier(16,25)(32,37)(48,49)
\qbezier(48,49)(59,57)(70,65)
\qbezier(70,65)(82,56)(94,47)
\qbezier(94,47)(109,36)(123,25)
\end{picture}
\end{center}
\caption{a) Triangular point loop.\qquad\qquad\quad
b) Triangular line loop.}
\label{fig:TriangleLoops}

\end{figure}

\begin{figure}[ht]

\begin{center}
\begin{picture}(235,85)
\put(-10,25){\line(1,0){224}}
\put(30,13){\vector(4,3){90}}
\put(80,80){\vector(4,-3){90}}
\put(46,25){\circle*{3}}
\put(153,25){\circle*{3}}
\put(100,65){\circle*{3}}
\put(40,33){\makebox(0,0){$C$}}
\put(170,33){\makebox(0,0){$B$}}
\put(96,78){\makebox(0,0){$A$}}
\put(112,48){\makebox(0,0){$c$}}
\put(88,48){\makebox(0,0){$b$}}
\put(100,33){\makebox(0,0){$a$}}
\put(78,49){\vector(4,3){3}}
\put(124,47){\vector(4,-3){3}}
\linethickness{1pt}
\qbezier(46,25)(62,37)(78,49)
\qbezier(78,49)(89,57)(100,65)
\qbezier(100,65)(112,56)(124,47)
\qbezier(124,47)(139,36)(153,25)
\qbezier(153,25)(168,25)(198,25)
\qbezier(199,25)(214,25)(229,25)
\put(199,25){\vector(1,0){3}}
\qbezier(-10,25)(4,25)(18,25)
\qbezier(19,25)(33,25)(46,25)
\put(19,25){\vector(1,0){3}}
\qbezier(50,17)(60,27)(50,35)
\put(50,36){\vector(-1,1){1}}
\qbezier(157,16)(167,26)(157,34)
\put(157,35){\vector(-1,1){1}}
\qbezier(103,57)(113,67)(103,75)
\put(103,76){\vector(-1,1){1}}
\end{picture}
\end{center}
\caption{Triangular  loop as a single curve.}
\label{fig:singlecurve}
\end{figure}
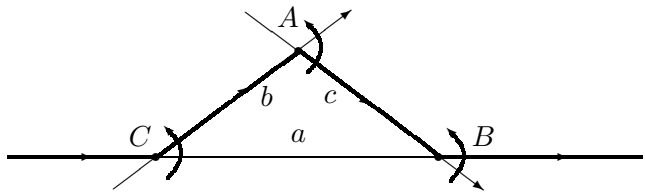

From now on (except in Section 6), we will deal exclusively with {\em
first-kind} triangular loops (i.e.\ time-like in the spaces with
kinematical interpretation). A (first-kind)  triangular point loop can
be considered as two different (first-kind) paths for a {\em point}
going from an initial position
$C$ to a final one $B$. One path will be the direct one along the
segment of the (first-kind) line $a$ determined by $C$ and $B$. The
other will be a two-step path made of two segments of (first-kind)
lines going from $C$ to an intermediary point $A$ along a line $b$ and
then from $A$ to $B$ along the line $c$ (see figure
\ref{fig:TriangleLoops}a); hereafter we will omit the reference to the
first-kind type of all lines.  For most purposes it is better to
look at the triangular point loop as a single (possibly open) 
poligonal curve (see figure \ref{fig:singlecurve}), obtained from
the line $a$ by replacing the segment $CB$ by the two geodesic
segments $CA$ and $AB$; this curve will be considered as an
oriented and cooriented curve, and will be only closed when the
geodesic $a$ itself is closed. The corners in the poligonal may
be smoothed (while maintaining the first-kind character) and thus
we will obtain a smooth  curve.

The previous view can be dualized, and the loop can be also
considered as a triangular {\em line loop}: the dual of the single
curve associated to the point loop is a moving   line which starts
at $a$, then rotates around $C$ going to $b$, then around
$A$ towards $c$ and finally comes back  to $a$ by means of a rotation
around  $B$ (see figure \ref{fig:TriangleLoops}b). This can be also
considered as an ordinary point loop in the dual space, because lines
in the given space are interpreted as points in the dual space.

In the kinematical interpretation of the  six spacetimes
with $\k_2\leq 0$, a triangular loop will therefore be
determined by two time-like future pointing  paths from
an initial spacetime point $C$ to a final one $B$, one
along the direct path and other through an intermediary
spacetime point $A$. Note  the assymmetric role the vertex
$A$ plays from the beginning. Each side of the loop
determines the generator of  {\em translations} along
the side, up to a non-zero scale factor which should be
splitted into a sign (corresponding to one of the two
possible  orientations of the line) and a positive scale
factor (corresponding to the choice of a unit length).
The restriction to first-kind sides means that the
following condition is satisfied:

\noindent
P1) The three  generators $P_a, P_b, P_c$ are either
equal or opposite to some conjugate by means of some
group transformation to the single fiducial generator of
translations $P_1$ along first-kind lines.

We shall now perform a fiducial choice of the still undetermined
factors in these  generators, which will be assumed fixed from
now on, according to a second condition:

\noindent P2) The positive sense of translation generated by
$P_a, P_b, P_c$ agrees with the orientation for the point loop
as a single curve.

The meaning of these conditions can be appreciated  more clearly
for the geometries with $\k_2\leq 0$, in the kinematical
interpretation, for which the first condition embodies the time-like
character of the three lines (here $P_1\equiv \stH$ generates
the future time translation along the fiducial time-like line),
and the second condition corresponds to the future character of
a time-like line loop. In the riemmanian cases
$(\k_2>0$) all geodesics can be considered to be simultaneously
of both first- and second-kind; then first condition is
automatic, while the second can always be clearly fulfilled.

Discussion of the trigonometry of triangular loops with sides of
different kind can be developed in full analogy with the pure
first-kind case discussed here; see Section 6.

The important fact is that for any triangular loop, a choice of
$P_a, P_b, P_c$ satisfying these two conditions is always
possible, and this is so simultaneously for the {\em nine CK
geometries}.  Now we will denote $a$ the (positive, unoriented)
distance between the points $C,B$ along the geodesic direct path
(the geodesic segment on $a$ which is missing on the curve in
figure \ref{fig:singlecurve}), and by $b, c$ the (positive,
unoriented) distances between the points $C,A$ and $A,B$ along the
geodesic segments on the lines $b$, $c$ on the curve. The use of the
same names for the lengths of the three loop sides and for the
baselines themselves is traditional and should cause no confusion.
In the kinematical cases,  the lengths $a, b, c$ will be the proper
times along the sides between their end events, and the angles
$A, B, C$ are the relative rapidities between the time-like
lines at each vertex (recall the rapidity is the natural angle
in minkowskian geometry \cite{LevyLeblond}). In kinematical
spaces the triangle  loop, seen as a single curve in figure
\ref{fig:singlecurve} is the wordline of the travelling twin in the
twin pseudoparadox.

On  the dual hand, the generators $J_A, J_B, J_C$ of {\em
rotations} around the vertices $A\equiv b\cap c, B\equiv c\cap
a, C\equiv a\cap b$ are again determined up to sign and a
positive scale factor, which we shall choose so as to satisfy
two conditions, dual to the previous ones:

\noindent
J1) The three  generators $J_A, J_B, J_C$ are conjugated
by means of some group transformation to the single
fiducial generator of rotations $J_{12}$.

\noindent
J2) The positive sense of rotation around each vertex is
the correct one determined by the given orientation and
coorientation of the loop as a curve.

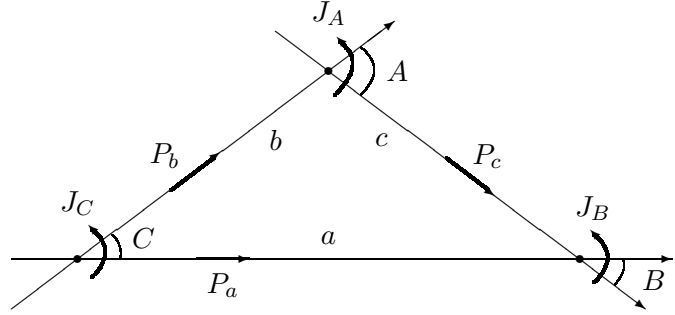
\begin{figure}[h]

\begin{center}
\begin{picture}(250,120)
\put(0,6){\vector(4,3){145}}
\put(0,25){\vector(1,0){250}}
\put(100,111){\vector(4,-3){140}}
\put(25,25){\circle*{3}}
\put(215,25){\circle*{3}}
\put(120,96){\circle*{3}}
\put(50,33){\makebox(0,0){$C$}}
\qbezier(41,25)(42,30)(38,34)
\put(146,96){\makebox(0,0){$A$}}
\qbezier(132,87)(142,97)(132,105)
\put(243,17){\makebox(0,0){$B$}}
\qbezier(228,15)(233,21)(231,25)
\put(120,33){\makebox(0,0){$a$}}
\put(100,70){\makebox(0,0){$b$}}
\put(140,70){\makebox(0,0){$c$}}
\put(80,15){\makebox(0,0){$P_a$}}
\put(58,64){\makebox(0,0){$P_b$}}
\put(180,64){\makebox(0,0){$P_c$}}
\put(25,46){\makebox(0,0){$J_C$}}
\put(220,44){\makebox(0,0){$J_B$}}
\put(120,118){\makebox(0,0){$J_A$}}
\linethickness{1pt}
\put(70,25){\vector(1,0){20}}
\qbezier(60,51)(68,57)(76,63)
\put(76,63){\vector(4,3){3}}
\qbezier(164,63)(172,57)(180,51)
\put(180,51){\vector(4,-3){3}}
\qbezier(30,18)(40,28)(30,36)
\put(30,37){\vector(-1,1){1}}
\qbezier(220,16)(230,26)(220,34)
\put(220,35){\vector(-1,1){1}}
\qbezier(122,87)(134,99)(124,107)
\put(124,108){\vector(-1,1){1}}
\end{picture}
\end{center}

\noindent
\caption{Pure first-kind triangle with three first-kind sides
$a$, $b$, $c$, two  inner angles   $B$, $C$ and an external
angle $A$, together with the generators of translations along
the sides and the generators of rotations around the vertices.}
\label{fig:TriangleCompatibility}

\end{figure}

The generators $P_a, P_b, P_c; J_A, J_B, J_C$ are not
independent. They are related by several {\em compatibility
conditions}:
\be
\begin{array}{ll}
P_b=\myexp{\CJC}P_a\myexp{-\CJC} \qquad&\qquad J_B
=\myexp{\cpc}J_A\myexp{-\cpc}
\cr
P_c=\myexp{-\AJA}P_b\myexp{\AJA} \qquad&\qquad
J_C=\myexp{-\apa}J_B\myexp{\apa}
\cr
P_a=\myexp{\BJB}P_c\myexp{-\BJB} \qquad&\qquad
J_A=\myexp{\bpb}J_C\myexp{-\bpb}   .
\end{array}
\label{OrthoTrig:compatibility}
\ee
which can be considered as giving an implicit group theoretical
definition for the three sides and the three angles; these
definitions are in agreement with the ones derived from differential
geometry with the metrics (\ref{eq:CKBiDMetric}) in the spaces of points
and lines. 

Our main contention is that all the trigonometry of the space is
{\em completely} contained in these equations, which have as  a
remarkable property  their explicit {\em  duality} under the
interchange $a, b, c
\leftrightarrow  A,   B,  C$ and $P  \leftrightarrow  J$; this
duality is a consequence of the fact that $\dual$
(\ref{eq:OrthoCKBiDDuality}) is an automorphism of the  family of
CK algebras which interchanges $P_1 \leftrightarrow -J_{12}$, and
therefore
$a, b, c \leftrightarrow -A,  -B, -C$.

The first equation in (\ref{OrthoTrig:compatibility}) gives the
translation generator $P_b$ as a conjugate of $P_a$ by means of 
a rotation around the vertex $C$; the same equation read
inversely gives $P_a$  as a conjugate of $P_b$ by means of the
inverse rotation around $C$. These expressions will be utilised
several times in the ongoing derivations, and we will refer to
them as $P_b(P_a)$ or $P_a(P_b)$; likewise the two remaining
equations for the translations will be referred to as 
$P_c(P_b)$ and $P_a(P_b)$. Similar shorthands $J_B(J_A)$, etc.\
will be  used to make reference to the  three equations relating
the rotation generators.

By cyclic substitution in the three equations $P_a(P_c),
P_c(P_b)$ and $P_b(P_a)$ we find the identity
\be
\myexp{\BJB} \myexp{-\AJA} \myexp{\CJC} P_a \myexp{-\CJC}
\myexp{\AJA}
\myexp{-\BJB}
= P_a
\label{OrthoTrig:cyclicPa}
\ee
as well as two similar equations for $P_b$ and $P_c$, obtained by
starting the substitution in either $P_b$ or
$P_c$.  Likewise, a dual completely parallel process allows us to
derive analogous identities for $J_A$ and $J_B$, which are
similar to the identity involving $J_C$:
\be
\myexp{-\apa}\myexp{\cpc}\myexp{\bpb} J_C
\myexp{-\bpb}\myexp{-\cpc}\myexp{\apa} = J_C.
\label{OrthoTrig:cyclicJC}
\ee
Equations (\ref{OrthoTrig:cyclicPa}) and  (\ref{OrthoTrig:cyclicJC}) can
be written alternatively as:
\be
\begin{array}{l}
\myexp{\BJB} \myexp{-\AJA} \myexp{\CJC}   \hbox{ must commute with } P_a
,\\
\myexp{-\apa} \myexp{\cpc}   \myexp{\bpb} \hbox{ must commute with } J_C .
\label{OrthoTrig:MustCommute}
\end{array}
\ee


\subsection{Loop excesses and  loop equations}
\label{Sec:iii:i}

The content of (\ref{OrthoTrig:MustCommute}) is transparent and could
indeed have been taken as the starting point, as the following
alternative reasoning shows: the product $\myexp{-\apa}
\myexp{\cpc} \myexp{\bpb}$ of the three translations along the
three sides of the triangle $C\overto{b} A \overto{c} B
\overto{-a} C$  moves the base point $C$ along the triangle and
returns it back to its original position, so it must necessarily
be a {\em rotation} around the  vertex $C$ by some angle
$-\Delta_C$:
\be
\myexp{-\apa} \myexp{\cpc}   \myexp{\bpb} = \myexp{-\Delta_C
J_C}.
\label{OrthoTrig:GBC}
\ee
Therefore the product $\myexp{-\apa} \myexp{\cpc} \myexp{\bpb} $
must  commute with $J_C$. Likewise, the product
$\myexp{\BJB} \myexp{-\AJA} \myexp{\CJC}$ of the three rotations
around the three  vertices, $a \overto{C} b \overto{-A} c
\overto{B} a$ must be a translation along the  side $a$ by an
amount $-\delta_a$:
\be
\myexp{\BJB} \myexp{-\AJA} \myexp{\CJC} = \myexp{-\delta_a P_a}.
\label{OrthoTrig:GBa}
\ee
The two quantities $\delta_a$ and $\Delta_C$ (as well as their
analogous $\delta_b, \delta_c$ and $\Delta_A$, $\Delta_B$ which
appear in equations similar to (\ref{OrthoTrig:GBC}) and
(\ref{OrthoTrig:GBa})) are so far unknown, but they are completely
determined by the triangle loop. To find them  we start with the
equation which gives $P_c(P_b)$ in the set
(\ref{OrthoTrig:compatibility}), replace
$P_c$ by $\myexp{-\cpc}P_c\myexp{\cpc}$ and then substitute 
$P_c(P_a)$ from the compatibility equations to obtain:
\be
\myexp{-\AJA} P_b \myexp{\AJA}= \myexp{-\cpc} \myexp{-\BJB} P_a
\myexp{\BJB}\myexp{\cpc} .
\ee
We introduce $J_B(J_C)$ and trivially simplify to obtain:
\be
\myexp{-\AJA} P_b \myexp{\AJA}= 
\myexp{-\cpc} \myexp{\apa} \myexp{-B J_C}
 P_a \myexp{B J_C}\myexp{-\apa}\myexp{\cpc}
\ee
which is equivalent to
\be
\myexp{B J_C}\myexp{-\apa}\myexp{\cpc}\myexp{-\AJA}
 P_b \myexp{\AJA}\myexp{-\cpc}\myexp{\apa} \myexp{-B J_C}= P_a.
\ee
Now we use $J_A(J_C)$, simplify, and finally substitute
$P_b(P_a)$. This gives:
\be
\myexp{B J_C} \myexp{-\apa}\myexp{\cpc}\myexp{\bpb}\myexp{-A J_C}
\myexp{C J_C}P_a \myexp{-C J_C}  \myexp{A J_C} \myexp{-\bpb}
\myexp{-\cpc}\myexp{\apa} \myexp{-B J_C} = P_a .
\label{OrthoTrig:papa}
\ee
Note that the three translations along the triangle  appear in a
single piece $\myexp{-\apa}\myexp{\cpc}\myexp{\bpb}$, while the 
three rotations are all  around the base point $C$. Now we can
go a bit further: (\ref{OrthoTrig:MustCommute}) implies that
$\myexp{-\apa}\myexp{\cpc}
\myexp{\bpb}$  must commute with $J_C$, so it will commute with
{\em any} rotation around $C$. Then we can commute the whole
piece $\myexp{-\apa}\myexp{\cpc}\myexp{\bpb}$ in
(\ref{OrthoTrig:papa}) with the rotations around $C$ and collect
these  altogether. Thus:
\be
\myexp{-\apa}\myexp{\cpc}\myexp{\bpb}
 \myexp{(- A + B + C) J_C} P_a  \myexp{-(- A + B + C) J_C}
\myexp{-\bpb}\myexp{-\cpc}\myexp{\apa}
= P_a
\ee
or
\be
\myexp{-\apa}\myexp{\cpc} \myexp{\bpb} \myexp{(- A + B + C) J_C}
\hbox{ must commute with } P_a .
\label{OrthoTrig:commPa}
\ee

Since we had already derived that the expression
$\myexp{-\apa}\myexp{\cpc} \myexp{\bpb} \myexp{X J_C}$ must
commute with $J_C$ for {\em any} angle $X$ (see
(\ref{OrthoTrig:MustCommute})), and this expression also commutes
with $P_a$ for the special value
$X=- A + B + C$, we immediately conclude that,
\be
\myexp{-\apa} \myexp{c P_c} \myexp{b P_b} 
\myexp{(- A + B + C) J_C}  = 1
\ee
since the identity is the only element of $SO_{\k_1,\k_2}(3)$
that commutes with two such  generators as $P_a$ and $J_C$.
This equation can be also written as:
\be
\myexp{-\apa} \myexp{c P_c} \myexp{b P_b} =
 \myexp{-(- A + B + C) J_C}
\label{OrthoTrig:GB}
\ee
and so it gives the unknown angle $\Delta_C=-A +B +C$ appearing
in (\ref{OrthoTrig:GBC}).  A very similar procedure (or direct use of
(\ref{OrthoTrig:compatibility}) in (\ref{OrthoTrig:GB})) allows us to
derive two analogous equations:
\be
\begin{array}{l}
\myexp{\bpb} \myexp{-\apa} \myexp{\cpc} 
   = \myexp{-(- A + B + C) J_A} \cr
\myexp{\cpc} \myexp{\bpb} \myexp{-\apa} 
   = \myexp{-(- A + B + C) J_B}
\end{array}
\label{OrthoTrig:GBab}\ee
hence we obtain
\be
\Delta_A = \Delta_B = \Delta_C = - A + B + C\equiv\Delta .
\label{eq:OrthoTrig:AngularExcess}
\ee
The quantity  $\Delta$ will be called the {\em  angular excess}
of the triangle loop, and fits very naturally into the view of
the point loop as a single curve which starts on the geodesic
$a$, and successively rotates by angles $C$, $-A$ and $B$ around
the three vertices of the triangle, so that $-A +B +C$ should be
looked as the (oriented) total angle turned by the line loop.
Equations (\ref{OrthoTrig:GB}) or (\ref{OrthoTrig:GBab}),  to be called
the {\em point loop equations}, simply state that the product of the
three translations along the oriented sides of the triangle loop
equals a  rotation around the base point of the loop, with an
angle equal to minus the {\em angular excess} of the triangle
loop.

The explicit duality of the starting equations
(\ref{OrthoTrig:compatibility}) under the interchange $a, b, c
\leftrightarrow A, B, C$ and $P \leftrightarrow J$ immediately
implies that the dual process leads, mutatis mutandis, to the
dual partners of equations  (\ref{OrthoTrig:GB}) and
(\ref{OrthoTrig:GBab}):
\be
\begin{array}{l}
\myexp{-\AJA} \myexp{\CJC} \myexp{\BJB} 
    =  \myexp{-(-a + b + c) P_c} \cr
\myexp{\BJB} \myexp{-\AJA} \myexp{\CJC} 
    =  \myexp{-(-a + b + c) P_a} \cr
\myexp{\CJC} \myexp{\BJB} \myexp{-\AJA} 
    =  \myexp{-(-a + b + c) P_b}
\label{OrthoTrig:GBpolarABC}
\end{array}
\ee
implying that
\be
\delta_a=\delta_b=\delta_c=-a+b+c\equiv \delta
\label{eq:OrthoTrig:LateralExcess}
\ee
which will be called the {\em lateral excess} of the triangle.
This appears as the (oriented)  total length of the point loop,
where $b$ and
$c$ are traversed in the same sense of the orientations chosen
for $P_b, P_c$, but $a$ is traversed backwards relative to
$P_a$. Therefore (\ref{OrthoTrig:GBpolarABC}), to be called {\em line
loop equations}, gives the  product of the three oriented 
rotations around the three vertices of a triangle as a
translation along the base line of the loop, by an amount equal
to minus the {\em lateral excess} of  the triangle loop.

Consequently, the canonical parameters of the `holonomy'
rotation, or of the  dual `holonomy' translation are independent
of the base point or line, and are therefore associated to the
triangle loop in an intrinsic way. The  excesses
$\Delta$ and $\delta$ are directly related   to other natural
quantities, the area and coarea of the triangular loop.


\subsection{The basic trigonometric identity}
\label{Sec:iii:ii}

Potentially, each of  the equations (\ref{OrthoTrig:GB}),
(\ref{OrthoTrig:GBab}) and (\ref{OrthoTrig:GBpolarABC}) contains all the
trigonometry of any CK space, that is, all  the relationships
between the three sides and the  three angles of the triangle.
However, sides and angles  appear in these equations not only
explicitly as canonical parameters, but also implicitly hidden
inside the translation and rotation generators. This prompts the
search for another relation, equivalent to the previous ones but
more suitable to display the trigonometric equations; this new
equation is indeed the bridge between the former equations and
the trigonometry of the space. The main idea is to  express {\em
all} the generators as suitable conjugates of {\em one}
translation generator  and {\em one} rotation generator, which
we will take as primitive {\em independent}  generators; the
choice is only restricted by the flag  condition that the center
of the rotation should lie on the axis of the translation.

A natural choice is to take $P_a$ and $J_C$ as `basic'
independent generators. Next by using the compatibility
conditions (\ref{OrthoTrig:compatibility}) we {\em define} the 
remaining triangle generators $P_b, J_A, P_c, J_B$ in term of
the previous ones and the values of  sides and angles as:
\be
\begin{array}{l}
P_b:=\myexp{\CJC}P_a\myexp{-\CJC} \cr
J_A:=\myexp{\bpb}J_C\myexp{-\bpb} \cr
P_c:=\myexp{-\AJA}P_b\myexp{\AJA} \cr
J_B:=\myexp{\cpc}J_A\myexp{-\cpc}
\end{array}
\ee
which after full expansion and simplification gives:
\be
\begin{array}{l}
P_b:=\myexp{\CJC}P_a \myexp{-\CJC}   \cr
J_A:=\myexp{\CJC}\myexp{b P_a}J_C  
         \myexp{-b P_a} \myexp{-\CJC}\cr
P_c:=\myexp{\CJC}\myexp{b P_a}\myexp{-A J_C} 
         P_a \myexp{A J_C}\myexp{-b P_a} \myexp{-\CJC}\cr
J_B:=\myexp{\CJC}\myexp{b P_a}\myexp{-A J_C} \myexp{c P_a} J_C
        \myexp{-c P_a}\myexp{A J_C}\myexp{-b P_a} \myexp{-\CJC}.
\end{array}
\label{OrthoTrig:GenDefinitions}
\ee
(Note the highly ordered pattern in these expressions). By direct
substitution in the equation (\ref{OrthoTrig:GB}) we obtain
\be
\myexp{-\apa} \cdot
\myexp{\CJC}\myexp{b P_a}\myexp{-A J_C} \myexp{c P_a} \myexp{A J_C}
\myexp{-bP_a}
\myexp{-\CJC}\cdot
 \myexp{\CJC} \myexp{b P_a}  \myexp{-\CJC}
     =  \myexp{-(- A + B + C) J_C}
\ee
which after obvious cancellations gives:
\be
\myexp{-a P_a} \myexp{C J_C} \myexp{b P_a} 
 \myexp{-A J_C} \myexp{c P_a} \myexp{B J_C} = 1.
\label{OrthoTrig:basicEq}
\ee
On the other hand, substitution of (\ref{OrthoTrig:GenDefinitions})
in the second equation in (\ref{OrthoTrig:GBpolarABC}) and
simplification gives
\be
\myexp{C J_C} \myexp{b P_a} \myexp{-A J_C} \myexp{c P_a}
\myexp{B J_C} \myexp{-a P_a}  = 1
\ee
which {\em coincides} with the previous  equation, because the
terms can be cyclically permuted. The same process  starting
from any of the two equations (\ref{OrthoTrig:GBab}) which are
associated to the  other two base points in the triangle, or any
of the equations in (\ref{OrthoTrig:GBpolarABC}) but  associated to
the other two base lines in the triangle, leads again to the
same  equation. This justifies to call (\ref{OrthoTrig:basicEq}) the
{\em basic trigonometric equation}.

It is possible to give an heuristic derivation of
(\ref{OrthoTrig:basicEq}). Consider a kind of {\em complete} group
motion associated to the triangle:
\be
\myexp{-\apa}\myexp{\BJB} \myexp{\cpc} \myexp{-\AJA}\myexp{\bpb}
\myexp{\CJC} .
\label{OrthoTrig:completeGB:aa}
\ee
It is clear that this group element leaves fixed {\em both} the
base line $a$ {\em and} the base point $C$. Therefore it should
be equal to the identity:
\be
\myexp{-\apa}\myexp{\BJB} \myexp{\cpc} \myexp{-\AJA}\myexp{\bpb}
\myexp{\CJC}
=1.
\label{OrthoTrig:completeGB:a}
\ee
Now substitution of (\ref{OrthoTrig:GenDefinitions}) in
(\ref{OrthoTrig:completeGB:a}) and obvious cancellations gives {\em
directly} the basic equation (\ref{OrthoTrig:basicEq}).  However,
this reasoning line does not display the relevance of loop
excesses. The equivalence between the equation
(\ref{OrthoTrig:completeGB:a}), which involves the three translation
generators $P_a, P_b, P_c$ with parameters
$-a, b, c$ and the three rotation generators $J_A, J_B, J_C$
with angles $-A, B, C$, and the equation (\ref{OrthoTrig:basicEq}),
where {\em only} the fiducial generators $P_a$ and $J_C$ appear
with the same parameters, is an apparently surprising but obvious
consequence of the compatibility conditions.

The results obtained so far can be summed up in the following:
\medskip

\noindent
{\bf Theorem 1.} Sides $a, b, c$ and angles $A, B, C$ of any
triangle loop are  linked by the single group identity  called
the {\em basic trigonometric identity}
\be
\myexp{-a P} \myexp{C J}  \myexp{b P} \myexp{-A J} \myexp{c P}
\myexp{B J}  = 1
\label{eq:OrthoTrig:BasicIdentTheora}
\ee
where $P, J$ are the generators of translations along {\em any}
fixed fiducial line $l$, and of rotations around any fixed
fiducial point $O$ on the line $l$.
\medskip

\noindent
{\em Proof}: A group motion can be used to move any triangle to a
canonical  position where the vertex $C$ lies on the fiducial
point $O$, and the side $a$ lies on the fiducial  line $l$. Then
the theorem statement is simply the equation 
(\ref{OrthoTrig:basicEq}).
\medskip

\noindent
{\bf Theorem 2.} Let  $P_a, P_b, P_c$ be the generators of {\em
translations}  along the three sides of a triangle  (whose
lengths are $a, b, c$), and $J_A, J_B, J_C$ the generators  of
{\em rotations} around the three vertices (with angles $A, B,
C$). Then we have   two sets of identities,  
called the {\em point loop} and the {\em line loop} equations
for the triangle:
\be
\begin{array}{l}
\myexp{-\apa} \myexp{\cpc} \myexp{\bpb} 
      =  \myexp{-(- A + B + C) J_C}
\qquad \myexp{\BJB} \myexp{-\AJA} \myexp{\CJC} 
      =  \myexp{-(-a + b + c) P_a} \cr
\myexp{\bpb}  \myexp{-\apa} \myexp{\cpc} 
      =  \myexp{-(- A + B + C) J_A}
\qquad \myexp{\CJC} \myexp{\BJB} \myexp{-\AJA} 
      =  \myexp{-(-a + b + c) P_b} \cr
\myexp{\cpc}  \myexp{\bpb} \myexp{-\apa} 
      =  \myexp{-(- A + B + C) J_B}
\qquad  \myexp{-\AJA} \myexp{\CJC} \myexp{\BJB}     
     =  \myexp{-(-a + b + c) P_c} .
\end{array}
\ee
Furthermore each of these identities is equivalent to the
identity in Theorem 1.
\medskip

As each point loop and each line loop equation is equivalent to
the basic identity, we conclude that the three point loop
equations, the three line loop equations and the basic equation
are all equivalent.

Several points are worth highlighting. First, each term in the
basic identity is either a translation along a fixed line $l$ or
a rotation around a fixed  point $O$, so any relation with the
original triangle translation or rotation generators $P_a, P_b,
P_c$ or $J_A, J_B, J_C$ is somewhat concealed. However, the
canonical parameters of these fiducial translations or rotations
are exactly the  sides and the angles of the triangle. In the
point loop or line loop equations, however, the transformations
involved are the translations  along the three sides or the
rotations around the three vertices.

Second, the structure of the equations can be easily remembered.
The point loop equations follow from a point travelling along
the triangle; it starts at the base point  $C$ and follows the
sides in the order $b,c,-a$, with $a$ negative as it is traversed
backwards (see figure \ref{fig:TriangleLoops}a). The line loop
equations follow from a line looping around the triangle; it 
starts in a base line (say $a$) and successively rotates by an
angle $C$, then by an angle $-A$ and finally by an angle $B$ around
the corresponding vertices thus ending up back on the starting
position
$a$ (see figure \ref{fig:TriangleLoops}b). The basic equation
follows from the pattern
$-a, C, b, -A, c, B$, which keeps track of both  sides and
vertices found when looping around the triangle; again $a, A$
appear with a minus sign, because both are traversed in negative
sense.

Third, the (three) point loop equations and the (three) line
loop equations are mutually dual sets. The single basic equation
is clearly self-dual.

Fourth, and worth some emphasizing, these equations hold in the
same explicit form for {\em all } 2D  real CK geometries, as no
{\em explicit} $\k_1, \k_2$ ever appear in them. In the well
known euclidean case, the angle sum theorem for a triangle
---which reads $A=B+C$  in terms of our external angle $A$---
is  usually taken as an elementary property of the geometry,
known beforehand trigonometry, and in this particular space the
point loop equation reduces to $\myexp{-\apa}
\myexp{\cpc}\myexp{\bpb} = 1$. This simply says that no holonomy
rotation is associated to the triangle at all, and conveys the
flatness of the euclidean plane. In our approach the holonomy
angle is {\em always} $-A+B+C$ (no matter of the values of $\k_1,
\k_2$), but whether or not this angle vanishes actually depends
on the values of
$\k_1, \k_2$; as we shall see the equations of trigonometry
themselves imply the vanishing of $-A+B+C$ for $\k_1=0$. In
other words, the euclidean relation $A=B+C$ as well as their
non-linear non-euclidean analogous will arise as part of the 
trigonometric relations.


\section{The basic equations of trigonometry in the nine
Cayley--Klein spaces}
\label{Sec:iv}

The most convenient way to obtain the trigonometric equations is
to start with the  basic trigonometry identity 
(\ref{eq:OrthoTrig:BasicIdentTheora}), in which from now on the two
generators $P$ and $J$ will be taken to be exactly $P_1$ and $J_{12}$.
For convenience, we write (\ref{eq:OrthoTrig:BasicIdentTheora}) as
\be
\myexp{-a P} \myexp{C J} \myexp{b P} 
   = \myexp{-B J} \myexp{-c P} \myexp{A J} .
\ee
By considering this identity in the fundamental 3D vector
representation of the motion group 
(\ref{eq:OrthoCKBiDOneParamSubgroups}) we obtain an
equality between $3 \times 3$ matrices, giving rise to nine
identities:
\be
\begin{array}{ll}
\mbox{1c}& \, \c(c) =\c(a)\c(b)+\k_1 \s(a)\s(b)\cc(C) \\
\mbox{1C}& \, \cc(C) =\cc(A)\cc(B)+\k_2 \ss(A)\ss(B)\c(c)\\
\mbox{2cA} \equiv \mbox{2aC}&\s(c) \ss(A)=\s(a)\ss(C)\\
\mbox{2cB} \equiv \mbox{2bC}&\s(c) \ss(B)=\s(b)\ss(C)\\
\mbox{3cA}& \s(c) \cc(A) = -\c(a)\s(b)+ \s(a)\c(b)\cc(C) \\
\mbox{3cB}&\s(c) \cc(B)=\c(b)\s(a)- \s(b)\c(a)\cc(C)\\
\mbox{3Ca}& \ss(C) \c(a)=-\cc(A)\ss(B) +\ss(A)\cc(B)\c(c)\\
\mbox{3Cb}&\ss(C) \c(b)=\cc(B)\ss(A)-\ss(B)\cc(A)\c(c)\\
\mbox{4AB}\equiv\mbox{4ab}&\k_2\ss(A) \ss(B)+\cc(A)\cc(B)\c(c)\\
&\qquad\qquad\qquad =\k_1\s(a)\s(b)+ \c(a)\c(b)\cc(C)
\end{array}\label{OrthoTrig:equations:c}
\ee
which are the equations for  the trigonometry of the space 
$\CKPointSpace=SO_{\k_1,\k_2}(3)/ SO_{\k_2}(2)$. The tag
assigned to each equation is  self-descriptive: all equations
are either self-dual (for instance 2cA $\equiv$ 2aC) or appear
in mutually dual pairs (as (1c, 1C) or (3cA, 3Ca)); this  could
have been expected due to the self-duality of the starting
equation.

A first elementary but remarkable property of these equations 
is the complete matching of sides and angles  with the two
labels $\k_1$ and $\k_2$; all sides (resp.\ angles) appear in the
equations through the trigonometric functions which have $\k_1$
(resp.\ $\k_2$) as a label. Then the association  between
labels  and  generators extends to the lengths and angles
themselves, and we will say that first-kind lengths have $\k_1$
(resp.\ angles $\k_2$) as a label. 

The two equations 2cA, 2cB taken together are the {\em sine
theorem}, whose general formulation for the nine CK  spaces
displays explicit self-duality (cf.\ (\ref{eq:OrthoTrig:Sine})).  Next,
the equation 1c is the {\em cosine theorem} for the side
$c$,  and 1C  is the {\em dual cosine theorem} for the angle $C$;
a pair of relationships whose mutual duality can be clearly seen
in this general approach. The two equations 3cA, 3cB can be
considered as two `addition' formulas for the side $c$ in terms
of the projections of the others, and the two formulas 3Ca, 3Cb
are their duals.  The single self-dual equation 4AB relates all
six trigonometric quantities.

To describe clearly the structure and dependence between these
equations, it is better to consider the group of equations
$(\ref{OrthoTrig:equations:c})$  altogether with two similar groups,
each equivalent as a set of equations to the previous one. These
can be obtained by starting from the basic identity written in
two alternative forms:
\be
\myexp{b P} \myexp{-A J} \myexp{c P} 
    \myexp{B J} \myexp{-a P} \myexp{C J} = 1
\qquad
\myexp{c P} \myexp{B J} \myexp{-a P} 
    \myexp{C J} \myexp{b P}\myexp{-A J} = 1.
\ee
Spliting them again as
\be
\myexp{b P} \myexp{-A J} \myexp{c P}
      = \myexp{-C J}\myexp{a P} \myexp{-B J}
\qquad
\myexp{c P} \myexp{B J} \myexp{-a P}
     =\myexp{A J}\myexp{-b P} \myexp{-C J}
\ee
and writing them in the fundamental  representation
(\ref{eq:OrthoCKBiDOneParamSubgroups}),
we obtain other two sets of equations very similar to
(\ref{OrthoTrig:equations:c}). We only write the two first equations
for each set:
\be
\begin{array}{ll}
\mbox{1a} &\qquad \c(a) =\c(b)\c(c)-\k_1 \s(b)\s(c)\cc(A)\\
\mbox{1A} &\qquad \cc(A) =\cc(B)\cc(C)-\k_2 \ss(B)\ss(C)\c(a)\\
\vdots &\qquad \qquad \vdots \\
\end{array}\label{OrthoTrig:equations:a}
\ee
\be
\begin{array}{ll}
\mbox{1b} &\qquad \c(b) =\c(a)\c(c)+\k_1 \s(a)\s(c)\cc(B)\\
\mbox{1B} &\qquad \cc(B) =\cc(A)\cc(C)+\k_2 \ss(A)\ss(C)\c(b)\\
\vdots &\qquad \qquad \vdots \\
\end{array}\label{OrthoTrig:equations:b}
\ee

All the equations (\ref{OrthoTrig:equations:c}),
(\ref{OrthoTrig:equations:a}) and (\ref{OrthoTrig:equations:b}) taken
altogether can be grouped into:

\noindent
$\bullet$ Three {\em cosine theorems} for sides:
\be
\begin{array}{ll}
\mbox{1a} &\qquad \c(a) =\c(b)\c(c)-\k_1 \s(b)\s(c)\cc(A) \\
\mbox{1b} &\qquad \c(b) =\c(a)\c(c)+\k_1 \s(a)\s(c)\cc(B) \\
\mbox{1c} &\qquad \c(c) =\c(a)\c(b)+\k_1 \s(a)\s(b)\cc(C)
\label{eq:OrthoTrig:Cosine}
\end{array}
\ee
and three {\em dual cosine theorems} for angles:
\be
\begin{array}{ll}
\mbox{1A} &\qquad \cc(A) =\cc(B)\cc(C)-\k_2 \ss(B)\ss(C)\c(a)\\
\mbox{1B} &\qquad \cc(B) =\cc(A)\cc(C)+\k_2 \ss(A)\ss(C)\c(b)\\
\mbox{1C} &\qquad \cc(C) =\cc(A)\cc(B)+\k_2 \ss(A)\ss(B)\c(c) .
\end{array}
\label{eq:OrthoTrig:DualCosine}
\ee

\noindent
$\bullet$ One self-dual {\em sine theorem}:
\be
2\qquad \quad \frac{\s(a)}{\ss(A)} = \frac{\s(b)}{\ss(B)}
=\frac{\s(c)}{\ss(C)}.
\label{eq:OrthoTrig:Sine}
\ee

\noindent
$\bullet$ Six `side addition' theorems which correspond to the
tags 3aB, 3aC, 3bA, 3bC, 3cA, 3cB and six dual `angle addition'
theorems 3Ab, 3Ac, 3Ba, 3Bc, 3Ca, 3Cb.

\noindent
$\bullet$ And three self-dual theorems 4AB $\equiv$ 4ab,  4AC
$\equiv$ 4ac, 4BC $\equiv$ 4bc.

We notice the signs in 1a, 1A, differing from those in 1b, 1B
and 1c, 1C. The same sign difference occurs in all other
equations, and can be traced back to the appearance of $-a$ and
$-A$ in the basic equation, as compared with $b, c$ and $B, C$.

These equations are a {\em complete} set of trigonometric
equations for any values of the constants $\k_1, \k_2$, but most
of them reduce to simpler, or even  trivial ones in  the
degenerate cases $\k_1=0$ or
$\k_2=0$.


\subsection{Alternative forms for the cosine theorems}
\label{Sec:iv:i}

The cosine theorems (\ref{eq:OrthoTrig:Cosine})  (resp.\
(\ref{eq:OrthoTrig:DualCosine})) give rise to trivial identities $1=1$
when $\k_1=0$ (resp.\ $\k_2=0$). This  can be circumvented by writing
these equations  in an alternative form.  Take  the cosine equation for
the side $c$ (1c of (\ref{eq:OrthoTrig:Cosine})) in the generic case with
$\k_1\neq 0$, and  write all cosines  in terms of versed sines by
introducing (\ref{eq:CKCosVersineRel}):
\be
1 -\k_1 \v(c) =
(1 -\k_1 \v(a)) (1 -\k_1 \v(b)) + \k_1 \s(a)\s(b)(1-\k_2 \vv(C)) .
\ee
By expanding and cancelling a common factor $\k_1$ we find that
\be
\v(c) = \v(a) + \v(b) -\k_1 \v(a) \v(b) -\s(a)\s(b)+
\k_2   \s(a)\s(b)\vv(C)  ,
\ee
which is rather simplified by means of  (\ref{eq:CKVersineSumDif}). The
remaining cosine theorems 1a, 1b and the dual  cosine equations
(\ref{eq:OrthoTrig:DualCosine}) allow a similar reformulation. Thus we 
obtain  the following alternative expressions:
\be
\begin{array}{ll}
\mbox{1'a} &\qquad \v(a) - \v(b+c)  = -\k_2 \s(b) \s(c)  \vv(A)    \\
\mbox{1'b} &\qquad \v(b) - \v(a-c)  = \k_2\s(a) \s(c)   \vv(B)   \\
\mbox{1'c} &\qquad \v(c) - \v(a-b)  =\k_2 \s(a) \s(b)  \vv(C)
\label{eq:OrthoTrig:Cosine:i}
\end{array}
\ee
\be
\begin{array}{ll}
\mbox{1'A} &\qquad \vv(A) - \vv(B+C)  = -\k_1\ss(B) \ss(C)    \v(a) \\
\mbox{1'B} &\qquad \vv(B) - \vv(A-C)  = \k_1\ss(A) \ss(C)   \v(b) \\
\mbox{1'C} &\qquad \vv(C) - \vv(A-B)  =\k_1 \ss(A) \ss(B)    \v(c)  .
\end{array}
\label{eq:OrthoTrig:DualCosine:i}
\ee
These relations  are   clearly equivalent to (\ref{eq:OrthoTrig:Cosine})
(resp.\  (\ref{eq:OrthoTrig:DualCosine})) when
$\k_1\neq 0$ (resp.\ $\k_2\neq 0$), but  do not reduce to 
trivial identities when $\k_1=0$   or $\k_2=0$. In this sense
they can be considered as a `good' form of cosine and dual
cosine equations.

Equations (\ref{eq:OrthoTrig:Cosine:i})  and
(\ref{eq:OrthoTrig:DualCosine:i}) still allow another alternative very
useful form. Consider the half sums of the three sides and of the three
angles (cf.\ (\ref{zzz:wa})),
\be
p=\frac{a+b+c}{2}
\qquad
P=\frac{A+B+C}{2}   \, .
\label{eq:OrthoTrig:Semisums}
\ee
By introducing the identities (\ref{zzz:wl}) and (\ref{zzz:wm}) applied
to the three sides   $a,b,c$ and angles $A, B, C$ into the
cosine theorems (\ref{eq:OrthoTrig:Cosine:i})  and
(\ref{eq:OrthoTrig:DualCosine:i}), we obtain
\be
\begin{array}{ll}
\mbox{1''a} &\quad 2 \s(p-a) \s(p)  = \k_2 \s(b)\s(c)  \vv(A)  \\
\mbox{1''b} &\quad 2 \s(p-a) \s(p-c) = \k_2  \s(a) \s(c)  \vv(B)  \\
\mbox{1''c} &\quad 2 \s(p-a) \s(p-b) = \k_2  \s(a) \s(b)  \vv(C)
\end{array}
\label{eq:OrthoTrig:Cosine:ii}
\ee
\be
\begin{array}{ll}
\mbox{1''A} &\quad  2 \ss(P-A) \ss(P) = \k_1 \ss(B) \ss(C) \v(a)  \\
\mbox{1''B} &\quad 2 \ss(P-A) \ss(P-C) =\k_1\ss(A) \ss(C) \v(b)  \\
\mbox{1''C} &\quad 2 \ss(P-A) \ss(P-B) =  \k_1 \ss(A) \ss(B) \v(c)  .
\end{array}
\label{eq:OrthoTrig:DualCosine:ii}
\ee


\subsection{Dependence and sets of basic equations}
\label{Sec:iv:ii}

The equations we have obtained so far contain the whole
trigonometry of the CK space
$\CKPointSpace\equiv SO_{\k_1,\k_2}(3)/ SO_{\k_2}(2)$ and hold
for any of the nine CK geometries by simply particularising the
real coefficients $\k_i$ to their values for each geometry. Not
all these equations can be independent: in any CK space, a
triangle is completely determined by {\em three} independent
quantities (e.g.\ $a, b, C$, because the side-angle-side (SAS)
congruence condition clearly holds in all the nine cases) so we
should expect {\em three} independent relations between the six
quantities $a, b, c; A, B, C$.

Let us first discuss the case with  $\k_1=0$ but $\k_2\neq 0$.
In these degenerate cases we obtain the well known trigonometry
of the euclidean plane ($\k_2>0$), and the less well known
lorentzian trigonometry of the $(1+1)$D minkowskian   spacetime
($\k_2<0$) \cite{BN}. The formulas (\ref{eq:OrthoTrig:Cosine}) reduce to
trivial identities $1=1$, but the alternative expressions
(\ref{eq:OrthoTrig:Cosine:i}) or   (\ref{eq:OrthoTrig:Cosine:ii}) lead
to the three ordinary flat euclidean or lorentzian cosine theorems, the
latter exactly as given in \cite{BN}. For instance, 1'c gives rise to
\be
\frac 12 c^2 - \frac 12 (a-b)^2 = \k_2 a b\, \vv(C)
    \quad\equiv\quad
c^2 = a^2 + b^2 - 2 a b\, \cc(C) .
\label{eq:OrthoTrig:CosineFlatTheorem}
\ee
All the remaining  equations of trigonometry do not reduce to
identities and are directly meaningful, yet simpler. By taking
into account the sine and cosine addition identities 
(\ref{eq:CKCosSumDif}) and (\ref{eq:CKSinSumDif}), we find that for
$\k_1=0$ the content of all the dual cosine theorems
(\ref{eq:OrthoTrig:DualCosine})  and all equations 3Ab, 3Ac, 3Ba, 3Bc,
3Ca,  3Cb,  4AB, 4AC, 4BC is the same and reduces to a triangular angle
addition in the form:
\be
A = B + C\quad \equiv \quad \Delta=0.
\label{eq:OrthoTrig:DualCosineFlatTheorem}
\ee
The equality $\Delta=0$ implies that for $\k_1 = 0$ the holonomy
(\ref{OrthoTrig:GB}) is equal to the identity, as it should in any
flat space. In these flat spaces where $\k_1=0$ the angles are
related by a `universal' linear equation not depending on the
sides. This universality is why  the equality $A=B+C$ is usually
taken as a property of euclidean geometry, and not as a
trigonometric  equation.  The sine theorem (\ref{eq:OrthoTrig:Sine})
reads now
\be
\frac{a}{\ss(A)} = \frac{b}{\ss(B)} = \frac{c}{\ss(C)}
\label{eq:OrthoTrig:FlatSineTheorem}
\ee
and the remaining equations 3aB, 3aC, 3bA, 3bC, 3cA, 3cB give 
each side as the sum of the projections of the other two. In
particular, relations 3cA and 3cB in (\ref{OrthoTrig:equations:c})
reduce to:
\be
 b =  a\,\cc(C)- c \,\cc(A) \qquad a= b\,\cc(C)+c \,\cc(B) .
\label{zzz:projeucl}
\ee
Notice that the three cosine theorems 1'a, 1'b, 1'c (as
(\ref{eq:OrthoTrig:CosineFlatTheorem}))  are still {\em independent}.
From these we can derive all the remaining non-trivial equations,
including the dual cosine
theorem (\ref{eq:OrthoTrig:DualCosineFlatTheorem}), the sine theorem
(\ref{eq:OrthoTrig:FlatSineTheorem}) (see below)  and the relations on
sum of projections as (\ref{zzz:projeucl}). Henceforth, when
$\k_1=0$ but $\k_2\neq0$, the canonical choice for three
independent equations is the three alternative cosine theorems
(\ref{eq:OrthoTrig:Cosine:i}) for the three sides.

A fully parallel dual discussion can be repeated for the case
$\k_2=0$ but $\k_1\neq 0$. Then the  equations
(\ref{eq:OrthoTrig:DualCosine}) give  rise to trivial identities, while
(\ref{eq:OrthoTrig:DualCosine:i}) or (\ref{eq:OrthoTrig:DualCosine:ii})
provide the  dual cosine theorems for the angles, which are the three
independent equations. The relations (\ref{eq:OrthoTrig:Cosine}), 3aB,
3aC, 3bA, etc.\  reduce to the same single equation:
\be
a=b+c  \quad \equiv \quad \delta=0 .
\label{eq:OrthoTrig:CosineCoFlatTheorem}
\ee

Finally, in the more contracted  case with $\k_1=\k_2=0$, that
is, the $(1+1)$D galilean geometry, the equations are simply
\be
 a=b+c  \qquad A=B+C  \qquad \frac aA = \frac bB = \frac cC
\label{eq:OrthoTrig:GalBasic}
\ee
which are fully linear.  The three independent equations are  a
single addition theorem for sides alone, another  for angles
alone, and a `sine' theorem stating the proportionality of the
three sides and angles.

All the results about the dependence of equations   can
be summed up in the following:

\medskip
\noindent
{\bf Theorem 3.} The full set of equations of  trigonometry
always contains, (i.e.\ for any value of $\k_1$, $\k_2$) exactly
{\em three} independent equations. Any other equation in the set
is a consequence of them. According to the values of $\k_1$,
$\k_2$ we find the following cases:

\noindent
$\bullet$ When $\k_1\neq 0$ and $\k_2\neq 0$,  the three
equations (\ref{eq:OrthoTrig:Cosine:i})  (or (\ref{eq:OrthoTrig:Cosine})
which for $\k_1\neq 0$ are equivalent to them) are three such
independent equations; the three
equations (\ref{eq:OrthoTrig:DualCosine:i}) (or
(\ref{eq:OrthoTrig:DualCosine}) which for $\k_2\neq 0$ are equivalent to
them)  are another choice of three such independent equations. All
trigonometry follows from either (\ref{eq:OrthoTrig:Cosine:i}) or
(\ref{eq:OrthoTrig:DualCosine:i}).

\noindent
$\bullet$ When $\k_1=0$ but $\k_2\neq 0$,   the three equations
(\ref{eq:OrthoTrig:DualCosine:i})   reduce to a single independent
equation (\ref{eq:OrthoTrig:DualCosineFlatTheorem}). The three equations
(\ref{eq:OrthoTrig:Cosine:i}) are independent, and
(\ref{eq:OrthoTrig:Cosine}) reduce to  trivial identities. All the
trigonometry  in this case follows from the three equations
(\ref{eq:OrthoTrig:Cosine:i}).

\noindent
$\bullet$ When $\k_1\neq 0$  but $\k_2=0$,  the three equations
(\ref{eq:OrthoTrig:Cosine:i}) reduce to a single   independent equation
(\ref{eq:OrthoTrig:CosineCoFlatTheorem}). The three equations
(\ref{eq:OrthoTrig:DualCosine:i}) are independent, and
(\ref{eq:OrthoTrig:DualCosine}) reduce to  trivial identities. All
the trigonometry in this case follows from the three equations
(\ref{eq:OrthoTrig:DualCosine:i}).

\noindent
$\bullet$ When  $\k_1=\k_2=0$,  the three equations
(\ref{eq:OrthoTrig:Cosine:i}) reduce to a single  independent equation,
and (\ref{eq:OrthoTrig:Cosine}) are trivial identities. The three
equations (\ref{eq:OrthoTrig:DualCosine:i})  reduce to a single
independent equation, and (\ref{eq:OrthoTrig:DualCosine})  are  trivial
identities. The trigonometry follows from
(\ref{eq:OrthoTrig:Cosine:i}),  (\ref{eq:OrthoTrig:DualCosine:i}) and
(\ref{eq:OrthoTrig:Sine}) which in this case are three  independent
equations (\ref{eq:OrthoTrig:GalBasic}).

\medskip

\noindent {\em Proof.} We consider the generic case with 
$\k_1\neq 0$,
$\k_2\neq 0$,  and the  cosine theorems  (\ref{eq:OrthoTrig:Cosine}) as
the three initial equations (they are independent {\em a fortiori}
since each of them involves one different angle). Let us obtain
the sine theorem (\ref{eq:OrthoTrig:Sine}); on the one hand, we compute:
\be
 \s^2(a)\ss^2(B)= \s^2(a) 
\frac{ \left(1- \cc^2(B) \right)} {\k_2}
=\frac{\s^2(a)}{\k_2}\left(1-
\frac{\left(\c(b)-\c(a)\c(c)\right)^2}{\k_1^2\s^2(a)\s^2(c)}
\right)
\ee
where we have introduced the equation 1b of 
(\ref{eq:OrthoTrig:Cosine}); by expanding and writing the sines in terms
of cosines in the numerator, we find:
\be
\s^2(a)\ss^2(B)=\frac{1-\c^2(a)-\c^2(b)-\c^2(c)+2\c(a)\c(b)\c(c)}
{\k_1^2\k_2\s^2(c)} .
\label{zzz:theor3a}
\ee
(Note that the numerator is the determinant of the Gramm matrix whose
elements are the scalar products, in the linear ambient space of
the vector model,  of the vectors corresponding to the
vertices).  On the other hand,  similar computations show that
 $\s^2(b)\ss^2(A)$ gives again the r.h.s.\ of (\ref{zzz:theor3a}), so
that
$\s(a)\ss(B)=\s(b)\ss(A)$. Likewise, we can prove that
$\s(a)\ss(C)=\s(c)\ss(A)$, completing  the sine theorem.  Next,
we deduce the dual cosine  theorems (\ref{eq:OrthoTrig:DualCosine}). By
taking into account the sine theorem (just proven) and
(\ref{zzz:theor3a}) it can be checked that
\be
\ss(A)\ss(B)=\frac{\s^2(a)\ss^2(B)}{\s(a)\s(b)}=
\frac{1-\c^2(a)-\c^2(b)-\c^2(c)+2\c(a)\c(b)\c(c)}
{\k_1^2\k_2 \s^2(c)\s(a)\s(b)}  .
\ee
The same expression can be obtained  starting from
\be
\frac{\cc(C)-\cc(A)\cc(B)}{\k_2\c(c)}
\ee
 and using the three equations  (\ref{eq:OrthoTrig:Cosine}) in order to
get rid of the angles, thus writing everything in terms of the
sides. Therefore the equation  1C of (\ref{eq:OrthoTrig:DualCosine})
follows; the two remaining dual cosine theorems can be obtained in the
same way. Once we have obtained (\ref{eq:OrthoTrig:Sine}) and
(\ref{eq:OrthoTrig:DualCosine})
 starting from (\ref{eq:OrthoTrig:Cosine}),  the 12 `addition' formulas
for sides and angles  (with tags 3aB, 3Ab, etc.) and the self-dual
equations  (with tags $4AB\equiv 4 ab$, etc.) can be
straightforwardly deduced from the former theorems. By duality,
it is clear that we can start from the cosine theorem for angles.

For the contracted case with $\k_1=0$ and $\k_2\ne 0$, as well as
for its dual with $\k_2=0$ and $\k_1\ne 0$, the proof follows
similar steps as in the generic case and we omit it. Finally, the
proof for  the last case with $\k_1=\k_2=0$  is trivial.

We display in table \ref{table:BasicTrigEqns} the cosine, dual
cosine and sine  theorems for each of the nine CK geometries
according to the values   of the curvatures $(\k_1,\k_2)$.  We
recall that   when $\k_1$ (resp. $\k_2$) is  {\em different}
from zero, by a suitable choice of length unit (resp.\ angle
unit), it can be reduced to either $+ 1$ or $-1$: in this table
we write the theorems with such an adapted `natural' choice of
units.


\begin{table}[h]

{\footnotesize
{ \noindent
\caption{Cosine,  sine and dual cosine
 theorems for the nine CK spaces.}
\label{table:BasicTrigEqns}
\smallskip
\noindent\hskip -30pt
\begin{tabular}{ccc}
\hline
&&\\[-8pt]
Elliptic\quad $(+1,+1)$&
Euclidean\quad $(0,+1)$&
Hyperbolic\quad $(-1,+1)$\\
$SO(3)/SO(2)$&$ISO(2)/SO(2)$&$SO(2,1)/SO(2)$\\[4pt]
$ \cos a=\cos b\cos c- \sin b\sin c\cos A$&
$a^2=b^2+c^2+ 2  b c\cos A$&
$\cosh a=\cosh b\cosh c+ \sinh b\sinh c\cos A$\\
$\cos b=\cos a\cos c+ \sin a\sin c\cos B$&
$b^2=a^2+c^2- 2  a c\cos B$&
$\cosh b=\cosh a\cosh c - \sinh a\sinh c\cos B$\\
$\cos c=\cos a\cos b+ \sin a\sin b\cos C$&
$c^2=a^2+b^2- 2  a b\cos C$&
$\cosh c=\cosh a\cosh b - \sinh a\sinh b\cos C$\\[2pt]
$\displaystyle{\frac{\sin a}{\sin A}=\frac{\sin b}{\sin B}
  =\frac{\sin c}{\sin C}}$&
$\displaystyle{\frac{a}{\sin A}=\frac{ b}{\sin B}
  =\frac{ c}{\sin C}}$&
$\displaystyle{\frac{\sinh a}{\sin A}=\frac{\sinh b}{\sin
B}=\frac{\sinh c}{\sin
C}}$\\[8pt]
$ \cos A=\cos B\cos C- \sin B\sin C\cos a$&
$A=B+C$&
$\cos A=\cos B\cos C- \sin B\sin C\cosh a$\\
$\cos B=\cos A\cos C+ \sin A\sin C\cos b$&
$B=A- C$&
$\cos B=\cos A\cos C + \sin A\sin C\cosh b$\\
$\cos C=\cos A\cos B+ \sin A\sin B\cos c$&
$C=A-B$&
$\cos C=\cos A\cos B + \sin A\sin B\cosh c$\\[4pt]
\hline
&&\\[-8pt]
Co-Euclidean\quad $(+1,0)$&
Galilean\quad $(0,0)$&
Co-Minkowskian\quad $(-1,0)$\\
Oscillating NH\quad $ISO(2)/\Re$&$IISO(1)/\Re$&
Expanding NH\quad $ISO(1,1)/\Re$\\[4pt]
$   a=b+c$&
$a=b+c$&
$a=b+c$\\
$b=a-c$&
$b=a-c$&
$b=a-c$\\
$c=a-b$&
$c=a-b$&
$c=a-b$\\[2pt]
$\displaystyle{\frac{\sin a}{ A}=\frac{\sin b}{ B}
    =\frac{\sin c}{ C}}$&
$\displaystyle{\frac{a}{ A}=\frac{ b}{ B}=\frac{ c}{ C}}$&
$\displaystyle{\frac{\sinh a}{ A}=\frac{\sinh b}{ B}
    =\frac{\sinh c}{
C}}$\\[8pt]
$A^2=B^2+C^2+ 2 B C\cos a$&
$A=B+C$&
$A^2=B^2+C^2+ 2 B C\cosh a$\\
$B^2=A^2+C^2- 2 A C\cos b$&
$B=A- C$&
$B^2=A^2+C^2 - 2 A C\cosh b$\\
$C^2=A^2+B^2- 2 A B \cos c$&
$C=A-B$&
$C^2=A^2+B^2- 2 A B \cosh c$\\[4pt]
\hline
&&\\[-8pt]
Co-Hyperbolic\quad $(+1,-1)$&
Minkowskian\quad $(0,-1)$&
Doubly Hyperbolic\quad $(-1,-1)$\\
Anti-de Sitter\quad $SO(2,1)/SO(1,1)$&$ISO(1,1)/SO(1,1)$&
De Sitter\quad $SO(2,1)/SO(1,1)$\\[4pt]
$ \cos a=\cos b\cos c- \sin b\sin c\cosh A$&
$a^2=b^2+c^2+ 2  b c\cosh A$&
$\cosh a=\cosh b\cosh c+ \sinh b\sinh c\cosh A$\\
$\cos b=\cos a\cos c+ \sin a\sin c\cosh B$&
$b^2=a^2+c^2- 2  a c\cosh B$&
$\cosh b=\cosh a\cosh c - \sinh a\sinh c\cosh B$\\
$\cos c=\cos a\cos b+ \sin a\sin b\cosh C$&
$c^2=a^2+b^2- 2  a b\cosh C$&
$\cosh c=\cosh a\cosh b - \sinh a\sinh b\cosh C$\\[2pt]
$\displaystyle{\frac{\sin a}{\sinh A}=\frac{\sin b}{\sinh B}
    =\frac{\sin c}{\sinh C}}$&
$\displaystyle{\frac{a}{\sinh A}=\frac{ b}{\sinh B}
    =\frac{ c}{\sinh C}}$&
$\displaystyle{\frac{\sinh a}{\sinh A}
   =\frac{\sinh b}{\sinh B}=\frac{\sinh c}{\sinh C}}$\\[8pt]
$\cosh A=\cosh B\cosh C+ \sinh B\sinh C\cos a$&
$A=B+C$&
$\cosh A=\cosh B\cosh C+ \sinh B\sinh C\cosh a$\\
$\cosh B=\cosh A\cosh C- \sinh A\sinh C\cos b$&
$B=A- C$&
$\cosh B=\cosh A\cosh C - \sinh A\sinh C\cosh b$\\
$\cosh C=\cosh A\cosh B- \sinh A\sinh B\cos c$&
$C=A-B$&
$\cosh C=\cosh A\cosh B - \sinh A\sinh B\cosh c$\\[4pt]
\hline
\end{tabular} }}
\end{table}


\subsection{Relation with the usual approach and with absolute
trigonometry}
\label{Sec:iv:iii}

While the equations we have obtained hold for {\em all}  the
nine 2D CK geometries, the spaces whose trigonometry is well
known are the three riemannian spaces of constant curvature
---the sphere, the  euclidean plane and the hyperbolic plane---
which are the  members of the CK family with $\k_2>0$. In these
three spaces  the usual `natural' choice of angle units
corresponds to making $\k_2=1$. Therefore by setting $\k_1=\k$
and $\k_2=1$, our generic relations give rise  to the equations
of the trigonometry of the three riemannian spaces of constant
curvature $\k$, and give directly the so-called `absolute' form
of trigonometry, which is valid for the three spaces
simultaneously \cite{Hsiang, Bonola, Martin}.

In this connection an elementary but  relevant point should be
kept in mind: spherical, euclidean and hyperbolic trigonometry
is usually formulated in terms of the {\em inner} angles. The
natural angles when the triangle is seen as a line loop are
however not the three inner angles; {\em one} of them must be an
external angle, like our
$A$. When $\k_2=1$, the  measure of a straight angle is equal to
$\pi$,  and the three internal angles
$\alpha$, $\beta$, $\gamma$ are related to our
$A$, $B$, $C$ as   $\alpha=\pi-A$, $\beta=B$, $\gamma=C$, so
that the triangle angular excess $\Delta$, usually defined as
$\alpha + \beta + \gamma - \pi$ appears here as
$-A+B+C$, and thus it does not involve $\pi$.  The quantity we
have denoted (see (\ref{eq:OrthoTrig:DualCosine:ii}))
\be 
P-A=\frac{\Delta}2=\frac{-A+B+C}{2}=\frac{\alpha + \beta +
\gamma}{2}-\frac{\pi}{2}
\ee
differs by $\pi/2$ from the half sum of internal angles, so our
sines and cosines of $P-A$ appear as cosines and sines of the
half  sum of internal angles. These elementary facts account for
all apparent discrepancies between the particularization of the
general equations given in this paper and the ones found in the
literature for the three constant curvature riemannian spaces
(for instance, \cite{Berger, Ratcliffe, RamsayRichtmyer}).
The use of the three {\em inner} angles makes
$\pi$ to enter unavoidably in the angle sum, the definition of
angular excess, the Gauss--Bonnet theorem for triangles, etc.,
and thus seems to preclude analogous relations in the cases with
a locally lorentzian metric, where
$\pi$ does not properly belong anymore. Incidentaly, in the three
riemannian geometries {\em any} angle can be taken as external,
but there is no longer such a freedom in the cases $\k_2\leq 0$,
where the choice of the external  angle $A$ in any first-kind
loop is dictated by the geometry itself, and is always opposite
to the largest side $a$.

Thus while the choice for angles in this paper can be
consistently done in all the nine CK spaces and affords the
equations in a natural form, it also introduces some minus signs
which are absent in the trigonometry of the three riemannian
spaces in their standard formulation with internal angles. 
Without any further reason  (as the inclusion of the {\em nine}
geometries provides) it would seem unwise to modify the
time-honoured form these equations have in the three familiar
constant curvature spaces; however this modification is an {\it
essential} step to make the equations meaningful for the nine
cases. This way the approach is more general than {\em absolute
trigonometry}, because it applies to the whole set of nine CK
spaces, and the explicit presence of $\k_2$ allows the
consideration of the scheme within a complete {\em duality},
which otherwise would be hidden, showing up only in  the
spherical case. For instance, the dual of the hyperbolic
geometry is the anti-de Sitter one, whose natural  metric is
lorentzian; simultaneous consideration of riemannian and
pseudoriemannian cases is therefore {\em essential} to fully
display duality.

When put in the right perspective, the possibility of
formulating  trigonometry for the {\em nine} CK spaces in a
single unified way where {\em all} equations are analogous and
directly meaningful in all cases is almost a triviality. But to
unfold this  view we have to abandon, by making it first
explicit, the implicit restriction
$\k_2=1$ which amounts to measuring angles in radians
and is universally enforced for the three riemannian cases. While
the curvature $\k_1$ allows to explicitly distinguishing between the
sphere,  euclidean and hyperbolic plane in the absolute
form of the trigonometric equations, the `dual' constant
$\k_2$ is restricted to a single particular (positive) value, so
it cannot be given the attention it deserves when it is allowed to
take on any real value. And further, the usual
definition of two-point homogeneity
\cite{Wang} excludes the degenerate riemannian and
pseudoriemannian spaces (where the isotropy subgroup is not
compact, and its action on the `unit sphere' is not transitive).
For instance the paper by Hsiang
\cite{Hsiang} is explicitly addressed to the study of
trigonometry of rank-one spaces, and covers in a single run the
three riemannian constant curvature spaces; however no reference
to the pseudoriemannian or degenerate riemannian planes is ever
made, nor it is  felt as missing in that paper, because the
restriction to two-point homogeneous spaces precludes this
consideration from the beginning. In some cases even a slightly
non-standard definition of  spaces of constant curvature is made
\cite{AlekVinSol} so as to embrace only the spaces whose
isotropy subgroup is compact (i.e, $\k_2>0$).

An interesting relation between pairs of  homogeneous symmetric
spaces is {\em Cartan duality}; it relates spherical and
hyperbolic geometry and is ultimately responsible for the fact
that hyperbolic geometry holds on the sphere of `imaginary'
radius; this was historically recognized by Lambert before the
Lobachewski hyperbolic geometry was satisfactorily settled, and
was considered by Lobachewski himself as an  unconclusive
evidence supporting the idea that his geometry was free of
contradictions. Each Cartan duality is related to an involutive
automorphism of the Lie algebra giving
rise to a non-compact Lie algebra if applied to a compact one.
Our approach is based on {\em two} commuting involutive
automorphisms, so it contains {\it two} Cartan dualities; each
is implemented by the change of sign in {\em one} of the two
constants $\k_1, \k_2$. In this sense, the hyperbolic space is
Cartan dual to the sphere, and also Cartan dual to the de Sitter
sphere.


\subsection{A compact notation}
\label{Sec:iv:iv}

In order not to burden the discussion with permanent reference
to the sign differences related to the side $a$ and angle $A$
(which are unavoidable if one wants a formulation valid in all
nine cases), the introduction of an auxiliary notation turns out
to be extremely convenient. We will denote the three sides as
$x_i$, $i=1, 2, 3$ and the three angles as $X_I$, $I=1, 2, 3$
according to
\be
x_1=-a  \qquad x_2=b  \qquad x_3=c  \qquad X_1=-A  \qquad X_2=B \qquad
X_3=C
\ee
(note the built-in minus sign in $x_1$ and $X_1$, which is
natural when the triangle is considered as a point loop with the
side $a$ traversed backwards, or as a side loop with the angle
$A$ rotated backwards). With this notation, the basic equation
(\ref{eq:OrthoTrig:BasicIdentTheora}) based in the vertex 2 is
\be
\myexp{x_1 P} \myexp{X_3 J} \myexp{x_2 P} \myexp{X_1 J}  \myexp{x_3 P}
\myexp{X_2 J}  = 1
\ee
and their alternative equivalent versions based in the vertex
$j$ can be written as:
\be
\myexp{x_i P} \myexp{X_K J} \myexp{x_j P} \myexp{X_I J} \myexp{x_k P}
\myexp{X_J J}  = 1
\ee
for {\em any} cyclic permutation $i=I,j=J,k=K$ of the  three
indices
$123$. In what follows  we will always adopt this convention
which makes all equations of trigonometry explicitly invariant
under any cyclic permutation of the `oriented' sides $x_i$ and
angles $X_I$; the use of capital indices will also help to 
distinguish between lengths and angles.

The triangular loop  lateral excess $\delta$
(\ref{eq:OrthoTrig:LateralExcess}) and angular excess $\Delta$
(\ref{eq:OrthoTrig:AngularExcess}) appear in the  present notation as
the {\em symmetric} sums of the three `oriented' sides or angles as:
\be
\delta =x_1+x_2+x_3 = -a + b + c  \qquad
\Delta = X_1+X_2+X_3 = -A + B + C .
\ee
It will be also convenient to replace the excesses by the
quantities
\be
\exc:= \delta/2 \qquad    \EXC:= \Delta/2
\ee
and to introduce three other quantities as well as their three
duals:
\be
\exc_i:= x_i -\exc  \qquad  \EXC_I:= X_I -\EXC  .
\ee
They verify the following equations
\be
\begin{array}{llll}
\exc_i- \exc_j=x_i-x_j&\quad
\exc_i+ \exc_j=-x_k&\quad
\exc + \exc_i=x_i &\quad
\exc -  \exc_i= x_j+x_k\cr
\EXC_I- \EXC_J=X_I-X_J&\quad
\EXC_I+ \EXC_J=-X_K&\quad
\EXC + \EXC_I=X_I &\quad
\EXC -  \EXC_I= X_J+X_K  \cr
\end{array}
\label{eq:OrthoTrig:RelsExcessSide}
\ee
and are related with the half sums $p$   and $P$
(\ref{eq:OrthoTrig:Semisums}) by:
\be
\begin{array}{llll}
\exc = p-a    &\quad \exc_1= -p &\quad \exc_2=  p-c &\quad
\exc_3=  p-b  \cr
  \EXC = P-A
 &\quad  \EXC_1= -P   &\quad
  \EXC_2=  P-C   &\quad
  \EXC_3=  P-B .
\label{eq:OrthoTrig:RelSemisumsExcesses}
\end{array}
\ee
Note that $\exc_1$ is different from zero and {\em negative},
while $\exc_2, \exc_3$ are (generically) different from zero and
{\em positive}, just like the three sides $x_1$ and $x_2, x_3$.
The same holds for the quantities related to angular excesses:
$\EXC_1$ is different from zero and {\em negative}, while
$\EXC_2, \EXC_3$ are (generically) different from zero and {\em
positive}. If $\k_2=0$ then the three $e_i$ reduce to the sides
$x_i$ ($\delta=\exc=0$). Dually, if
$\k_1=0$ then the three $E_I$ reduce to the angles $X_I$
($\Delta=\EXC=0$).

In terms of these quantities, the  equations of trigonometry we
have obtained, including the alternative forms for the cosine
theorems (\ref{eq:OrthoTrig:Cosine:ii})  and
(\ref{eq:OrthoTrig:DualCosine:ii}), turn out to be
\be
\begin{array}{ll}
\mbox{1i}&  \c(x_i) =\c(x_j)\c(x_k)-\k_1 \s(x_j)\s(x_k)\cc(X_I)
\cr
\mbox{1I}& \cc(X_I) =\cc(X_J)\cc(X_K)-\k_2
\ss(X_J)\ss(X_K)\c(x_i)
\cr
\mbox{1''i}&  2 \s(\exc) \s(\exc_i) = - \k_2 \s(x_j) \s(x_k)
\vv(X_I) \cr
\mbox{1''I}&  2 \ss(\EXC)\ss(\EXC_I) = -\k_1 \ss(X_J)
\ss(X_K) \v(x_i) \cr
\mbox{2}&
\displaystyle{ \frac{\s(x_i)}{\ss(X_I)} = \frac{\s(x_j)}{\ss(X_J)}
=\frac{\s(x_k)}{\ss(X_K)} }\cr
\mbox{3iJ}&  \s(x_i) \cc(X_J) =
-\c(x_j)\s(x_k)- \s(x_j)\c(x_k)\cc(X_I) \cr
\mbox{3Ij}&  \ss(X_I)
\c(x_j)=-\cc(X_J)\ss(X_K) -\ss(X_J)\cc(X_K)\c(x_i) \cr
\mbox{4IJ}\equiv\mbox{4ij}&
\k_2\ss(X_I) \ss(X_J)-\cc(X_I)\cc(X_J)\c(x_k) \cr
&\qquad\qquad\quad =
\k_1\s(x_i)\s(x_j)-\c(x_i)\c(x_j)\cc(X_K)
\end{array}
\label{eq:OrthoTrig:BasicTrigEqsCompact}
\ee
and we now have a completely uniform pattern, without casual
signs. When $\k_2=0$, the three equations 1''i clearly imply
$\s(\exc)=0$, as already commented. This means that when the
constant $\k_2\to 0$, the sine of the {\em lateral  excess} of
the triangle also goes to zero, but the quotient $\s(\exc)/\k_2$
remains finite, and is given by:
\be
\frac{\s(\exc)}{\k_2}   =
- \frac{\s(x_j) \s(x_k) \vv(X_I)}{2 \s(\exc_i)} .
\label{OrthoTrig:sincoarea}
\ee
Dually, when $\k_1=0$, the  equations 1''I lead to
$\ss(\EXC)=0$; the sine of the {\em angular  excess} of the
triangle goes to zero when $\k_1 \to 0$, but the quotient
$\ss(\EXC)/\k_1$ remains finite:
\be
\frac{\ss(\EXC)}{\k_1}
 = - \frac{\ss(X_J) \ss(X_K) \v(x_i)}{2 \ss(\EXC_I)} .
\label{OrthoTrig:sinarea}
\ee


\subsection{Area and coarea and the dualities length/area
and angle/coarea}
\label{Sec:iv:v}

The previous expressions  have shown the natural appearance in
this group theoretical approach of the combinations 
${\s(\exc)}/{\k_2}$ and ${\ss(\EXC)}/{\k_1}$  which  are well
defined in all the nine cases, no matter of the values of $\k_1$
and $\k_2$. This is  because $\s(\exc)$ goes to zero linearly
with the `line curvature' $\k_2$, and therefore has a kind of
{\em residue} with a well defined value even in the (limiting)
case $\k_2=0$. Dually, the same phenomenon happens for
$\ss(\EXC)$, which vanishes when $\k_1=0$.

By construction, the angular excess of a triangle loop is
additive under decomposition of a triangle loop into two.  Thus
it is obvious that ${\ss(\EXC)}/{\k_1}$ is related to the
triangular loop {\em area}. This relation is very well known in
the two riemannian spherical and hyperbolic geometries, where
the standard expression for the absolute value $\area$ of the
area enclosed by the triangle loop is  easy to derive from the
Gauss--Bonnet theorem and is related to the angular excess by 
$\k_1 \area = \Delta$. This suggests a purely {\em group
theoretical}   definition of area (and its dual quantity, coarea
$\coarea$) for triangle loops, which extends the earlier group
theoretical definitions of length and angle. This group
theoretical definition holds for all the nine 2D CK spaces, no
matter of the values of $\k_1$ or $\k_2$, hence applying also to
the pseudoriemannian and degenerate riemannian CK spaces. This
should  be in full agreement with the standard definition for
area based on differential geometry.
All these requirements are satisfied by the following {\em
definitions} of  area and coarea for a triangle:
\be
\area := \frac{\Delta}{\k_1}
\qquad \coarea:= \frac{\delta}{\k_2} .
\label{zzz:arearelation}
\ee

All appearances of ${\ss(\EXC)}/{\k_1}$ in the equations of
trigonometry could be rewritten in terms of trigonometric
functions of the area of the loop, and dually for
${\s(\exc)}/{\k_2}$ and coarea. In this rewriting, the label
naturally associated to the area is $\k_1^2 \k_2$, while the
coarea label is $\k_1 \k_2^2$; this makes sense as area should
be to the product $P_1 P_2$ (resp. coarea to $J_{12} P_2$)
what length is to $P_1$ with label $\k_1$ and angle to $J_{12}$
with label $\k_2$. For  the two basic sine and cosine functions
of area and coarea we have:
\bea
&&\carea(\area):= \cc(\Delta) = \cc(2\EXC)  \qquad
\sarea(\area):= \frac{\ss(\Delta)}{\k_1} = \frac{\ss(2\EXC)}{\k_1} \cr
&&\ccoarea(\coarea):= \c(\delta) = \c(2\exc)  \qquad
\scoarea(\coarea):= \frac{\s(\delta)}{\k_2} = \frac{\s(2\exc)}{\k_2}.
\label{eq:OrthoTrig:SinArea}
\eea
Therefore the two quotients (\ref{OrthoTrig:sinarea}) and
(\ref{OrthoTrig:sincoarea}) are the sine of half the area and half
the coarea of the triangle loop, each with its canonical label:
\be
\frac{\ss(\EXC)}{\k_1} = \sarea(\area/2)  \qquad
\frac{\s(\exc)}{\k_2} = \scoarea(\coarea/2) .
\label{OrthoTrig:ExcessToArea}
\ee

A length/area and angle/coarea dualities for the sphere have
been recently discussed by  Arnol'd \cite{ArnGeomSphCur} in
a paper devoted to the geometry of spherical curves. These
`dualities' are indeed a {\em general} property for all the nine
CK geometries, and follow directly from the fundamental
self-duality of the whole scheme of CK spaces (between lengths
and angles), together with the `transference' from angles
$\EXC$ (or $\Delta$) to areas $\area/2$ (or from lengths
$\exc$ (or $\delta$) to coareas
$\coarea/2$) implicitly contained in the equations
(\ref{OrthoTrig:ExcessToArea}).
However, while these dualities are present in all CK geometries,
they are only clearly visible for the sphere, where by a
suitable choice of length and angle units the two constants
$\k_1$ and $\k_2$ can be reduced to $1$. In this spherical case
the labels of either length, angle, area or coarea are all
equal, so the transference from angle to area (or length to
coarea) amounts to a simple equality between numerical values,
therefore displaying in a single geometry the full richness of
the CK scheme. In this case all trigonometrical quantities
(either lengths, angles, areas or coareas) appear in all
trigonometric formulas as arguments of {\em circular}
trigonometric functions.

In other CK geometries where some of the constants are negative,
unveiling these dualities requires explicit use of a transference
similar to those in (\ref{OrthoTrig:ExcessToArea}). The leading idea
in Arnol'd paper is to link to each spherical curve its `dual' and
its `derivative' curves determined from the given curve by moving a
quadrant along the normal or the tangent to the curve. This triple
has very interesting properties, giving rise to a kind of {\em
triality}, and it is within its study that the length/area and
angle/coarea dualities appear. This spherical triality must have an
interesting `hyperbolic' version where the three related curves
---the given one, its dual and its derivative--- lives in a
different space, say the hyperbolic plane, the co-hyperbolic or
anti-de Sitter sphere and the douby hyperbolic or de Sitter sphere.
This triality seems worth studying, mainly in view of the strong
current interest in the anti-de Sitter spacetime and should be
related to the conformal field theory on the sphere at infinity in
hyperbolical geometry.


\subsection{The trigonometric equations in the 
{\em minimal}  form}
\label{Sec:iv:vi}

By adding the two new triangular loop quantities, area $\area$
and coarea
$\coarea$, to sides and angles, all the  basic equations can be
written in a {\em minimal} form, with {\em no explicit}
constants $\k_1, \k_2$. These equations (as well as  {\em all}
equations derived from them), are directly meaningful for all
the nine cases and do not reduce to trivial identities.
Furthermore the versed sine of the angle $X_I$ in
(\ref{OrthoTrig:sincoarea}) and side $x_i$ in (\ref{OrthoTrig:sinarea})
can be expressed in terms of the sine of half the angle or the
side using the relation (\ref{eq:CKSinHalf}). Thus the equations read
\be
\begin{array}{ll}
\mbox{1''i}&\quad \scoarea(\coarea/2) \s(\exc_i) = -  \s(x_j)
\s(x_k)
\ss^2(X_I/2) \cr
\mbox{1''I} &\quad \sarea(\area/2)\ss(\EXC_I) = - \ss(X_J)
\ss(X_K) \s^2(x_i/2) \cr
2&\quad
\displaystyle {\frac{\s(x_i)}{\ss(X_I)}
= \frac{\s(x_j)}{\ss(X_J)} =\frac{\s(x_k)}{\ss(X_K)}} .
\end{array}
\label{OrthoTrig:minimal}
\ee

Two things are worth remarking. First, the corresponding
equations for any two particular spaces only differ by the {\it
implicit} appearances of the constants (curvatures)
$\k_1, \k_2$ (and also $\k_1^2\k_2, \k_1\k_2^2$) as labels of the
trigonometric functions of sides, angles (and also area, coarea)
respectively. This is reminiscent to the minimal coupling idea
in  general relativity:  no explicitly dependent {\em curvature}
terms should be introduced in the basic free equations when
formulating the corresponding equation for a curved spacetime,
but only those introduced (implicitly) through the conexion. In
(\ref{OrthoTrig:minimal}), any trigonometric equation in a space with
curvatures $\k_1,
\k_2$ different from zero can be obtained from the corresponding
flat one by introducing the corresponding implicit label in the
trigonometric functions of sides, angles, area and coarea, but
without any explicitly dependent $\k_1, \k_2$ term. Secondly,
the only trigonometric function involved in these equations is
the sine, which reduces to the variable itself when the label
equals to zero. Therefore the implementation of this kind of
`minimal coupling' to obtain the general  equations
(\ref{OrthoTrig:minimal}) consists simply  in replacing each term
entering the `purely flat' $\k_1=0,
\k_2=0$ equations
\be
\begin{array}{lll}
\mbox{1''i}&\quad (\coarea/2) \exc_i = - x_j x_k (X_I/2)^2 & \cr
\mbox{1''I} &\quad (\area/2) \EXC_I = -  X_J X_K (x_i/2)^2 & \cr
2&\quad
\displaystyle {\frac{x_i}{X_I} = \frac{x_j}{X_J} =\frac{x_k}{X_K}}
& \end{array}
\label{OrthoTrig:minimal:fiducial}
\ee
by their sines, each with the corresponding label.


\section{A trigonometric bestiarium}
\label{Sec:v}

Starting from the set of basic equations (\ref{OrthoTrig:minimal}),
we can easily derive a complete trigonometric {\em bestiarium}.
All equations are written in a way which is simultaneously
meaningful for {\em all} the nine geometries, and by allowing
area and coarea to enter the basic equations no  {\em explicit}
constants $\k_1$, $\k_2$ ever appear. Should area and coarea be
avoided in favour of $\EXC$ or $\exc$, then {\em all} explicit
appearances of  $\k_1$, $\k_2$ can be  reduced (see
(\ref{OrthoTrig:ExcessToArea})) to the combinations
${\s(\exc)}/{\k_2}$ or ${\ss(\EXC)}/{\k_1}$, where the joint
appearance of both curvatures $\k_1$ and $\k_2$ (one as a label
and the other explicitly) reminds that the length $\exc$ is not
directly a distance between two points or the angle $\EXC$ is
not an angle between two lines, but both come from the triangle
as a whole.

The minus signs explicitly appearing in (\ref{OrthoTrig:minimal}),
and in most of the ensuing equations are artifacts following our
definitions on $x_i, e_i$ and $X_I, E_I$. These signs will
disappear when the equations are rewritten in terms of the
positive quantities
$a, b, c, p, \coarea$ and $A, B, C, P, \area$, but then the
casual signs related to the external angle will spoil the
uniform appearance of the equations.


\subsection{Equations of  Euler, Gauss--Delambre--Mollweide and
  Napier}
\label{Sec:v:i}

Equation  1''i of (\ref{OrthoTrig:minimal}) gives directly
\be
\ss^2\left(\frac{X_I}{2}\right) = -\frac{  \scoarea( {\coarea}/2)
\s(\exc_i)}{\s(x_j)
\s(x_k)} .
\label{OrthoTrig:EulerSinI}
\ee
Consequently, the corresponding relation for the cosine is
derived from it by applying (\ref{eq:CKCosSinIdentity})  and
(\ref{zzz:wp}):
\be
 \cc^2\left(\frac{X_I}{2}\right) =
\frac{\s(\exc_j) \s(\exc_k)}{\s(x_j) \s(x_k)}
\label{OrthoTrig:EulerCosI}
\ee
and the quotient of the above relations gives
\be
\TT^2\left(\frac{X_I}{2}\right) = -\frac{  \scoarea( {\coarea}/2)
\s(\exc_i)}{\s(\exc_j)
\s(\exc_k)} .
\label{OrthoTrig:EulerTanI}
\ee
In the same way their duals can be deduced:
\be
\begin{array}{l}
\displaystyle \s^2\left(\frac{x_i}{2}\right) =  -\frac{
\sarea({\area}/2)
\ss(\EXC_I)}{\ss(X_J)
\ss(X_K)}  \qquad
\displaystyle \c^2\left(\frac{x_i}{2}\right) = \frac{\ss(\EXC_J)
\ss(\EXC_K)}{\ss(X_J) \ss(X_K)}  \cr
\displaystyle \T^2\left(\frac{x_i}{2}\right) = - \frac{
\sarea({\area}/2)
\ss(\EXC_I)}{\ss(\EXC_J)   \ss(\EXC_K)} .
\end{array}
\label{OrthoTrig:EulerSinCosTanSides}
\ee
These relations were obtained by Euler for the sphere. From them
and by using (\ref{eq:CKSinTwice}), expressions for the sine of the  
angles  (and not their half) in terms of the sides can be
deduced,  as well as for the sides  in terms of the angles (dual
of the previous ones):
\bea
&&\ss(X_I) = -\frac{2}{\s(x_j) \s(x_k)} \left\{
-\scoarea({\coarea}/2)
\s(\exc_i)\s(\exc_j) \s(\exc_k) \right\} ^{1/2}\cr
&&\s(x_i) = -\frac{2}{\ss(X_J) \ss(X_K)} \left\{- \sarea({\area}/2)
\ss(\EXC_I)\ss(\EXC_J) \ss(\EXC_K) \right\} ^{1/2} \!\! .
\label{OrthoTrig:DoubleEuler}
\eea
By dividing both expressions and taking into account the sine
theorem we find
\be
\frac{\s(x_i)}{\ss(X_I)} = \frac{\s(x_j)}{\ss(X_J)}
 =\frac{\s(x_k)}{\ss(X_K)} =
\frac{ \left\{ -\scoarea({\coarea}/2) \s(\exc_i)\s(\exc_j)
\s(\exc_k)\right\}^{1/2} }
{  \left\{ -\sarea( {\area}/2) \ss(\EXC_I)\ss(\EXC_J) \ss(\EXC_K)
\right\}^{1/2} }.
\ee

On the other hand, by combining the formulas
(\ref{OrthoTrig:EulerSinI})--(\ref{OrthoTrig:EulerTanI})  in different
ways we obtain
\bea
&& \cc\left(\half{X_I}\right) \cc\left(\half{X_J}\right)
\cc\left(\half{X_K}\right) =
\frac{\s(\exc_i)\s(\exc_j) \s(\exc_k)}{\s(x_i) \s(x_j) \s(x_k)}\cr
&&\ss\left(\half{X_I}\right) \ss\left(\half{X_J}\right)
\cc\left(\half{X_K}\right) =
-\frac{\s(\exc_i)\s(\exc_j)\scoarea({\coarea}/2)}{\s(x_i)
\s(x_j)
\s(x_k)}\cr
&& \TT\left(\half{X_I}\right)\TT\left(\half{X_J}\right) =
-\frac{\scoarea({\coarea}/2)}{\s(\exc_k)}
\qquad\qquad
\frac{\TT\left(\half{X_I}\right)}{\TT\left(\half{X_J}\right)} =
\frac{\s(\exc_i)}{\s(\exc_j)}
\label{OrthoTrig:PreDelambre}
\eea
and in the same way their duals.  Starting from these equations
we can deduce the general forms, valid for all the nine
geometries, of the Gauss--Delambre--Mollweide  analogies. For
instance, let us consider the identity (cf.\ (\ref{eq:CKSinSumDif}))
\be
\ss\left(\half{X_I + X_J}\right)=
\ss\left(\half{X_I}\right)
\cc\left(\half{X_J}\right)+
\ss\left(\half{X_J}\right)
\cc\left(\half{X_I}\right) .
\ee
We apply the second relation of (\ref{OrthoTrig:PreDelambre}) and group
common terms
\be
\ss\left(\half{X_I + X_J}\right)=
-\frac{  \scoarea( {\coarea}/2) \s(\exc_k)}{\s(x_i) \s(x_j)}\
\frac{\left(\s(\exc_i) + 
\s(\exc_j)\right)}{\s(x_k)\ss\left(\half{X_K}\right)  }
\ee
and by introducing the Euler equation  (\ref{OrthoTrig:EulerSinI}) and
using (\ref{eq:CKSinTwice}), (\ref{eq:CKSinSumDifToProd}) and
(\ref{eq:OrthoTrig:RelsExcessSide}) we find:
\be
\ss\left(\half{X_I + X_J}\right)=
-\frac{\ss\left(\half{X_K}\right)}{\c\left(\half{x_k}\right)}
\c\left(\half{x_i-x_j}\right) .
\ee
Likewise, we can obtain the full set of
Gauss--Delambre--Mollweide analogies:
\bea
&& \frac{\ss\left(\half{X_I + X_J}\right)} {\ss\left(\half{X_K}\right)}
= - \frac{ \c\left(\half{x_i - x_j}\right)} {
\c\left(\half{x_k}\right)}  \qquad\qquad
\frac{\cc\left(\half{X_I + X_J}\right)} {\cc\left(\half{X_K}\right)}
= \frac{ \c\left(\half{x_i + x_j}\right)} {
\c\left(\half{x_k}\right)} \cr
&&\frac{\ss\left(\half{X_I - X_J}\right)} {\ss\left(\half{X_K}\right)}
=  \frac{ \s\left(\half{x_i - x_j}\right)} {
\s\left(\half{x_k}\right)}  \qquad\qquad
\frac{\cc\left(\half{X_I - X_J}\right)} {\cc\left(\half{X_K}\right)}
= -  \frac{ \s\left(\half{x_i + x_j}\right)} {
\s\left(\half{x_k}\right)} .
\label{eq:OrthoTrig:CKDelambre}
\eea
By quotient of these expressions we obtain the Napier  analogies:
\bea
&& \frac{\TT\left(\half{X_I + X_J}\right)} {\TT\left(\half{X_K}\right)}
= - \frac{ \c\left(\half{x_i - x_j}\right)} {
\c\left(\half{x_i + x_j}\right)}  \qquad\qquad
\frac{\TT\left(\half{X_I - X_J}\right)} {\TT\left(\half{X_K}\right)}
=- \frac{ \s\left(\half{x_i - x_j}\right)} {
\s\left(\half{x_i + x_j}\right)} \cr
&&\frac{\T\left(\half{x_i+x_j}\right)} {\T\left(\half{x_k}\right)}
=  -\frac{ \cc\left(\half{X_I-X_J}\right)} {
\cc\left(\half{X_I+X_J}\right)}  \qquad\qquad
\frac{\T\left(\half{x_i-x_j}\right)} {\T\left(\half{x_k}\right)}
= -  \frac{ \ss\left(\half{X_I-X_J}\right)} {
\ss\left(\half{X_I+X_J}\right)}  
\label{eq:OrthoTrig:CKNeper}
\eea
and by further division, we obtain the self-dual equations:
\be
\frac{\TT\left(\half{X_I + X_J}\right)} {\TT\left(\half{X_I -
X_J}\right)} =   \frac{ \T\left(\half{x_i + x_j}\right)} {
\T\left(\half{x_i - x_j}\right)}  .
\ee

These relations cover the general versions of practically every
equation found in  textbooks on spherical or hyperbolic
trigonometry.  Others, which we refrain from writing, are very
easily obtained from them.


\subsection{Equations  for area and coarea}
\label{Sec:v:ii}

There is another interesting and not so widely known group of
equations which can be formulated involving only the tangents of
half the quantities $\exc_i$, $\EXC_I$ and $\exc$, $\EXC$. The
excesses $\delta=2\exc$, $\Delta=2\EXC$ always appear in the
equations to be given below under the apparently indeterminate
but finite forms ${{\T(\half{\exc})}}/{\k_2}$ and
${{\TT(\half{\EXC})}}/{\k_1}$.  By taking into account the
relations (\ref{eq:OrthoTrig:SinArea}), these quotients   can be
expressed in terms of area and coarea as
\be
\tarea\left(\frac{\area}4\right)=\frac{{\TT(\frac{\EXC}{2})}}{\k_1}
\qquad\qquad
\tcoarea\left(\frac{\coarea}4\right)
=\frac{{\T(\frac{\exc}{2})}}{\k_2} .
\label{zzz:besta}
\ee

We consider now the first group of the
Gauss--Delambre--Mollweide  analogies  (\ref{eq:OrthoTrig:CKDelambre})
which by means of (\ref{eq:OrthoTrig:RelsExcessSide}) can be written as
\be
\frac{\ss\left(\half{\EXC -\EXC_K}\right)}
{\ss\left(\half{\EXC +\EXC_K}\right)} =
- \frac{ \c\left(\half{\exc_i - \exc_j}\right)}
{ \c\left(\half{\exc_i + \exc_j}\right)}
\ee
which is equivalent to
\be
\frac{\ss\left(\half{\EXC -\EXC_K}\right)-\ss\left(\half{\EXC
+\EXC_K}\right)}
{\ss\left(\half{\EXC -\EXC_K}\right)+\ss\left(\half{\EXC
+\EXC_K}\right)} =
\frac{- \c\left(\half{\exc_i - \exc_j}\right)-
\c\left(\half{\exc_i + \exc_j}\right)}
{- \c\left(\half{\exc_i - \exc_j}\right)+
\c\left(\half{\exc_i + \exc_j}\right)} .
\ee
We apply the relations 
(\ref{eq:CKCosSumToProd})--(\ref{eq:CKSinSumDifToProd}) and introduce
tangents; thus we find  a new equation:
\be
\frac{\TT\left(\frac{\EXC_K}{2}\right)}
{\TT\left(\frac{\EXC}{2}\right)}=-\frac{1}
{\k_1\T\left(\half{\exc_i}\right) \T\left(\half{\exc_j}\right)} .
\ee
The remaining analogies allow similar derivations. The final
set of equations coming from (\ref{eq:OrthoTrig:CKDelambre}) is given by
\bea
&&\frac{{\TT(\half{\EXC})}}{\k_1}   =
- {\T\left(\half{\exc_i}\right)}{\T\left(\half{\exc_j}\right)}
{\TT\left(\frac{\EXC_K}{2}\right)}
\qquad\quad
{\frac{\TT(\frac{\EXC}{2})}{\k_1}}\,{\TT\left(\frac{\EXC_I}{2}\right)}=
{\frac{\T(\frac{\exc}{2})}{\k_2}}\,{\T\left(\frac{\exc_i}{2}\right)}
\cr
&&\frac{\T(\frac{\exc_i}{2})}{\TT(\frac{\EXC_I}{2})} =
\frac{\T(\frac{\exc_j}{2})}{\TT(\frac{\EXC_J}{2})}
\qquad\qquad \frac{{\T(\half{\exc})}}{\k_2} =
- {\TT\left(\half{\EXC_I}\right)}{\TT\left(\half{\EXC_J}\right)}
{\T\left(\frac{\exc_k}{2}\right)} .
  \label{zzz:bestb}
\eea
These expressions lead to a self-dual  equation
reminiscent  to the sine theorem:
\be
\frac{\T(\frac{\exc_i}{2})}{\TT(\frac{\EXC_I}{2})} =
\frac{\T(\frac{\exc_j}{2})}{\TT(\frac{\EXC_J}{2})} =
\frac{\T(\frac{\exc_k}{2})}{\TT(\frac{\EXC_K}{2})} =
\frac{\frac{\TT(\frac{\EXC}{2})}{\k_1}}
     {\frac{\T(\frac{\exc}{2})}{\k_2}} =
\frac{ \tarea(\area/4)  } {\tcoarea(\coarea/4)  } .
\label{zzz:bestc}
\ee
Furthermore, the above relations  give rise to other
expressions similar to Euler's equations:
\be
\TT^2\left(\frac{\EXC_I}{2}\right) = - \frac{ \tcoarea(\coarea/4)
\T(\frac{\exc_i}{2})}
 {\T(\frac{\exc_j}{2}) \T(\frac{\exc_k}{2}) }  \qquad\qquad
\T^2\left(\frac{\exc_i}{2}\right) = - \frac{ \tarea(\area/4)
\TT(\frac{\EXC_I}{2})}
 {\TT(\frac{\EXC_J}{2}) \TT(\frac{\EXC_K}{2}) }
\label{zzz:bestd}
\ee
and a  pair of mutually dual equations:
\bea
&&\tarea^2\left(\frac{\area}4\right) =-
\frac{\T(\frac{\exc}{2})}{\k_2}
\T\left(\frac{\exc_i}{2}\right) \T\left(\frac{\exc_j}{2}\right)
\T\left(\frac{\exc_k}{2}\right) \cr
&&\tcoarea^2\left(\frac{\coarea}4\right)  =-
\frac{\TT(\frac{\EXC}{2})}{\k_1}
\TT\left(\frac{\EXC_I}{2}\right) \TT\left(\frac{\EXC_J}{2}\right)
\TT\left(\frac{\EXC_K}{2}\right) .
\label{zzz:beste}
\eea

A large number of expressions for area and coarea of  a triangle
loop in terms of sides and angles can be easily obtained by
simple analitical transformations involving the identities 
given in the Appendix. In what follows we deduce only some
equations for area (or angular excess (\ref{eq:OrthoTrig:SinArea})), but
clearly any expression can be dualized and remains true, so the
reader can easily write down the dual formulas for the coarea of
a triangle loop.

We first give  formulas for trigonometric functions of half
the area:
\bea
&&\carea\left(\frac{\area}{2}\right) =
\frac{1 +  \c(x_i)+  \c(x_j)+  \c(x_k)}{ 4
\c(\half{x_i}) \c(\half{x_j}) \c(\half{x_k}) } \label{zzz:bestgg}\\
&&\sarea\left(\half{\area}\right) =  -
\frac{\ss(X_I)\s(x_j)\s(x_k)}{4 \c(\half{x_i}) \c(\half{x_j})
\c(\half{x_k})}
= - \frac{ \ss(X_I) \s(\frac{x_j}{2}) \s(\frac{x_k}{2})}{\c(\half{x_i})
} \cr
&&\qquad\qquad\qquad=
\frac{ \left\{- \frac{\s({\exc})}{\k_2} \s(\exc_i)\s(\exc_j)
\s(\exc_k)
\right\} ^{1/2}}{ 2\c(\half{x_i})\c(\half{x_j})\c(\half{x_k}) }
\label{zzz:bestg}\\
&&\tarea\left(\half{\area}\right) =
- \frac{ \ss(X_I)\s(x_j)\s(x_k) }{ 1 +  \c(x_i)+  \c(x_j)+  \c(x_k) } \cr
&&\qquad\qquad\qquad =2
\frac{ \left\{- \frac{\s({\exc})}{\k_2} \s(\exc_i)\s(\exc_j)
\s(\exc_k)
\right\} ^{1/2} }{  1 +  \c(x_i)+  \c(x_j)+  \c(x_k)  } .
\label{zzz:besti}
\eea

In order to prove these equations we write
\bea
&&\sarea\left(\half{\area}\right) =
\frac{\ss(\EXC)}{\k_1}=\frac 1{\k_1} \ss\left(\frac{X_I}2+
\frac{X_J+X_K}2\right)\cr
&& =\frac 1{\k_1}\left\{
\ss\left(\frac{X_I}2 \right)\cc\left(\frac{X_J+X_K}2\right)
+\ss\left(\frac{X_J+X_K}2\right)\cc\left(\frac{X_I}2\right)\right\};
\eea
by introducing the analogies (\ref{eq:OrthoTrig:CKDelambre}) and grouping
common terms we find that
\be
=\frac{\ss\left(\frac{X_I}2 \right)\cc\left(\frac{X_I}2 \right)}
{\k_1\c\left(\frac{x_i}2 \right)}\left\{
\c\left(\frac{x_j+x_k}2\right)-
\c\left(\frac{x_j-x_k}2\right)\right\}
\ee
and by applying (\ref{eq:CKCosDifToProd})  and (\ref{eq:CKSinTwice}) we
prove the second equation of (\ref{zzz:bestg}):
\be
=- 2\frac{\ss\left(\frac{X_I}2 \right)\cc\left(\frac{X_I}2 \right)
\s\left(\frac{x_j}2 \right)    \s\left(\frac{x_k}2 \right) }
{\c\left(\frac{x_i}2 \right)}
=-  \frac{\ss\left({X_I}\right)
\s\left(\frac{x_j}2\right) \s\left(\frac{x_k}2 \right)}
{\c\left(\frac{x_i}2 \right)} .
\ee
The relation (\ref{eq:CKSinTwice}) allows us  to find the  first
equation of (\ref{zzz:bestg}), while (\ref{OrthoTrig:DoubleEuler}) leads
to the third one. The cosine equation (\ref{zzz:bestgg}) is obtained by
using (\ref{eq:CKCosSinIdentity}) and the first identity of
(\ref{eq:CKSinTriSide}) starting  with the third equation of
(\ref{zzz:bestg}). By taking quotients in (\ref{zzz:bestgg}) and
(\ref{zzz:bestg}) we find  (\ref{zzz:besti}). Furthermore,  by
introducing in the first expression of (\ref{zzz:besti}) the cosine
theorem for $\c(x_i)$, and applying the relations (\ref{eq:CKSinTwice})
and (\ref{eq:CKCosHalf}), we obtain:
\be
\tarea\left(\half{\area}\right) =
- \frac{ \ss(X_I) \T(\half{x_j}) \T(\half{x_k}) }{ 1 -\k_1 \cc(X_I)
\T(\half{x_j})
\T(\half{x_k})  } .
\ee

The  equation (\ref{zzz:bestgg}) gives rise to a relation for the
cosine of $\area$ by applying  (\ref{eq:CKCosHalf}),  while
(\ref{zzz:bestgg}) and the first equation of (\ref{zzz:bestg}) give rise
to a relation for the sine by means of  (\ref{eq:CKSinTwice}):
\bea
&&\carea(\area)=
\frac{ \big[ 1 +  \c(x_i)+  \c(x_j)+  \c(x_k)  \big]^2 -
8\c^2(\half{x_i}) \c^2(\half{x_j}) \c^2(\half{x_k})  }
{8\c^2(\half{x_i}) \c^2(\half{x_j}) \c^2(\half{x_k}) }\cr
&&\sarea(\area)= -
\frac{ \s(x_i) \s(x_j) \ss(X_K) \big[ 1 +  \c(x_i)+  \c(x_j)+  \c(x_k)
\big] }
{8\c^2(\half{x_i}) \c^2(\half{x_j}) \c^2(\half{x_k})} .
\label{zzz:bestf}
\eea
Finally,  the relation (\ref{zzz:bestgg})  leads to two functions of
$\area/4$ by considering (\ref{eq:CKCosHalf})  and (\ref{zzz:wwd}),
and   by making use of (\ref{eq:CKSinHalf}) and (\ref{zzz:wwc}). The
expressions so obtained are:
\bea
&&\carea^2\left(\frac{\area}{4}\right)  =
\frac{\c(\frac{\exc}{2})
\c(\half{\exc_i})\c(\half{\exc_j})\c(\half{\exc_k})}{
\c(\half{x_i})\c(\half{x_j})\c(\half{x_k}) } \cr
&&\sarea^2\left(\frac{\area}{4}\right) =
- \frac{ \frac{\s(\frac{\exc}{2})}{\k_2}
\s(\half{\exc_i})\s(\half{\exc_j})\s(\half{\exc_k})}{
\c(\half{x_i})\c(\half{x_j})\c(\half{x_k}) } .
\label{zzz:besth}
\eea
Notice that the quotient of these relations allows us to
recover   (\ref{zzz:beste}).


\subsection{Some historical comments}
\label{Sec:v:iii}

Trigonometry, motivated mainly in the spherical case by
astronomy, has a very interesting and well documented history. A
good authoritative reference for its historical  development is
the book by Rozenfel'd \cite{Ros}; Ratcliffe \cite{Ratcliffe}
contains also historical notes. A standard reference for
spherical trigonometry covering all the spherical versions of
the equations we have discussed here and much more ---which
suitably reformulated also extends without exception to the nine
geometries--- is the book by Todhunter--Leathem
\cite{Todhunter}. Hyperbolic trigonometry was first
satisfactorily settled by Lobachewski and Bolyai and is an
essential tool when studying hyperbolic manifolds; see e.g.
\cite{Berger}--\cite{AlekVinSol}.

The consideration of cosine and dual cosine in the  forms 1i,
1I as the basic equations of spherical trigonometric dates back
to Euler, but the `alternative' form 1'i is much older and
essentially is the one given by Regiomontanus, ellaborating on
earlier results. The spherical cosine equations
themselves are usually ascribed to Albategnius, and the first
explicit appearance of the dual cosine equations is ascribed to
Vieta, even though spherical polarity was clear to \alTusi in
the thirteenth century. The sine theorem, in another equivalent
form, appears in Menelaus.  The formulas
 (\ref{OrthoTrig:EulerSinI})--(\ref{OrthoTrig:EulerSinCosTanSides}) are
the general ($\k_1, \k_2$) versions of the spherical formulas due to
Euler, while (\ref{eq:OrthoTrig:CKDelambre}) are the general versions of
Gauss--Delambre--Mollweide analogies (in the old meaning of
proportion), and (\ref{eq:OrthoTrig:CKNeper})   are the general version
of the Napier analogies. The first formula for the sine in
(\ref{zzz:bestg}) is known in the spherical case as Cagnoli's
theorem, and the  expression for the area (or excess) of a
spherical triangle in terms of the sides (\ref{zzz:beste}) is due
to L'Huillier, extending the euclidean Heron--Archimedes area
formula. Most formulas for spherical triangle area in terms of
sides and/or angles were obtained by Euler. Other formulas for
the trigonometric functions of one half or one fourth the area
are also known in the spherical case, but bear no name; some are
due to L'Huillier and Serret. It is interesting to remark that
suitably reformulated {\em all} these formulas hold for the nine
CK plane geometries, and the full richness in the trigonometry of
each individual space shows up at full only when studying the 
whole set of the nine spaces.


\subsection{Existence conditions}
\label{Sec:v:iv}

Browsing through the equations in Section 5.1, it appears
clearly that the two quantities
\bea
&&\!\!\!\!\!\!\!\! \left\{
- \sarea({\area}/2) \ss(\EXC_I)\ss(\EXC_J) \ss(\EXC_K) \right\} ^{1/2}
\!\!=
\left\{
-\frac{\ss(\EXC)}{\k_1}\ss(\EXC_I)\ss(\EXC_J) \ss(\EXC_K)\right\}^{1/2}
\cr
&&\!\!\!\!\!\!\!\! \left\{
-\scoarea({\coarea}/2) \s(\exc_i)\s(\exc_j) \s(\exc_k) \right\} ^{1/2}
\!\!=
\left\{
-\frac{\s(\exc)}{\k_2} \s(\exc_i)\s(\exc_j) \s(\exc_k) \right\} ^{1/2}
\eea
must be real in order to the triangle to exist, so the
quantities under the square root must be non-negative. The
sides and angles of a triangle in the CK space with
constants $\k_1, \k_2$ must satisfy the inequalities:
\be
- \frac{\ss(\EXC)}{\k_1} \ss(\EXC_I)\ss(\EXC_J) \ss(\EXC_K) \geq 0
\qquad
-\frac{\s(\exc)}{\k_2} \s(\exc_i)\s(\exc_j) \s(\exc_k) \geq 0 .
\ee
In terms of the explicit notation 
(\ref{eq:OrthoTrig:RelSemisumsExcesses}), the relation relative to
sides  reads:
\be
\frac{\s(p-a)}{\k_2} \s(p)\s(p-b) \s(p-c) \geq 0
\ee
and by using the identities (\ref{zzz:wl}) and (\ref{zzz:wm}),
this can also be rewritten as:
\be
\frac{1}{\k_2} \left\{ \v(b+c) - \v(a) \right\}  \left\{
\v(a)-\v(b-c) \right\} \geq 0 .
\ee
This inequality must be satisfied by the sides of the triangle
loop. When $\k_2=0$ we already know that $a=b+c$, so it
suffices to discuss the cases $\k_2 \neq 0$. When $\k_2$ is
positive, both terms in brackets must have the same sign, and if
$\k_2$ is negative, each term must have a different sign. A
completely similar inequality holds, mutatis mutandis, for
angles.

The complete discussion of the inequalities that sides and
angles must satisfy can be done starting from this, but
requires attention to subcases when any of the
constants $\k_1$, $\k_2$ is positive. The derivative of the
versed sine function
$V_{\k}(x)$  for a generic label $\k$ is the sine:
$\frac{d}{dx}V_{\k}(x)=S_{\k}(x)$; hence
$V_{\k}(x)$  is increasing for all values of $x$ when the label
is zero or negative, but for positive $\k$, when $x$ is larger
than a quadrant, $x \geq \frac{\pi}{2 \sqrt{\k}}$ (see
(\ref{eq:OrthoTrig:OrthoValue}) below),   the versed sine is decreasing.
If when $\k_1>0$ (resp.\ $\k_2>0$) we restrict sides (resp.\ angles) to
less than a quadrant, and label $a$ the largest side (resp.\ $A$ the
largest external angle) then it is easy to transform  the basic
inequalities for sides and angles to give:
\be
\begin{array}{ll}
B+C > A >|B-C| \quad \hbox{when}\quad \k_1 >0 &\qquad
B+C < A         \quad\hbox{when}\quad \k_1 <0  \\
b+c > a >|b-c|  \quad\hbox{when}\quad \k_2 >0 &\qquad
b+c < a         \quad\hbox{when}\quad \k_2 <0
\label{OrthoTrig:Inequalities}
\end{array}
\ee
which reduce to the known conditions in the three riemannian cases.
These inequalities can be also and more clearly expressed in
terms of the (half) excesses $E$ and $e$, and read:
\be
\begin{array}{ll}
E > 0  \quad\hbox{when}\quad \k_1 >0 &\qquad
E < 0  \quad\hbox{when}\quad \k_1 <0  \\
e > 0  \quad\hbox{when}\quad \k_2 >0 &\qquad
e < 0  \quad\hbox{when}\quad \k_2 <0.
\label{OrthoTrig:InequalitiesBis}
\end{array}
\ee


\section{Other types of triangles}
\label{Sec:vi}

In the general CK space $SO_{\k_1,\k_2}(3)/SO_{\k_2}(2)$ there
are four  types of triangles according to the kind of the
sides: (i) the three sides are first-kind (time-like), (ii) two
first-kind and one second-kind, (iii) one first-kind and two
second-kind, and (iv) three  second-kind (space-like).  Of
course when $\k_2 >0$,  first- and second-kind lines coincide so
that the four types merge into one. Triangles with isotropic
sides are limiting non-generic cases.
In this Section  we present the main equations which
characterize the trigonometry of  pure  second-kind triangles
of type (iv) and `mixed'  triangles of type (ii); for the
latter we will only consider orthogonal triangles.


\subsection{Second-kind triangles}
\label{Sec:vi:i}

With rather obvious modifications the method we have developed
for the pure first-kind triangles of type (i)  will give the
equations of the pure second-kind ones of type (iv). For the
latter all sides are second-kind lengths (i.e., {\em space
lengths} in the kinematical cases) and the translations along
sides will be conjugated to the generator $P_2$ instead of
$P_1$. Henceforth, the label of sides will be the product
$\k_1\k_2$ and not $\k_1$ (see (\ref{eq:Ortho:CKspaceSecKindLines}) and 
(\ref{eq:OrthoCKBiDOneParamSubgroups})),
which in turn means that the equations for the second-kind
triangles can be directly obtained by replacing  $\k_1$ by
$\k_1\k_2$ in {\em all} relations  for  first-kind triangles.
Consequently,  the general appearance of these new equations is
very similar to the pure first-kind ones, but now the side
lengths are of second-kind. As an example we only write the
cosine, dual cosine and sine theorems for  a {\em pure
second-kind triangle}:
\be
\begin{array}{l}
\ccc(x_i) =\ccc(x_j)\ccc(x_k)-\k_1\k_2 \sss(x_j)\sss(x_k)\cc(X_I)\cr
\cc(X_I) =\cc(X_J)\cc(X_K)-\k_2\ss(X_J)\ss(X_K)\ccc(x_i)\cr
\displaystyle{ \frac{\sss(x_i)}{\ss(X_I)} =
\frac{\sss(x_j)}{\ss(X_J)} =\frac{\sss(x_k)}{\ss(X_K)} }
\end{array}
\ee
to be compared with the first-kind equations 1i, 1I and 2 of
(\ref{eq:OrthoTrig:BasicTrigEqsCompact}).

For $\k_2>0$  both translation generators are conjugated, so the
trigonometry for pure first-kind and second-kind triangles is
essentially identical, differing only by the fact that the sides
should be measured according to $P_2$. For $\k_2 <0$ the sign in the
cosine equations for sides will change, and equations for triangles
with second-kind sides will be different from the first-kind ones.
In the $(1+1)$D case, the difference between anti-de Sitter and de
Sitter spacetimes reduces simply to the interchange between time-like
and space-like lines, so the trigonometry of pure space-like
triangles in de Sitter spacetime (resp.\ anti-de Sitter) coincides
with the trigonometry of time-like ones in anti-de Sitter spacetime
(resp.\ de Sitter). In the $(1+1)$D minkowskian  spacetime the
interchange of time-like and space-like lines is a symmetry, and
again the only difference is the factor
$c=1/\sqrt{-\k_2}$ between the length of any segment  measured
in `time' or in `space' units. In our approach, this symmetry is
encoded by the interchange
$\k_1\leftrightarrow \k_1\k_2$ ($P_1\leftrightarrow P_2$) with
$\k_2$ ($J_{12}$) invariant.


\subsection{Orthogonal triangles}
\label{Sec:vi:ii}

The  two remaining types of triangles, (ii) and (iii), are
`mixed' triangles with sides of two different kind and
are something different to the pure ones. They are important
because triangles with two orthogonal sides belong necessarily
to these types (in the kinematical cases, a direction
orthogonal to a time-like direction must be space-like).

The equations of the trigonometry of such triangles could be
obtained by suitably adapting the method introduced in Section
3. But there is also an alternative way, which allows us to
formally derive the equations of any type of triangles from the
pure first-kind ones; we will sketch this procedure. It is
based in the consideration of the angle between two orthogonal
directions; this is a {\em real} angle when $\k_2>0$ but is only a
formal device (an {\em ideal} angle) when
$\k_2\leq0$. This  {\em orthogonal angle} or {\em quadrant of
angle} will be  defined by
\be
\orto_{\k_2} := \frac{\pi}{2\sqrt{\k_2}}
\label{eq:OrthoTrig:OrthoValue}
\ee
which is well defined whenever $\k_2\ne 0$ and has a real value
when $\k_2>0$; if $\k_2=0$ then $\orto_{\k_2}=\infty$. This
verifies
\be
\cc(\orto_{\k_2})=0\qquad \ss(\orto_{\k_2})=\frac{1}{\sqrt{\k_2}} .
\label{eq:OrthoTrig:OrthoCompSinCosValue}
\ee
Of course, a quadrant of length could be analogously defined;
this will be real when $\k_1>0$ as in the case of the sphere,
but will be only a formal device when $\k_1<0$, e.g.\ for hyperbolic
geometry.

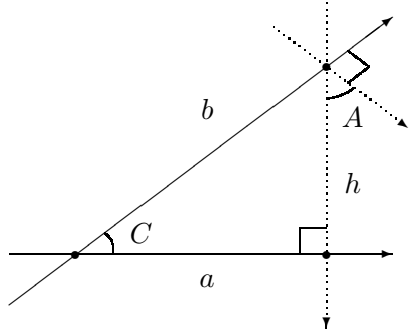
\begin{figure}[ht]

\begin{center}
\begin{picture}(145,120)
\put(25,25){\circle*{3}}
\put(120,25){\circle*{3}}
\put(120,96){\circle*{3}}
\put(50,33){\makebox(0,0){$C$}}
\put(130,77){\makebox(0,0){$A$}}
\put(130,52){\makebox(0,0){$h$}}
\put(75,15){\makebox(0,0){$a$}}
\put(75,80){\makebox(0,0){$b$}}
\put(0,6){\vector(4,3){145}}
\put(0,25){\vector(1,0){145}}
\qbezier[50](120,0)(120,60)(120,120)
\put(120,0){\vector(0,-1){3}}
\qbezier[25](100,111)(124,93)(148,75)
\put(148,75){\vector(4,-3){3}}
\put(110,25){\line(0,1){10}}
\put(110,35){\line(1,0){10}}
\qbezier(120,84)(127,85)(130,88)
\qbezier(39,25)(40,30)(36,33)
\qbezier(128,102)(132,99)(136,96)
\qbezier(128,90)(132,93)(136,96)
\end{picture}
\end{center}
\noindent
\caption{Orthogonal triangle with two first-kind sides
$a$ and $b$, one second-kind side $h$, an inner angle $C$ and an
external angle $A$.}
\label{fig:OrthogTriangle}

\end{figure}

We restrict here to the equations of trigonometry for
{\em orthogonal triangles} with two first-kind sides  $a$ and
$b$ and one second-kind side $h$ orthogonal to the side $a$.
This kind of triangle has an inner angle $C$ at the vertex
$a\cap b$ and an 'external' angle $A$  (between the side $h$ and
the line orthogonal to the side $b$ as shown in figure
\ref{fig:OrthogTriangle}) at vertex $b\cap h$.

For the generic situation with $\k_2\ne 0$, this orthogonal
triangle can be obtained as a formal case  of  a pure first-kind
one  (compare figure \ref{fig:OrthogTriangle} to figure
\ref{fig:TriangleCompatibility}) where out of the  six quantities $a,
b, c, A, B, C$ associated to a   first-kind  triangle $A$, $B$ and
$c$ should be understood through the replacements $A\to
A+\orto_{\k_2}$,
$B\to
\orto_{\k_2}$ and $c$ (with label $\k_1$) $\to h$ (with label 
$\k_1\k_2$), while $a$, $b$ and $C$ are unchanged. By taking
into account  the properties (\ref{eq:OrthoTrig:OrthoCompSinCosValue}),
we find that this transformation can be described by the following
substitution:
\bea
&& a\to a \qquad b\to b \qquad C\to C\cr
&&\c(c)\to \ccc(h) \qquad \cc(A)\to -\sqrt{\k_2}\,\ss(A)
\qquad \cc(B)\to 0\cr
&&\s(c)\to \sqrt{\k_2}\,\sss(h) \qquad \ss(A)\to
 \frac{1}{\sqrt{\k_2}}\,\cc(A)
\qquad \ss(B)\to  \frac{1}{\sqrt{\k_2}} .
\label{eq:OrthoTrig:OrthoSubstitution}
\eea

The trigonometric relations for the orthogonal triangle can be
straightforwardly obtained from those corresponding to the
first-kind one (given in Section 4) by simply applying the above
transformations. Hence the  theorems
(\ref{eq:OrthoTrig:Cosine})--(\ref{eq:OrthoTrig:Sine}) give rise to

\noindent
$\bullet$ Three {\em cosine theorems} for sides:
\be
\begin{array}{l}
\c(a) =\c(b)\ccc(h)+\k_1\k_2 \s(b)\sss(h)\ss(A) \\
\c(b) =\c(a)\ccc(h)  \\
\ccc(h) =\c(a)\c(b)+\k_1 \s(a)\s(b)\cc(C) .
\label{eq:OrthoTrig:OrthoCosine}
\end{array}
\ee

\noindent
$\bullet$ Three {\em cosine theorems} for angles:
\be
\begin{array}{l}
\ss(A) = \ss(C)\c(a)\\
0 =-\ss(A)\cc(C)+ \cc(A)\ss(C)\c(b)\\
\cc(C) =  \cc(A) \ccc(h) .
\end{array}
\label{eq:OrthoTrig:OrthoDualCosine}
\ee

\noindent
$\bullet$ One  {\em sine theorem}:
\be
\frac{\s(a)}{\cc(A)} =  \s(b)
=\frac{\sss(h)}{\ss(C)}.
\label{eq:OrthoTrig:OrthoSine}
\ee

We stress that although  the maps (\ref{eq:OrthoTrig:OrthoSubstitution})
are well defined only for the generic case with $\k_2\ne 0$, in the
resulting equations
(\ref{eq:OrthoTrig:OrthoCosine})--(\ref{eq:OrthoTrig:OrthoSine})  the
contraction $\k_2=0$ is always well defined.  We also remark that the
three relations (\ref{eq:OrthoTrig:OrthoCosine}) and the third one of 
(\ref{eq:OrthoTrig:OrthoDualCosine})   can be expressed in terms of
versed sines avoiding the trivial identities $1=1$ when $\k_1$ or $\k_2$
is equal to zero. For instance, when $\k_1=0$ the second relation of
(\ref{eq:OrthoTrig:OrthoCosine}) reduces simply to $b^2=a^2+\k_2 h^2$
(the `pythagorean' theorem for sides), while the third equation of
(\ref{eq:OrthoTrig:OrthoDualCosine})  gives
$C^2=A^2+\k_1 h^2$ when $\k_2=0$.

Other interesting relations, which can be obtained from
(\ref{eq:OrthoTrig:OrthoCosine})--(\ref{eq:OrthoTrig:OrthoSine}),  read
\be
\begin{array}{l}
\displaystyle{\cc(C)=\frac{\T(a)}{\T(b)}\qquad
\TT(C)=\frac{\TTT(h)}{\s(a)}}\cr
\displaystyle{\ss(A)=\frac{\TTT(h)}{\T(b)}\qquad
\TT(A)=\frac{\sss(h)}{\T(a)}} .
\end{array}
\label{zzz:ortog}
\ee

\begin{table}[t]
{\footnotesize
\noindent
\caption{Equations of trigonometry for
orthogonal triangles for the nine CK spaces.}
\label{table:OrthogTrigEqns}
\smallskip
\noindent\hfill
\begin{tabular}{ccc}
\hline
&&\\[-8pt]
Elliptic\quad $(+1,+1)$&
Euclidean\quad $(0,+1)$&
Hyperbolic\quad $(-1,+1)$\\
$SO(3)/SO(2)$&$ISO(2)/SO(2)$&$SO(2,1)/SO(2)$\\[4pt]
$\cos b =\cos a \cos h $&$b^2=a^2+h^2$ &
$\cosh b =\cosh a \cosh h$\\
$\cos C =\cos A \cos h $&$ C=A$&
$\cos C =\cos A \cosh h$\\
$ \sin h = \sin b \sin C $&$ h={b}\sin C$&$
 \sinh h ={\sinh b}\sin C $\\
${\tan h} ={\tan b}\sin A $& $$& $\tanh h = \tanh b \sin A$\\
${\sin a} ={\sin b}\cos A$& $$& ${\sinh a} ={\sinh b}\cos A$\\
$\tan a ={\tan b}\cos C$&$a={b}\cos C$&${\tanh a} ={\tanh b}\cos C$\\[4pt]
$\displaystyle{\sin \area =\frac {\sin a \sin h} {1+\cos b}}$ &
$\displaystyle{\area =\frac 12 a\,h}$ &
$\displaystyle{\sin \area =\frac {\sinh a \sinh h} {1+\cosh b}}$\\[8pt]
\hline
&&\\[-8pt]
Co-Euclidean\quad $(+1,0)$&
Galilean\quad $(0,0)$&
Co-Minkowskian\quad $(-1,0)$\\
Oscillating NH\quad $ISO(2)/\Re$&$IISO(1)/\Re$&
Expanding NH\quad $ISO(1,1)/\Re$\\[4pt]
$b = a $&$b= a$&$b = a$\\
$C^2 = A^2+h^2 $&$C= A$ &$ C^2 = A^2-h^2$\\
${h} = C\sin b $&$h= b\, C$&$ h = C\sinh b$\\
$ h = A\tan b $& $$ & $h =  A\tanh b$\\[4pt]
$\displaystyle{\area =\frac {h\sin a } {1+\cos b}}$ &
$\displaystyle{\area =\frac 12 a\,h}$ &
$\displaystyle{\area =\frac {h\sinh a } {1+\cosh b}}$\\[8pt]
\hline
&&\\[-8pt]
Co-Hyperbolic\quad $(+1,-1)$&
Minkowskian\quad $(0,-1)$&
Doubly Hyperbolic\quad $(-1,-1)$\\
Anti-de Sitter\  $SO(2,1)/SO(1,1)$&$ISO(1,1)/SO(1,1)$&
De Sitter\  $SO(2,1)/SO(1,1)$\\[4pt]
$\cos b =\cos a \cosh h$&$b^2=a^2-h^2$&$
\cosh b =\cosh a \cos h$\\
$\cosh C =\cosh A \cosh h$&$C=A$&$
\cosh C =\cosh A \cos h$\\
$\sinh h ={\sin b}\sinh C$&$
 h ={b}\sinh C$&$\sin h ={\sinh b}\sinh C$\\
$\tanh h ={\tan b}\sinh A$&$$&$ \tan h ={\tanh b}\sinh A $\\
$\sin a ={\sin b}\cosh A$& $$ &$\sinh a ={\sinh b}
\cosh A$\\
${\tan a} ={\tan b}\cosh C $&$
a ={b}\cosh C$&$\tanh a ={\tanh b}\cosh C$\\[4pt]
$\displaystyle{\sinh \area =\frac {\sin a \sinh h} {1+\cos b}}$ &
$\displaystyle{\area =\frac 12 a\,h}$ &
$\displaystyle{\sinh \area =\frac {\sinh a \sin h} {1+\cosh b}}$\\[8pt]
\hline
\end{tabular}\hfill}
\end{table}

The dependence between all these relations is described by:

\medskip
\noindent
{\bf Theorem 4.} For an orthogonal triangle in any of the nine 2D
CK geometries,  there are always {\em three} independent
equations. Any other trigonometric relation follows from them.
When $\k_1\neq0$ and $\k_2\neq0$, a possible choice for these
three equations is:
\be
\c(b) =\c(a)\ccc(h)\qquad
\cc(C) =  \cc(A) \ccc(h)\qquad
\sss(h)= \s(b)\ss(C)   .
\label{zzz:ortoh}
\ee
When $\k_1=0$ (resp.\ $\k_2=0$)  the first (resp.\ second)
equation should be replaced by $b^2=a^2+\k_2 h^2$ (resp.\ 
$C^2=A^2+\k_1 h^2$).
\medskip

The proof is a matter of simple algebraic manipulations and is
omitted.

The last interesting quantity to introduce is the area; starting
again from its definition for a first-kind triangle, $\area
=(B+C-A)/\k_1$, and applying the transformation of the angles we
find for the area for an orthogonal triangle: 
\be
\area = \frac{C-A}{\k_1}  .
\label{zzz:ortoi}
\ee
An expression for the sine of the area can be easily deduced:
\bea
&&\sarea(\area)= \frac {1}{\k_1}\ss(C-A) =
 \frac{1}{\k_1}
\left(\ss(C)\cc(A)-\ss(A)\cc(C)\right)\cr
&&  =\frac{\s(a)\sss(h)}
{\k_1\s^2(b)} \left( 1-\frac {\c^2(b)}{\ccc(h)\c(a)} \right)
=\frac{\s(a)\sss(h)}
{1-\c^2(b)}( 1-{\c(b)})
\eea
which finally gives
\be
\sarea(\area)=\frac{\s(a)\sss(h)}{1+\c(b)} .
\label{eq:OrthoTrig:OrthoSineArea}
\ee
When $\k_1=0$, we recover  the flat space familiar value
$\area=\frac 12{ah}$. In the same way we find
\be
\carea(\area)=\frac {\c(a) +\ccc(h)}{1+\c(b)} .
\ee

We show in  table \ref{table:OrthogTrigEqns} the main
trigonometric equations and the area relation 
(\ref{eq:OrthoTrig:OrthoSineArea}) for these orthogonal triangles
particularised in  each of the nine 2D CK spaces to the standard values
of $\k_i$ equal to $+1$, $0$ and $-1$; the first three  equations are
independent. Note that self-duality does not hold for orthogonal
triangles; the dual of  an orthogonal triangle would be a triangle with
a  side whose length equals to a quadrant.


\section{On the trigonometry of homogeneous spacetimes}
\label{Sec:vii}

Once familiarity is gained with the labelled cosine, sine and
tangent functions, the general way of writing the equations of
trigonometry becomes clearer than the conventional formulation.
However in order to facilitate the reading of these general
equations, we present in table \ref{table:EqnsSixSpacetimes} a
sample of equations for the area in the six homogeneous
spacetimes:   the generalised   formula  of Cagnoli (first
equation in (\ref{zzz:bestg})) and that of Heron--L'Huillier
(\ref{zzz:beste}) as well as the  equation (\ref{zzz:bestc}) that
relates area and coarea. We use the conventional units, and keep 
explicitly the {\em universe (time) radius} $\tiempo$ and the {\em
relativistic constant} $c$; recall that the  CK
constants in the kinematical  interpretation
(\ref{eq:OrthoCKBiDKimematInt}) of the CK space of points as $(1+1)$D
spacetime  are $\k_1=\pm 1/\tiempo^2$ and $\k_2=- 1/c^2$ (see Section
2.3).

To avoid confusion of the relativistic
constant $c$ with a side, and also  to stress that these sides are
{\em proper times}, we will denote in this Section the (time) side
lengths by $\ta, \tb, \tc$, while angles in the $(1+1)$D
spacetime  are kinematically rapidities, and will be denoted
according to the traditional notation as  $\xA, \xB,
\xC$. Should these equations be extended to include the $\k_2>0$ case
(say let $\k_2=1/c^2$), the $c$ would play the role of a
conversion constant between radians and the chosen angular
measure.


\begin{table}[t]
{\footnotesize
{ \noindent
\caption{Some equations involving the area $\area$ and coarea
$\coarea$ of a time-like triangle  in the six homogeneous
spacetimes with (time) radius $\tiempo$ (curvature $\k_1=\pm
1/\tiempo^2$) and relativistic constant $c$ ($\k_2=-1/c^2$).
Here the three sides are the (proper) time intervals $\ta, \tb,
\tc$ and $\tp:=(\ta+\tb+\tc)/2$. The angles are the relative
rapidities $\xA, \xB, \xC$ and
$\xP:=(\xA+\xB+\xC)/2$.}
\label{table:EqnsSixSpacetimes}
\medskip
\noindent\hskip -40pt
\begin{tabular}{cc}
\hline
&\\[-8pt]
Relative-time spacetimes&Absolute-time spacetimes\\[4pt]
\hline
&\\[-8pt]
Anti-de Sitter\quad $(+1/\tiempo^2,-1/c^2)$&Oscillating NH\quad
$(+1/\tiempo^2,0)$\quad $c=\infty$\\[4pt]
$\displaystyle{
\sinh\left(\frac{\area}{2\tiempo^2 c}\right)=
\frac{\sin(\frac{\ta}{\tiempo})\sin(\frac{\tb}{\tiempo})
\sinh(\frac{\xC}{c})}
{4\cos(\frac{\ta}{2\tiempo})\cos(\frac{\tb}{2\tiempo})
\cos(\frac{\tc}{2\tiempo})} }$&
$\displaystyle{
\area= \frac{\tiempo^2\sin(\frac{\ta}{\tiempo})\sin(\frac{\tb}{\tiempo})
\xC}{2\cos(\frac{\ta}{2\tiempo})\cos(\frac{\tb}{2\tiempo})
\cos(\frac{\tc}{2\tiempo})} }$\\[8pt]
$\displaystyle{
\tanh^2\left(\frac{\area}{4\tiempo^2 c}\right)=-
\tan\left(\frac{\tp}{2\tiempo}\right)
\tan\left(\!\frac{\tp\closeminus\ta}{2\tiempo}\!\right)
\tan\left(\!\frac{\tp\closeminus\tb}{2\tiempo}\!\right)
\tan\left(\!\frac{\tp\closeminus\tc}{2\tiempo}\!\right) }$&
$\displaystyle{
\area^2=4\coarea\tiempo^3 \tan\left(\!\frac{\tp}{2\tiempo}\!\right)
\tan\left(\!\frac{\tp-\tb}{2\tiempo}\!\right)
\tan\left(\!\frac{\tp-\tc}{2\tiempo}\!\right) }$\\[8pt]
$\displaystyle{
\frac{\tanh (\frac{\area}{4\tiempo^2 c})}
{\tan (\frac{\coarea}{4\tiempo c^2})}
=\frac{\tan(\frac{\tp}{2\tiempo})}{\tanh(\frac{\xP}{2c})}
=\frac{\tan(\frac{\tp-\tb}{2\tiempo})}{\tanh(\frac{\xP-\xB}{2c})}
=\frac{\tan(\frac{\tp-\tc}{2\tiempo})}{\tanh(\frac{\xP-\xC}{2c})} }$&
$\displaystyle{ \frac{ \area }{ \coarea }
=\frac{\tiempo\tan(\frac{\tp}{2\tiempo})}{\frac{\xP}{2}}
=\frac{\tiempo\tan(\frac{\tp-\tb}{2\tiempo})}{\frac{\xP-\xB}{2}}
=\frac{\tiempo\tan(\frac{\tp-\tc}{2\tiempo})}{\frac{\xP-\xC}{2}}}$\\[12pt]
\hline
&\\[-8pt]
Minkowskian\quad $(0,-1/c^2)$\quad $\tiempo=\infty$&
Galilean\quad $(0,0)$\quad $\tiempo=\infty$, $c=\infty$\\[4pt]
$\displaystyle{\area=\frac{1}{2} \ta \tb\, c\sinh(\frac{\xC}{c})  }$&
$\displaystyle{\area=\frac{1}{2} \ta \tb  \xC  }$\\[4pt]
$\displaystyle{\area^2= -c^2  \tp(\tp-\ta) (\tp-\tb)(\tp-\tc) }$&
$\displaystyle{\area^2=\frac 12 \coarea  \tp (\tp-\tb)(\tp-\tc) }$\\[8pt]
$\displaystyle{
\frac{ \area }{ \coarea }
=\frac{\frac{\tp}{2}}{c\tanh(\frac{\xP}{2c})}
=\frac{\frac{\tp-\tb}{2}}{c\tanh(\frac{\xP-\xB}{2c})}
=\frac{\frac{\tp-\tc}{2}}{c\tanh(\frac{\xP-\xC}{2c})} }$&
$\displaystyle{\frac{ \area }{ \coarea }
=\frac{\tp}{\xP}
=\frac{ \tp-\tb } { \xP-\xB }=\frac{ \tp-\tc } { \xP-\xC }  }$\\[12pt]
\hline
&\\[-8pt]
De Sitter\quad $(-1/\tiempo^2,-1/c^2)$&
Expanding NH\quad $(-1/\tiempo^2,0)$\quad $c=\infty$\\[4pt]
$\displaystyle{ \sinh\left(\frac{\area}{2\tiempo^2 c}\right)=
\frac{\sinh(\frac{\ta}{\tiempo})\sinh\left(\frac{\tb}{\tiempo}\right)
\sinh(\frac{\xC}{c})}
{4\cosh(\frac{\ta}{2\tiempo})\cosh(\frac{\tb}{2\tiempo  })
\cosh(\frac{\tc}{2\tiempo})} }$&
$\displaystyle{\area=
\frac{\tiempo^2\sinh(\frac{\ta}{\tiempo})\sinh(\frac{\tb}{\tiempo})
\xC}{2\cosh(\frac{\ta}{2\tiempo})\cosh(\frac{\tb}{2\tiempo})
\cosh(\frac{\tc}{2\tiempo})} }$\\[8pt]
$\displaystyle{
\tanh^2\left(\frac{\area}{4\tiempo^2 c}\right)=
-\tanh\left(\frac{\tp}{2\tiempo}\right)
\tanh\left(\!\frac{\tp\closeminus\ta}{2\tiempo}\!\right)
\tanh\left(\!\frac{\tp\closeminus\tb}{2\tiempo}\!\right)
\tanh\left(\!\frac{\tp\closeminus\tc}{2\tiempo}\!\right) }$&
$\displaystyle{
\area^2=4\coarea\tiempo^3 \tanh\left(\frac{\tp}{2\tiempo}\right)
\tanh\left(\!\frac{\tp-\tb}{2\tiempo}\!\right)
\tanh\left(\!\frac{\tp-\tc}{2\tiempo}\!\right) }$\\[8pt]
$\displaystyle{
\frac{\tanh (\frac{\area}{4\tiempo^2 c})}
{\tanh (\frac{\coarea}{4\tiempo c^2})}
=\frac{\tanh(\frac{\tp}{2\tiempo})}{\tanh(\frac{\xP}{2c})}
=\frac{\tanh(\frac{\tp-\tb}{2\tiempo})}{\tanh(\frac{\xP-\xB}{2c})}
=\frac{\tanh(\frac{\tp-\tc}{2\tiempo})}{\tanh(\frac{\xP-\xC}{2c})} }$&
$\displaystyle{ \frac{ \area }{ \coarea }
=\frac{\tiempo\tanh(\frac{\tp}{2\tiempo})}{\frac{\xP}{2}}
=\frac{\tiempo\tanh(\frac{\tp-\tb}{2\tiempo})}{\frac{\xP-\xB}{2}}
=\frac{\tiempo\tanh(\frac{\tp-\tc}{2\tiempo})}{\frac{\xP-\xC}{2}}}$
\\[12pt]
\hline
\end{tabular} }}
\end{table}

A suitable view of the general trigonometric equations is as a
kind of {\em deformation} of the  purely linear equations
(\ref{eq:OrthoTrig:GalBasic}). The deformation  is governed by two
constants $\k_1,
\k_2$,  whose geometrical role is that of spacetime curvatures
and/or signatures of the metric. From this viewpoint, it is clear
that the good `totally flat' reference 2D geometry is not the
euclidean one, but should be instead the galilean geometry, whose
trigonometric equations are {\em purely} linear; all others being
better described by its departures, governed by $\k_1,
\k_2$, relative to the galilean one. The kinematical
interpretation of the three basic trigonometric equations in
galilean spacetime, where rapidities, defined as the canonical
parameter of subgroups of pure inertial transformations, are equal to
ordinary velocities, is clear:
\be
 \ta=\tb+\tc  \qquad \xA=\xB+\xC  \qquad \frac {\ta}{\xA} = \frac
{\tb}{\xB} = \frac {\tc}{\xC}  .
\label{eq:OrthoTrig:GalSpaceTimeBasic}
\ee
The first equation
means that the (proper) time interval along any future time-like
curve depends only on the endpoints, and corresponds to the
{\em absolute time}; the same equation holds in both
Newton--Hooke cases (see table \ref{table:BasicTrigEqns}). The
second is the additivity, in galilean spacetime, of the relative
rapidities of three {\em non-concurrent} free motions; this also
holds in the minkowskian case, but not in the four curved
spacetimes. This relation should not be confused with the
additivity of relative rapidities for coplanar and concurrent
free motions, which holds in {\em all} cases because rapidities
are defined as canonical parameters of the one-parameter subgroup
generated by $\stK$. The third equation states that relative
rapidities and time interval lengths in any triangle in galilean
spacetime are proportional; this is  an absolutely elementary
property of classical spacetime and holds only in this case. For
the area and coarea of a galilean triangle we have:
\be
\area=\frac{1}{2}  \xA \tb \tc   \qquad 
\coarea=\frac{1}{2}  \ta  \xB \xC .
\ee
We recall that  coarea of a triangular loop in the six
kinematical spaces is (proportional) to the difference of
actions for a free particle following either of the two
worldlines
$CB$ and $CAB$ which determine the triangle loop
\cite{ActGeoInt1, ActGeoInt2}.

These purely linear equations allow a {\em deformation} in two
different senses
\cite{expansion}, either by endowing spacetime with curvature
$\k_1 \neq 0$ (obtaining the two Newton--Hooke spacetimes), or
keeping it flat but introducing curvature, necessarily negative
if causality must be preserved, in the space of time-like lines,
described by the constant $\k_2 <0 $ (obtaining the minkowskian
spacetime of special relativity). If both processes are
simultaneously made, we obtain the two de Sitter spacetimes.

This structural `unfolding' of the complete CK scheme starting
from its most degenerate case runs in a striking parallel with
the historical development. Spacetime is nearly flat at the time
and length human scales: this {\em fact} lies behind classical
physics. With the present hindsight, we can say that to assume a
flat (i.e., $\k_1=0$) homogeneous model for the $(1+1)$D spacetime
involved in 1D kinematics was natural. At the human speed (or
even solar system) scale, the curvature in the space of uniform
motions is also negligible, so to assume again a {\em flat}
space of motions (or time-like lines) ---embodied in the
equality $\k_2=0$ for the 1D kinematic group---, was the only
practical choice. Both assumptions greatly simplified (or
rather, {\em allowed}) the linear mathematical description of
classical physics. However, even at the homogeneous level of
approximation, Nature  does not seem to be characterized by
these non-generic choices. Relativity can be described as the
discovery of a {\em  negative} curvature in the space of 1D
motions, and then the all-important relativistic constant $c$
appears simply as related to the value of the curvature of this
space of motions by $\k_2=-1/c^2$. Special relativity still keeps
a flat spacetime, another approximation which is abandoned, in a
way much more general than by assuming it homogeneous, in the
context of general relativity; if homogeneity is still kept,  the
possibilities are $\k_1 = \pm1/\tiempo^2$.

Before spacetime geometry was under consideration, a similar
situation happened for the physical 3D {\em space} geometry,
whose  characterization among the mathematical possibilities was
at the root of Riemann's program.  General 3D CK spaces are
parametrized by {\em three} CK constants, say $\mu_1$, $\mu_2$,
$\mu_3$ and correspond to a space with  constant
curvature $\mu_1$  and whose metric is reducible at
each point to $\mbox{diag}(1,\mu_2, \mu_2\mu_3)$; restriction to
a locally euclidean space is embodied into the choices $\mu_ 2>0,
\mu_3>0$.  In the $(1+3)$D homogeneous spacetime
$SO_{\k_1,\k_2,\k_3,\k_4}(5)/ SO_{\k_2,\k_3,\k_4}(4)$ (see Section
2.1), whose labels are $\k_1=\pm 1/\tiempo^2$, $\k_2=-1/c^2$,
$\k_3=1$, $\k_4=1$,  the three-space orthogonal to the fiducial
time-like line through the origin (and
hence all three-spaces orthogonal to any time-like line) can be
identified with the homogenous space
$SO_{\k_1\k_2,\k_3,\k_4}(4)/SO_{\k_3,\k_4}(3)$.  Hence the constants
$\mu_1$, $\mu_2$, $\mu_3$   are given by $\mu_1=\k_1\k_2$, $\mu_2=\k_3$,
$\mu_3=\k_4$ so that the  curvature of the three-space orthogonal to a
given time-like direction is the product $\k_1\k_2=\mp 1/(\tiempo^2
c^2)$. This means that in the non-relativistic spacetimes, even if
spacetime is curved, the  three-space is {\em flat}. In the three
relativistic cases, three-space is only flat in the
minkowskian case, but is  curved in the anti-de Sitter (space
curvature $-1/(\tiempo^2 c^2)$) and de Sitter spacetimes (space
curvature $1/(\tiempo^2 c^2)$). In both cases the universe radius
$R$, with dimensions of space-length, is given by $R= c \tiempo$.

Here we have also a clear  example of {\em hidden universal
constants}, in the sense given to this term by L\'evy-Leblond
\cite{LevyLeblonUnivConst}: both $\k_3$ and
$\k_4$ are not usually considered as  universal constants only
because  the fact that they are {\em non-zero and positive}
allow to make them apparently dissapear by reducing them to the
value $1$ (thus making plane angles and dihedral space angles
apparently dimensionless).  Once performed, this reduction
forbids further consideration of these constants and the
possibility they {\em could} be either zero or negative in other
conceivable but still homogeneous spaces is simply out of sight.
The character of $\k_1\k_2$ as a possible universal constant was
understood much earlier:  Lobachewski explored the possibility
of physical three-space geometry being hyperbolic (i.e.\
negatively curved), and tried to give experimental bounds to a 
constant he called $k$ (in modern terms, the  curvature would be
$-1/k^2$) for our physical space under the assumption, later
disproven by Einstein's theory, that light travels along
geodesics in this physical three-space; the argument is based on
the existence of a {\em minimum} parallax (for a given baseline)
even for infinitely distant stars \cite{Bonola, Milnor}. The
founding fathers of hyperbolic geometry could have hardly
imagined that the geometry they were discovering/inventing was
indeed realized by Nature and to a good approximation, not
as the geometry of space itself, but as  the geometry of
the space of uniform motions. 

The equations we have developed can be used to discuss many
questions. For instance, a recent paper by Jing-Ling Chen and 
Mo-Lin Ge \cite{ChenMoLinGe} identifies the Wigner angle of the
rotation appearing in the product of pure Lorentz transformations
to the defect of a triangle in hyperbolic geometry; this and similar
results follow also directly from  our approach,  because in the
$(1+3)$D spacetime labeled by $\k_1$, $\k_2=-1/c^2$, $\k_3=1$,
$\k_4=1$ the geometry of the  3D set of time-like lines through
a fixed spacetime point is characterized by the three last
labels $\k_2$, $\k_3$, $\k_4$, so it is $\{ -1/c^2, 1, 1\} $.
This geometry is hyperbolic in the three relativistic cases
degenerating to euclidean in the three non-relativistic ones.
Similar results involving triangle defects appear also in
relation to geometrical phases. Triangles in anti-de Sitter and
de Sitter spaces also display properties similar to the
`parallelism angle' found in hyperbolic geometry, and are
related to the existence of horizons. 


\section{Concluding remarks}
\label{Sec:viii}

As far as we know, the approach we have  given to trigonometry of
the real  CK spaces (the symmetric rank-one homogeneous spaces of
real type and their limiting spaces) is new. Furthermore and  even in
this case of rank-one spaces of real type, we have also obtained
some  results apparently unknown on the trigonometry of several
spacetimes. 

In spite of these remarks, the value of any new method for studying
such a venerable body of knowledge as trigonometry could be
considered with some scepticism, especially if this method were only
applicable to the family of spaces we have discussed in this paper.
This is definitely not the case and we feel that the main value
of this paper is to display in this simplest case the potentialities
this group-theoretical approach to trigonometry has for the
study of many other interesting spaces whose trigonometry is
still unknown. 

Along this line, the trigonometry of complex, quaternionic and
octonionic type CK spaces will be discussed in a companion
forthcoming paper \cite{HermTrig}; only the complex case is
really relevant, as the others reduce directly to the complex
case. The CK family of complex spaces has as their `elliptic'
member $\k_1>0, \k_2>0$ the hermitian elliptic space, which
coincides with the complex projective space
$\Ce P^2$. Therefore the mathematical structure underlying the
quantum state space appears as a particular space in the complex
CK family. Thus, and although primarily mathematical, the study
of trigonometry in hermitian complex spaces has a very direct and
deep  connection with physics, the link being the geometry of the
quantum space of states. Unlike the real spherical or hyperbolic
trigonometry, this `hermitian' trigonometry is   not widely known
yet, though several sets of basic equations in the  elliptic and
hyperbolic cases are established \cite{Hsiang},
\cite{Coo}--\cite{RosGeoLieGroups}. The method we have developed
here  affords a new overall view to hermitian trigonometry
\cite{HermTrigBoya, HermTrigBurgos} and leads to a description of
triangles by sides, angles, lateral phases and angular phases,
which exhibits explicitly the duality in the whole family and leads
to self-duality in the `elliptic' case.  This self-duality
is completely hidden in the equations derived by
\cite{Hsiang,ShiPetRoz,  Bre, RosGeoLieGroups},  mainly due to their
choice for the basic trigonometric quantities. 

These new equations can be taken as a starting point  allowing
the complete exploration of the still rather unknown whole zoo of
hermitian trigonometric equations, at a very modest price. The
essential resemblance between the method for the real and complex
spaces makes the exploration of the new territory easier by
consciously exploiting the analogies, while at the same time
pointing out the differences with the familiar real trigonometry.

The next natural objective is the study of trigonometry of higher
rank grassmannians, either real, complex or quaternionic. This is
still largely unknown in the general case (see however
\cite{Hangan}). We hope the method outlined in this paper should
be able to produce in a direct  form the equations of trigonometry
for grassmannians, thus making a further step towards a general
approach to trigonometry of any symmetric homogeneous space.

This goal will  require first to group all symmetric homogeneous
spaces into CK families, and then to study trigonometry for each
family. Work on this line is also in progress
\cite{MSClasLieGroups}, and opens the possibility of realizing
{\em all} simple Lie algebras, even the special linear ones,
$SO^*(2n)$, $SU^*(2n)$ and the exceptional ones, as `unitary'
algebras, leaving invariant an `hermitian' (relative to some
antiinvolution)  form over a tensor product of two
pseudo-division algebras. This realization should allow a quick
checking on whether or not some extension of the ideas outlined
here afford the equations of trigonometry for any homogeneous
space in an explicit and simple enough way. In this area only
very few results are known; for instance,  the trigonometry of
the rank-two spaces $SU(3)$ and  $SL(3, \Ce)/SU(3)$ is discussed
by \cite{Aslaksen1, Aslaksen2}, and these  rely heavily on the
use of the Weyl theorems on invariant theory and
characterization of invariants by means of traces of products of
matrices. A purely group theoretical derivation would be
interesting. 


\section*{Acknowledgments}
\label{Sec:Ack}

This work was partially supported  by  DGICYT, Spain  (Project
PB94/1115) and Junta de Castilla y Le\'on, Spain  (Project  
CO2/399).


\section*{Appendix: Some relations for the trigonometric 
functions}
\label{Sec:Ap}
\setcounter{equation}{0}
\renewcommand{\theequation}{A.\arabic{equation}}

The main  identities for the trigonometric functions defined in
 (\ref{eq:OrthoCKCosine})--(\ref{eq:OrthoCKTangentVersine}) depending on
a (curvature) label
$\k$ and involving one or two arbitrary arguments
$x$,  $y$ are given by \cite{Vulpi}:
\be
 C_{\k}^2(x)+\k S_{\k}^2(x)=1
\label{eq:CKCosSinIdentity}
\ee
\be
 C_{\k}(x)=1-\k V_{\k}(x)
\label{eq:CKCosVersineRel}
\ee
\bea
&&C_\k(2x) = C^2_\k(x)- \k S^2_\k(x)\label{eq:CKCosTwice}\\
&&  S_\k(2x)
=2 S_\k(x) C_\k(x)
\label{eq:CKSinTwice}
\eea
\bea
&& C^2_\k\biggl(\frac{x}{2}\biggr)=\frac {{ C_\k(x)+1}}{ 2}
\label{eq:CKCosHalf}\\
&& S^2_\k\biggl(\frac {x}{2}\biggr)
=\frac {{1- C_\k(x)}}{ 2\k}=\frac {1}{ 2} V_\k(x) \label{eq:CKSinHalf}\\
&& T_\k\biggl(\frac {x}{ 2}\biggr)
=\frac{{1- C_\k(x)}}{ \k S_\k(x)}=
\frac {{ S_\k(x)}}{{ C_\k(x)+1}}
\label{eq:CKTanHalf}
\eea
\bea
&& C_\k(x\pm y) = C_\k(x) C_\k(y)\mp \k S_\k(y) S_\k(x) 
\label{eq:CKCosSumDif}\\
&& S_\k(x\pm y)  = S_\k(x) C_\k(y)\pm  S_\k(y) C_\k(x) 
\label{eq:CKSinSumDif}\\
&& V_\k(x\pm y) = V_{\k}(x)+ V_{\k}(y)-\k  V_{\k}(x) V_{\k}(y)\pm
  S_\k(x) S_\k(y) \label{eq:CKVersineSumDif}\\
&& T_\k(x\pm y) =\frac{{ T_\k(x)\pm  T_\k(y)}}{ {1\mp
\k T_\k(x) T_\k(y)}}
\label{eq:CKTanSumDif}
\eea
\bea
&&C_\k(x)+C_\k(y)=2 C_\k\left(\frac{x+y}2
\right) C_\k\left(\frac{x-y}2\right) \label{eq:CKCosSumToProd}\\
&& C_\k(x)- C_\k(y) =-2\k S_\k\left(\frac{x+y} 2
\right) S_\k\left(\frac{x-y}2\right)\label{eq:CKCosDifToProd}  \\
&&  S_\k(x)\pm  S_\k(y) =2 S_\k\left(\frac{x\pm y}2
 \right) C_\k\left(\frac{x\mp y} 2 \right)\label{eq:CKSinSumDifToProd} \\
&& V_\k(x)+ V_\k(y) =2\left\{ V_\k\left(\frac{x+y}
2 \right)+ V_\k\left(\frac{x-y}2\right) \right.\cr
&&\qquad\qquad\qquad\qquad\qquad
-\left.  \k V_\k\left(\frac{x+y}2\right) V_\k\left(\frac{x-y}2
\right)\right\} \label{zzz:zy} \\
&& V_\k(x)-  V_\k(y) =2 S_\k\left(\frac{x+y}
2 \right) S_\k\left(\frac{x-y} 2 \right) .
\label{zzz:zz}
\eea

Let $x$,  $y$,  $z$ three arbitrary real numbers.  We consider
their half sum $p$ and some  related quantities defined by:
\be
 p=\frac{x+y+z}{2}\quad  p-x=\frac{y+z-x}{2}\quad  
 p-y=\frac{x+z-y}{2} \quad p-z=\frac{x+y-z}{ 2}.   \label{zzz:wa}
\ee
Then we find the following identities involving three arbitrary
arguments:
\bea
&&C_\k(x+y)+ C_\k(z) =2 C_{\k}(p) C_{\k}(p-z)      \label{zzz:wb}\\
&& C_\k(x-y)+ C_\k(z) =2 C_{\k}(p-x) C_{\k}(p-y)   \label{zzz:wc}\\
&& C_\k(x+y)- C_\k(z) =-2\k S_{\k}(p) S_{\k}(p-z)  \label{zzz:wd}\\
&& C_\k(x-y)- C_\k(z) =2\k S_{\k}(p-x) S_{\k}(p-y) \label{zzz:we}
\eea
\bea
&& S_\k(x+y)+ S_\k(z) =2 S_{\k}(p) C_{\k}(p-z)    \label{zzz:wf}\\
&& S_\k(x-y)+ S_\k(z) =2 S_{\k}(p-y) C_{\k}(p-x)  \label{zzz:wg}\\
&& S_\k(x+y)- S_\k(z) =2 S_{\k}(p-z) C_{\k}(p)    \label{zzz:wh}\\
&& S_\k(x-y)- S_\k(z) =-2 S_{\k}(p-x) C_{\k}(p-y) \label{zzz:wi}
\eea
\bea
&& V_\k(x+y)+ V_\k(z) =
2\bigl\{ V_\k(p)+ V_\k(p-z) -\k V_\k(p) V_\k(p-z)\bigr\} 
                                                  \label{zzz:wj}\\
&& V_\k(x-y)+ V_\k(z) =
2\bigl\{ V_\k(p-x)+ V_\k(p-y) -\k V_\k(p-x) V_\k(p-y)\bigr\} 
                                                  \label{zzz:wk}\\
&& V_\k(x+y)- V_\k(z) =2 S_{\k}(p) S_{\k}(p-z)    \label{zzz:wl}\\
&& V_\k(x-y)- V_\k(z) =-2 S_{\k}(p-x) S_{\k}(p-y) \label{zzz:wm}
\eea
\bea
&& C_{\k}(x) S_{\k}(y) =
     C_{\k}(p) S_{\k}(p-z)+ S_{\k}(p-x) C_{\k}(p-y) \label{zzz:wn}\\
&& C_{\k}(x) S_{\k}(y) =
     S_{\k}(p) C_{\k}(p-z)-C_{\k}(p-x) S_{\k}(p-y) \label{zzz:wo}
\eea
\bea
&& S_{\k}(x) S_{\k}(y)
= S_{\k}(p) S_{\k}(p-z)+ S_{\k}(p-x) S_{\k}(p-y)
\label{zzz:wp}\\
 && \k S_{\k}(x) S_{\k}(y)
=- C_{\k}(p) C_{\k}(p-z)+ C_{\k}(p-x) C_{\k}(p-y)
\label{zzz:wq}
\eea
\bea
&& 4 C_{\k}\left(\mifrac x2\right)
C_{\k}\left(\mifrac y2\right) C_{\k}\left(\mifrac z2\right)
 -\left[ 1+ C_{\k}(x)+ C_{\k}(y)+ C_{\k}(z)\right] \cr
&&\qquad\qquad\quad =8\k^2 S_{\k}\left(\mifrac p2\right)
 S_{\k}\left(\mifrac {p-x}2\right) S_{\k}\left(\mifrac {p-y}2\right)
S_{\k}\left(\mifrac {p-z}2\right)  \label{zzz:wwc}\\
&&4 C_{\k}\left(\mifrac x2\right)
C_{\k}\left(\mifrac y2\right) C_{\k}\left(\mifrac z2\right)
+\left[
1+ C_{\k}(x)+ C_{\k}(y)+ C_{\k}(z)\right] \cr
&&\qquad\qquad\quad =
 8 C_{\k}\left(\mifrac p2\right)
 C_{\k}\left(\mifrac {p-x}2\right) C_{\k}\left(\mifrac {p-y}2\right)
C_{\k}\left(\mifrac {p-z}2\right)
\label{zzz:wwd}
\eea
\bea
&& 4\k^2 S_{\k}(p) S_{\k}({p-x}) S_{\k}({p-y}) S_{\k}({p-z})  \cr
&&\qquad\qquad\quad
    = 16 C^2_{\k}\left(\mifrac x2\right) C^2_{\k}\left(\mifrac
y2\right) C^2_{\k}\left(\mifrac z2\right)
         -\bigl[1+ C_{\k}(x)+ C_{\k}(y)+ C_{\k}(z)\bigr]^2     \cr
&&\qquad\qquad\quad = 1- C^2_{\k}(x)- C^2_{\k}(y)- C^2_{\k}(z)
        + 2 C_{\k}(x) C_{\k}(y) C_{\k}(z) \label{eq:CKSinTriSide}\\
&& 4 C_{\k}(p) C_{\k}({p-x}) C_{\k}({p-y}) C_{\k}({p-z}) \cr
&&\qquad\qquad\quad = -1+ C^2_{\k}(x)+ C^2_{\k}(y)+ C^2_{\k}(z)
     +2 C_{\k}(x) C_{\k}(y) C_{\k}(z)   .  \label{zzz:wwh}
\eea

\newpage



\end{document}